\documentclass[twoside,axodraw,epsfig]{book}
\usepackage{axodraw}
\usepackage{epsfig}
\usepackage{latexsym}
\newcommand{\fge}{[\bar{\eta},\eta,J_\mu]}
\newcommand{\bp}{\bar{\psi}}
\newcommand{\fgp}{[\bp,\psi,A_\mu]}

\newcommand{\tbyt}{2\times 2}
\newcommand{\fbyf}{4\times 4}

\newcommand{\etal}{\emph{et. al. }}

\newcommand{\slsh}[1]{{\not \! #1}}
\newcommand{\be}{\begin{equation}}
\newcommand{\ee}{\end{equation}}
\newcommand{\bea}{\begin{eqnarray}}
\newcommand{\eea}{\end{eqnarray}}
\newcommand{\M}{{\cal M}}
\newcommand{\nn}{\nonumber}
\newcommand{\Sf}[1]{S_F(#1)}
\newcommand{\kp}{k \cdot p}

\def\ponfigura#1{
       \centerline{ \psfig{file=#1, angle=-90, width=3.05in}}
	}
\def\sexp{ \mbox{\large{\textsl{e}}} }

\newcommand{\tm}{(t)}
\newcommand{\x}{(x)}
\newcommand{\xp}{(t,-x,y)}
\newcommand{\xnp}{(t,x,y)}
\newcommand{\xti}{(-t,\vec x)}
\newcommand{\xnt}{(t,\vec x)}
\newcommand{\abs}[1]{\left| #1\right|}
\newcommand{\pw}{p \cdot w}
\newcommand{\kw}{k \cdot w}
\newcommand{\ktop}{\stackrel{k\leftrightarrow p}{=}}
\newcommand{\bra}[1]{\left. \langle #1 \right|}
\newcommand{\ket}[1]{\left| #1 \rangle\right.}

\begin{document}
\title{Gauge Invariance and Construction of the Fermion-Boson Vertex in QED3}
\date{September, 2003}
\author{Alfredo Raya Monta\~no}
\maketitle
\chapter*{Preface}
The study of phisics at the level of the fundamental constituents
of the universe has represented a challenge for the human mind.
The clasification of such objects by their properties is very
helpful for understanding the interactions among them.  One of these
properties, their mass, remains as one of the most intriguing,
since, up to date, there is no theory that can explain its
origin.  The manner in which we understand the interactions at the
fundamental level is by the exchange of mediators in the Standar
Model (SM) of Particle Physics.  All  the processes are then
described in terms of Feynman diagrams, and in order to compare
with the experimental results, one must calculate the amplitudes
for a given event, adding the contributions of all the possible ways in
which the event can take place.  It is true that this model
satisfactorily reproduces the dynamics of the particles through
the symmetries of the mathematical objects that describe such
particles, but it assumes that they are massless.  On the other hand,
in the different high energy laboratories around the world,
masses for the fundamentel constituents of the universe have been
measured, in some cases very accurately.  This reflects the fact
of the breaking of the symmetry of the SM.  In order to
conciliate this discrepancy between theory and experiment, the
quest for a self consistent theoretical model and its
experimental corroboration that explains the origin of the
masses of the fundamental blocks of the universe has begun.

The most famous as well as popular of these models is the so-called
Higgs mechanism, which explains that the mass of the particles
emerges by their interaction with the Higgs boson,
a phenomenom that involves the Spontaneous Breaking of the
Chiral Symmetry in the SM.  The search for this boson in  particle
accelerators in the US and in Europe has impulsed the
technological de\-ve\-lop\-ment, due to the technical difficulties
and low budgets to produce and detect it; and the theoretical advance,
which offers simpler channels towards its discovery. At the same
time, it has promoted the increase of the number of students and
scientists dedicated to this branch of physics.  In spite of the
eficiency on which the SM describes the phenomenology of the
fundamental particles, there are still some problems.

The Higgs boson has not  been discovered experimentally yet.  The
SM assumes that this boson is fundamental, but it does not exist
a fundamental scalar in nature, at least, it has not been
observed yet.  On the other hand, when we try to extend the SM to
include some other forces by constructing a Grand Unification
Theory, we face the hierarchy problem, and in order to
solve it, one must perform an unnatural fine tunning of the Higgs
parameters.  Supersymmetry or SUSY takes care of this problem at
the expense of doubling the spectrum of particles.  Fundamental
scalars are still there and SUSY keeps on eluding us.

As an alternative, models have been propossed in which the
exsistence of a scalar particle is no longer necessary for the
symmetry breaking, since it is given in the presence of
condensates.  Some of these models are Technicolor and its
Extended versions, Top Condensate and Top Color.  Predictions
from these models are obtained by non perturbative calculations,
that even in the simplest cases are very hard to perform and they
are based in severly incorrect assumptions.  For example, in the
Top Condensate Models in the context of the Schwinger-Dyson
Equations (SDE)-which we shortly will talk about-, the QDC
corrections are only calculated in Landau gauge.  If one repeats
the same excercise in other gauges, one finds that the top quark mass
depends upon the gauge, which is physically senseless.  Trying to
solve this situation in complicated theories like QCD is a
formidable problem.  In this thesis we discuss similar topics in
a simple model, QED in 2+1 dimensions.

We decided to study the Dynamical Breaking of Chiral Symmetry
in the context of Schwinger-Dyson equations.  These equations are
an infinite set of integral relations among the Green's functions
of a Quantum Field Theory, and they provide the analitic structure
that such functions must have.  Their implementation is extended
to the High Energy Physics as well as Nuclear Physics, therefore,
the solutions for these equations are of interest for a wide
sector of physicists around the world.

Solutions to the SDE are obtained by truncating the infinite
tower of integral equations, and the main task for the
solution of these equations should be to incorporate all of the
gauge identities of the theory in order to calculate gauge parameter
independent physical observables.  To pursue this objective,
a\-na\-ly\-ti\-cal studies, up to where the complexity of the theory
allows, are performed.  Such studies are either based on the use
of ans\"atze for the Greens function or in their spectral
representation.  There are also succesful attempts to describe
the dynamics of the fundamental particles by discretising the
space-time, i. e., assumming a lattice where the dynamics takes
place.  Then, the different processes are calculated with the
unavoidable use of computers, which the more powerful, the more useful
in this context.

Even though there have been attempts to solve the DSE in QCD and
in QED in a four dimensional space-time, such studies face
a difficulty inheret to four-dimensional theories:  Ultraviolet
divergences.  Added to this, the initial assumptions
to truncate the infinite tower of SDE have not allowed to clearly
identify the sources of the gauge dependence of physical
observables, nor the role which play the gauge invariance constraints
in the restauration of the gauge
independence of these quantities, like the Ward Identities for QED
of the Slavnov-Taylor Identities for QCD, or some others, like the
Landau-Khalatnikov-Fradkin transformations in QED, which describe
the precise manner in which the Green's function vary under a
gauge transformation.

Therefore, we decided to focus or research in Quantum
Electrodynamics on a plane, or QED3, where, besides of enjoying of
the benefits which theories of phenomena occurring on planar
surfaces offer, we can identify exactly the role of the initial
assuptions made to solve the SDE and the gauge invariance
constraints, since QED3 lacks of ultraviolet divergences.

The special features of the electromagnetic dynamics on a plane
are stu\-died in Chapter~1, while the derivation of the SDE as well
as the Ward-Green-Takahashi Identities (WGTI) and the
Landau-Khalatnikov-Fradkin (LKF) transformations is carried out in
Chapter~2.  For a better understanding of the phenomenom of the Dynamical
Generation of Fermion Masses, in Chapter~3 we solve the SDE for the Fermion Propagator with
the only assuption that the fermions interact among them in the
simplest known way, that is, assuming that the vertex of the
intaraction is only the bare one.  This allows us to identify the
sources of the gauge dependence for the two relevant
physical observables, the Euclidean mass and the Chiral
Condensate.  This scenario also allows us to study the role of
the WGTI in the restauration of the gauge independence of the
above mentioned physical observables.  The conclusion that the
imposition of the WGTI is a necesary, but not sufficient condition to guarantee the 
gauge independence of the physical observables is translated as
the necessity for the incorporation of other gauge invariance
constraints.  The LKF transformations are the next ingredient to
consider, but their implementation is not as simple as the
WGTI.  In Chapter~4 we describe the manner to implement these
transformations as a requirement that the solutions for the SDE
must fulfill.  Being then only left with the initial
assumption about the vertex, in Chapter~5 we advocated our
attention to remove it by constructing the vertex for the
electromagnetic interaction making use of Perturbation Theory as a
guide.  The main reason for this is that in Perturbation
Theory, gauge identities (WGTI and LKF transformations) and the
gauge independence of physical observables are satisfied order by
order.  Then if our initial assumption for the fermion-boson vertex
reduces to its Feynman expansion in the weak coupling regime, we
stand our best chance to obtain the correct gauge behavior in the
nonperturbative regime.  Our construction is the most ambitious
of its kind, because, besides con\-si\-de\-ring the gauge invariance
constraints of QED3, it deals with massive fermions in the
interaction.  Our vertex provides the first insight of the
nonperturbative interaction, because when we write an effective
Ward identity for the transverse part of the vertex (i.e., a
relation between the transverse vertex and the fermion
propagator), we are left with an interaction explicitly
independend of the electromagnetic coupling.  Consequences of
this construction as well as the possible paths for the
implementation of this vertex in similar studies are discussed in
Chapter~6, where besides we offer our conclusions.  We keep the
hope that the technique we developed can be applied in future to
alternative models to the SM which help us to discover
the origin of masses for the fundamental particles, where more reliable calculations are
required with improved vertex ans\"atze.
\newpage
From this work there have been published the following papers~:
\begin{enumerate}
\item \emph{Constructing the fermion-boson vertex in
three-dimensional QED.} \textsf{A. Bashir y A. Raya}.  Phys.  Rev. {\bf
D64} 105001 (2001).

\item \emph{Gauge dependence of mass and the condensate in
chirally asymmetric phase of QED3.} \textsf{A. Bashir, A. Huet y
A. Raya}. Phys. Rev. {\bf D66} 025029 (2002).

\item \emph{Landau-Khalatnikov-Fradkin transformation and the
fermion propagator in Quantum Electrodynamics}.  \textsf{A.Bashir y A.
Raya}. Phys. Rev. {\bf D66} 105005 (2002).
\end{enumerate}

\vfill
This dissertation was originally written and defended in spanish.

\tableofcontents
\chapter{Quantum Electrodynamics on a Plane}
\pagestyle{myheadings}
\markboth{Quantum Electrodynamics on a Plane}{Quantum Electrodynamics on a Plane}
\section{Introduction}
\noindent

Even though we live in a four-dimensional space-time --three of
its dimensions being spatial and one temporal--, many theories
are also interesting when they are formulated in other
dimensions.  Phenomena like the Dynamical Breaking of Chiral
Symmetry in Quantum Field Theories are more easily understood
when such theories restric their dynamics to less than four
dimensions.  Three-dimensional models are very useful to
understand phenomena that happen on planar surfaces, where
precisely one counts with two spatial dimensions and one
temporal.  They are also helpful to study the high temperature
behavior of four-dimensional theories, since these theories have
as infinite-temperature limit their three-dimensional
counterparts~\cite{ht4d1,ht4d2}.  In such cases, a
three-dimensional theory is interesting not only as an academic
model, but because they have a direct physical interpretation.
QED3 is one of such models.
 This is the
subject of our interest in this thesis and we start from an
introduction to it in this chapter.

In QED3, the phenomenom of Dynamical Mass Generation has been
studied as an alternative mechanism to that of Higgs in order to
explain how fundamental particles acquire such property~(an
excelent review can be found in~\cite{review}).  Although the
Higgs mechanism offers a solution to this question, it is partial
in the sense that it relates particles to their masses, but it
says nothing about the value of these masses.  Besides, if the
Higgs boson is detected in experiments at LEP and/or Tevatron, it
can take a while before determining whether it is fundamental or
not.

The theory we study has a coupling $e^2$ with dimensions of mass,
which allows the dynamically generated mass to be related to this
coupling, and it is no longer necessary to introduce by hand a
cut-off mass scale for this model, since $e^2$ defines this
scale.  Even more, this theory is superrenormalizable, and lacks
of ultraviolet divergences, one of the most serious obstacles to
study physical phenomena in four-dimensional theories.  The order
of divergence of QED3 is
\cite{finap}~:
\be
w=3-f-\frac{1}{2}b-\frac{1}{2}n\;,
\ee
where $f$ is the number of external fermions in a diagram, $b$
the number of external photons and $n$ the perturbative order of
the diagram.  It can be shown that the Green's functions become
finite in the ultraviolet region.  However, it does not mean that
the theory is free from divergences.  In our case, the problem
is that there are infrared divergent integrals, at least in the
massless theory.  That is why one is forced to start from certain
assumptions that remove these divergences to solve the
Schwinger-Dyson Equations (SDE).

Therefore, we have a very interesting model, from which we can
learn pretty much about the Dynamical Generation of Masses and
the analytic structure of the Fermion Propagator in
general~\cite{maris}, but which is mathematically more easily
treatable than in four dimensions.  Results we obtained can serve
as a guide to more complicated theories, like Grand Unification
Theories.

This theory, besides, posseses a direct physical relevance.  QED3
with a dinamically generated mass has been propossed as a model
for two-dimensional superconductivity, specially for the
discovery of superconducting quasiplanar oxides at high
temperature, like $La_2CuO_4$ y $YBa_2Cu_6$~\cite{mavro}.  There
are other models that describe bidemensional superconductivity at
high temperature, like anyon models, but they are problematic,
because they lead to parity violation, which is not observed in
nature.

There are also lattice studies on the structure of the Fermion
and Photon Propagators which allow us to understand the Dynamical
Generation of Masses.  In this kind of studies, the search for
the relevant critical exponents is more easily carried out that
in analytical studies in the continuum.  Also, there are studies
of QED3 at finite temperature, due to the potentially predictive
power of this model to explain early universe phenomena, and it
is also considered as the first step towards the comprehension of
the hadronic structure, since in its unquenched version, i. e.,
considering vacuum polarization effects, the theory exhibits
confinement~\cite{BR1}.  This can be seen in a heuristical manner
by considering the classical potential\footnote{Although the
arguments of Green's functions are the momenta squared, in order
to simplify the notation we will consider that $G=G(p)$.}
\bea
V(x)&\equiv&\int_{-\infty}^\infty dx_0\int\frac{d^3q}{(2\pi)^3}
\sexp^{i(q\cdot x+q_0x_0)}e^2\Delta^T(q)\nn\\
&=&\int\frac{d^2q}{(2\pi)^2}
\sexp^{iq\cdot x}e^2\Delta^T(q)\;,
\eea
where
\be
\Delta^T(q)=\frac{1}{q^2[1+\Pi(q)]}\;
\ee
is related to the transverse part of the photon propagator as~:
\be
\Delta_{\mu\nu}^T(q)=\left( -g_{\mu\nu}+\frac{q_\mu
q_\nu}{q^2}\right)\frac{\Delta^T(q)}{q^2}\;.
\ee
Setting $\Pi(q)=0$, one obtains
\be
V(x)=\frac{e^2}{2\pi} \ln{e^2 x}\;,
\ee
which is a logarithmically confinement potential.  $e^2$ is
obviously the electromagnetic coupling.

In this chapter we will show the dynamics of the
electromagnetically in\-te\-rac\-ting objects on a plane.  We start by
explicitly showing the three-dimensional equivalent to the
Lorentz Group.  Next we study the Dirac equation explicitly,
using $\tbyt$ and $\fbyf$ representations for the Dirac matrices.
Then, we study the chiral symmetry, that we will be considering
throughout the next chapters, and the discrete symmetries of
QED3.  In the end, we shall discuss why we prefer to study QED3
in its representation $4\times 4$ in this thesis.

\section{Lorentz Group}
\noindent
Lorentz Group can be represented by $3\times 3$ matrices, and there are 9 parameters to classify
them. However, the relation
\be
\Lambda_\alpha^\mu g_{\mu\nu}\Lambda^\nu_\beta =g_{\alpha\beta}
\ee
fixes 6 of these parameters, leaving only three, corresponding to two boosts, each along the corresponding axis, and one rotation.

Explicit representation for the boosts are~:
\bea
B_1(\gamma)&=& \left( \begin{array}{ccc}
\gamma &\gamma v &0\\  \gamma v& \gamma & 0\\
0& 0& 1\end{array}\right),\\ \nn
 B_2(\gamma)&=&\left( \begin{array}{ccc}
 \gamma & 0 & \gamma v
 \\  0& 1 & 0\\
 \gamma v& 0& \gamma  \end{array}\right).
\eea
Obviously $\gamma=\sqrt{1-v^2}$. To obtain the infinitesimal generators, we perform the parametrization
\be
 \gamma =\cosh{\psi}.
\ee
Then,
\bea
 B_1 (\psi)&=& \left( \begin{array}{ccc}
 \cosh{\psi} & \sinh{\psi} & 0\\
\sinh{\psi}& \cosh{\psi} & 0\\
 0& 0& 1\end{array}\right),\\ \nn
 B_2 (\psi)&=& \left( \begin{array}{ccc}
 \cosh{\psi} & 0 & \sinh{\psi} \\
 0& 1 & 0\\
 \sinh{\psi}& 0& \cosh{\psi} \end{array}\right).
\eea
These infinitesimal generators satisfy~:
\be
K_i=-i \left.\frac{d B_i}{d\psi}\right|_{\psi=0}\;,
\ee
which explicitly correspond to the matrices
\bea
 K_1&=&\left(\begin{array}{ccc}
 0& -i & 0\\  -i& 0& 0\\
 0& 0& 0\end{array}\right),\\ \nn
 K_2&=&\left(\begin{array}{ccc}
 0& 0 & -i\\ 0& 0& 0\\
 -i& 0& 0\end{array}\right).
\eea
A rotation on the plane is described by the matrix
\be
R(\theta )=\left( \begin{array}{ccc}
 1& 0& 0\\  0& \cos{\theta} &  \sin{\theta}\\
 0& -\sin{\theta }& \cos{\theta} \end{array}\right).
\ee
This matrix has the infinitesimal generator
\be
J=-i\left.\frac{dR}{d\theta}\right|_{\theta =0}=\left(\begin{array}{ccc}
 0& 0& 0\\ 0& 0& -i\\  0& i& 0\end{array}\right).
\ee
Now, these three generators are in one-to-one correspondence with
the pa\-ra\-me\-ters of the Lorentz Group.  They obey the following
commutation relations~:
\bea
\left[ K_1, K_2 \right] &=& i J \nn\\
\left[ J , K_1 \right] &=& -i K_2 \nn\\
\left[ J, K_2 \right] &=& +i K_1 \;.\label{lorentzcr}
\eea
As it is well known~(see, for instance, \cite{ryder}), pure
Lorentz transformations do not for a group.  Let us observe that
$J$ is a Hermitian matrix, but neither $K_1$ nor $K_2$ are.
However, with the definitions
\be
A_1= iK_1,\hspace{1cm}A_2=-iK_2\hspace{1cm}A_3=J,
\ee
relations~(\ref{lorentzcr}) can be re-written as~:
\be
[A_i,A_j]=i\epsilon_{ijk}A_k,\;,
\ee
where
\begin{eqnarray*}
\epsilon_{ijk}&=&\left\{
\begin{array}{cl}
+1& \mbox{if} \;(ijk) \; \mbox{is an even permutation of
(123)}\\
-1 &\mbox{if} \;(ijk) \; \mbox{is an odd permutation of
(123)}\\
0 &\mbox{else.}\end{array}\right.
\end{eqnarray*}
Therefore, we conclude that the Lorentz Group on a plane is isomorphic to the $SU(2)$ group.
 Owing to the reduced number of degrees of freedom
on a plane, Dirac equation can also be studied in its matrix representation
of dimensions less than 4. We discuss this matter in the next section.

\section{Dirac Equation and the Lagrangian}
In order to solve the Dirac Equation, let us recall that Paul Adrien Maurice's idea~(see, for instance~\cite{hanmar}) was precisely to write down a Hamiltonian with linear dependence on $\partial_0$, and consequently on $\partial_i$, in such a way that its most general form was
\be
H\psi =(\alpha\cdot P +\beta m)\psi\;,
\ee
with the requirement that it must fulfill the relativistic relation
\be
H^2\psi = (P^2+m^2)\psi.
\ee
Dirac found that $\alpha$ and $\beta$ are no numbers, but matrices with the following characteristics~:
\begin{itemize}
\item $\alpha_1$, $\alpha_2$ y $\alpha_3$ commute with each other.

\item $\alpha_1^2=\alpha_2^2=\alpha_3^2=\beta^2=1$.
\end{itemize}
Dirac equation can be written then as~:
\be
(i\slsh{\partial} - m)\psi=0\;,\label{dirac}
\ee
with $\slsh{\partial}=\gamma^\mu\partial_\mu$, where
\be
\gamma^\mu\equiv (\beta,\beta\alpha).
\ee
These matrices satisfy the Clifford's algebra~:
\be
\{\gamma^\mu,\gamma^\nu\}=2g^{\mu\nu}.
\ee
Once the representation for the $\gamma$ matrices has been chosen, one can proceed to solve the Dirac equation.

\subsection{$\tbyt$ Representation}
In QED3, only three $\gamma$ matrices are necessary to write down
the Dirac Equation.  The problem is that these matrices should be
trated in a special manner when the number of dimensions of the
space-time is other than four.  In the three-dimensional case, a
basis for the Clifford's algebra is given by the
monimials~\cite{muta}~:
\be
1,\: \gamma^0,\:\gamma^1,\:\gamma^2,\:
\gamma^0\gamma^1,\:\gamma^0\gamma^2,\:
\gamma^1\gamma^2,\:\gamma^0\gamma^1\gamma^2\;.
\ee
If $\Gamma_i$ represents any element of this set of monomials, it follows that the matrix
$\gamma=\gamma^0\gamma^1\gamma^2$ satisfies
\be
[\gamma ,\Gamma_i ]=0\;,
\ee
Then, by Schur's lemma,
\be
\gamma =cI\;,
\ee
where $c$ is a constant. We also have that
\be
(\gamma )^2=-1\;,
\ee
therefore,
\be
c=\pm i\;.
\ee
We can see that we have two inequivalent representations for the Clifford's algebra, depending upon the choice of the two signs for $\gamma$. In the case of $\tbyt$ matrices, Pauli's $\sigma$ matrices represent the Dirac matrices. We start with the metric
\be
g^{\mu\nu}=\left(\begin{array}{ccc}
 1 &  0 &  0\\
 0 &  -1 &  0\\
 0&  0&  -1\end{array}\right)
\ee
and proceed to identify~\cite{finap}~:
\be
\gamma^0=\sigma_3, \hspace{1cm} \gamma^1=i\sigma_1,\hspace{1cm}
\gamma^2=i\sigma_2\;.
\ee
The (anti-)commutation relations among these matrices are then~:
\bea
 \left\{\gamma^\mu,\gamma^\nu \right\}&=& 2g^{\mu\nu}\;, \\
 \gamma^\mu\gamma^\nu&=& g^{\mu\nu}-i\epsilon^{\mu\nu\alpha}
\gamma_\alpha\;.
\eea
These relations imply
\bea
 \gamma_\mu\gamma^\mu&=& 3\;,\\
 \gamma_\mu \slsh{p}\gamma^\mu&=& -\slsh{p}\;,\\
 \epsilon_{\mu\nu\alpha}\epsilon^{\beta\gamma\alpha}&=& \delta_\mu
^\beta \delta_\nu^\gamma-\delta_\mu^\gamma\delta_\nu^\beta\;,\\
 \epsilon^{\alpha\beta\gamma}\epsilon_{\alpha\beta\gamma}&=& 3\;,
\eea
and we have the traceology~:
\bea
 Tr\left[\gamma^\mu \right]&=& 0\;,\\
 Tr\left[\gamma^\mu\gamma^\nu \right]&=& 2g^{\mu\nu}\;,\\
 Tr\left[\gamma^\mu\gamma^\nu\gamma^\rho\right]&=& -2i\epsilon^
{\mu\nu\rho}\;,\\
 Tr\left[\gamma^\mu\gamma^\nu\gamma^\rho\gamma^\sigma\right]&=& 2
\left(g^{\mu\nu}
g^{\rho\sigma}-g^{\mu\rho}g^{\nu\sigma}+g^{\mu\sigma}g^{\nu\rho}
\right)\;.
\eea
Let us notice that the Lorentz indices run from 0 to 2. In our case
\be
\gamma^0=\beta=\sigma_3,\hspace{1cm}\gamma^1=\beta\alpha_1=i\sigma_1,
\hspace{1cm}\gamma^2=\beta\alpha_2=i\sigma_2.
\ee
To solve the equation, let us proceed in the standard way, that is, let us consider first the particle at rest. The corresponding equation reads
\begin{eqnarray*}
 [i\gamma^0\partial_0 -m ]\psi(t)&=& 0\;,\\
\left[  i\left(\begin{array}{cc}
1& 0\\  0& -1
\end{array}\right)
\partial_0 -m\left(
\begin{array}{cc}
1& 0\\ 0& 1\end{array}
\right)\right]
\left(\begin{array}{c}
\psi_A(t)\\
 \psi_B(t)\end{array}\right)&=&  0\;,
\end{eqnarray*}
or,
\be
\left(\begin{array}{c} \frac{\partial\psi_A(t)}{\partial t}\\
-\frac{\partial\psi_B(t)}{\partial t}
\end{array}\right)=
 -i m\left(
\begin{array}{c}
 \psi_A(t)\\  \psi_B(t)
\end{array}\right) \;.
\ee
So, we find the solutions~:
\be
\psi_A(t)=\sexp^{-i m t}\psi_A(0),\hspace{1cm}\psi_B(t)=\sexp^{i m
t}\psi_B(0)\;,
\ee
such that the wavefunction can be written as~:
\bea
 \psi(t)&=& \left(\begin{array}{c}
 \sexp^{-i m t}\psi_A(0)\\  \sexp^{i m t}\psi_B(0)
\end{array}\right)\;,
\eea
with
\bea
 \psi(0)&=& \left(\begin{array}{c}
 \psi_A(0)\\  \psi_B(0)
 \end{array}\right)\;.
\eea
If we look for independent solutions of the form
\be
\psi_1(0)=
\left(\begin{array}{c} 1\\  0\end{array}\right),\hspace{1cm}
\psi_2(0)=\left(\begin{array}{c} 0\\  1\end{array}\right),
\ee
we find that the solutions for a particle at rest are~:
\bea
 \psi_1(t)=\left(\begin{array}{c} 1\\  0\end{array}
\right)  \sexp^{-i m t},
\hspace{1cm}E>0,\\ \nn
 \psi_2(t)=\left(\begin{array}{c} 0\\  1\end{array}\right)
 \sexp^{i m t},
\hspace{1cm}E<0.
\eea
For a moving particle, we look for solutions of the form
\be
\psi (x)=\sexp^{-i x\cdot p}u(p)\;.
\ee
Let us observe that
\be
\partial_\mu\psi (x)=\partial_\mu [\sexp^{-i x\cdot p}]u(p)=-i p_\mu
\sexp^{-i x\cdot p}u(p)=-i p_\mu \psi (x)\;,
\ee
therefore, we can write the Dirac equation as follows~:
\be
(\gamma^\mu p_\mu-m)u(p)=0\;.
\ee
Since
\bea
 \gamma^\mu p_\mu&=& \gamma^0 p_0-\vec \gamma\cdot \vec p\nn\\
&=&\left(\begin{array}{cc} E& -(p^2+i p^1)\\ \
 p^2-i p^1&  -E\end{array}\right) \;,
\eea
in matrix form we have
\be
\left(\begin{array}{cc} E-m& -(p^2+i p^1)
\\  p^2-i p^1& -E-m\end{array}\right)
\left(\begin{array}{c} u_A(p)\\  u_B(p)
\end{array}\right)=0\;,
\ee
which leads us to the following system of equations for the spinor componets~:
\bea
 (E-m)u_A(p)-(p^2+i p^1)u_B(p)&=&  0\;,\nn \\
 (p^2-i p^1)u_A(p)-(E+m)u_B(p)&=& 0\;.
\eea
To solve this system, we choose $u_A(p)=1$, such that
\bea
 u_B(p)&=& \left[ \frac{p^2-i p^1}{E+m}\right]\;,\\ \nn
 \Rightarrow \psi_1(x)&=& \left(\begin{array}{c} 1\\
 \frac{p^2-i p^1}{E+m}
\end{array}\right) \sexp^{-i p\cdot x}\;.
\eea
This is the positive energy solution. If now we choose
 $u_B(p)=1$,
\bea
 u_A(p)&=& \left[ \frac{p^2+i p^1}{E-m}\right]\;,\\ \nn
 \Rightarrow \psi_2(x)&=& \left(\begin{array}{c}
 \frac{p^2+i p^1}{E-m}\\ 1
\end{array}\right) \sexp^{-i p\cdot x}\;,
\eea
which is the negative energy counterpart.

\subsection{$\fbyf$ Representation}
\noindent
In this case, we choose the Weyl or chiral representation for the $\gamma$ matrices~\cite{hanmar}~:
\be
\gamma^0=\left(\begin{array}{cc} \sigma_3& 0\\
 0& -\sigma_3\end{array}\right)\;,
\hspace{1cm}
\gamma^1=\left(\begin{array}{cc} i\sigma_1& 0\\
 0& -i\sigma_1\end{array}\right)\;,
\hspace{1cm}
\gamma^2=\left(\begin{array}{cc} i\sigma_2& 0\\
 0& -i\sigma_2\end{array}\right)\;.
\ee
To solve the Dirac equation, once again we start with the equation for the particle at rest, which reads
\bea
\left(\left( \begin{array}{cc} i\sigma_3& 0\\
0& -i\sigma_3\end{array}\right)
 \partial_0-m\left(\begin{array}{cc}
 1& 0\\  0& 1\end{array}\right) \right)
\left(\begin{array}{c} \psi_A\tm\\  \psi_B\tm
\end{array}\right)&=&  0\;,\\ \nn
\left(\begin{array}{cc} i\sigma_3\frac{\partial}{\partial t}-mc&0\\
 0& -i\sigma_3\frac{\partial}{\partial t} -mc\end{array}\right)
 \left(
\begin{array}{c} \psi_A\tm\\  \psi_B\tm\end{array}\right)&=&  0\;,
\eea
and that leads su to the following system of equations for the spinors~:
\bea
 i\sigma_3\frac{\partial\psi_A}{\partial t}&=& m\psi_A\tm\\ \nn
 i\sigma_3\frac{\partial\psi_B}{\partial t}&=& -m\psi_B\tm\;.
\eea
If we decompose the spinors into their components in the form~:
\be
\psi_A\tm=\left(\begin{array}{c} \psi_{A_1}\tm\\
\psi_{A_2}\tm\end{array}\right)\;,
\hspace{1cm}
\psi_B\tm=\left(\begin{array}{c} \psi_{B_1}\tm\\
\psi_{B_2}\tm\end{array}\right)\;,
\ee
we obtain now the following system of equations~:
\bea
 i\frac{\partial\psi_{A_1}\tm}{\partial t}&=& -m\psi_{A_1}\tm\;,\\ \nn
 i\frac{\partial\psi_{A_2}\tm}{\partial t}&=& m\psi_{A_2}\tm\;,\\ \nn
 i\frac{\partial\psi_{B_1}\tm}{\partial t}&=& m\psi_{B_1}\tm\;,\\ \nn
 i\frac{\partial\psi_{B_2}\tm}{\partial t}&=& -m\psi_{B_2}\tm\;,
\eea
whose solutions are
\bea
 \psi_{A_1}(t)&=& \sexp^{-i m t}\psi_{A_1}(0)\;,\\ \nn
 \psi_{A_2}(t)&=& \sexp^{i m t}\psi_{A_2}(0)\;,\\ \nn
 \psi_{B_1}(t)&=& \sexp^{i m t}\psi_{B_1}(0)\;,\\ \nn
 \psi_{B_2}(t)&=& \sexp^{-i m t}\psi_{B_2}(0)\;.
\eea
Then, the wavefunction $\psi(t)$ can be written as
\be
\psi(t)=\left(\begin{array}{c}  \sexp^{-i m t}\psi_{A_1}(0)\\
 \sexp^{i m t}\psi_{A_2}(0)\\  \sexp^{i m t}\psi_{B_1}(0)\\
 \sexp^{-i m t}\psi_{B_2}(0)\end{array}\right)\;.
\ee
We can, as we did before, choose independent solutions such that the particle at rest can be described as
\bea
 \psi_1(t)&=& \left(\begin{array}{c}
 1\\  0\\  0\\  0
\end{array}\right) \sexp^{-i m t}\;,\hspace{1cm}
 \psi_2(t)\;=\; \left(\begin{array}{c}
 0\\  1\\  0\\  0
\end{array}\right) \sexp^{i m t}\;,\\ \nn
 \psi_3(t)&=& \left(\begin{array}{c}
 0\\  0\\  1\\  0
\end{array} \right)	 \sexp^{i m t}\;,\hspace{1cm}
\; \psi_4(t)\;=\;\left(\begin{array}{c}
 0\\  0\\  0\\ 1
 \end{array}\right) \sexp^{-i m t}\;.
\eea
When the particle is moving, the Dirac equation is expressed as~:
\be
\left(\begin{array}{cc} \sigma_3-i\sigma^1p^1-i\sigma_2p^2-m& 0\\
 0& -\sigma_3+i\sigma^1p^1+i\sigma_2p^2-m\end{array}\right)
 \left( \begin{array}{c} u_A(p)\\  u_B(p)\end{array}\right)=0\;,
\ee
from where we obtain the system of equations
\bea
 [\sigma_3-i\sigma^1p^1-i\sigma_2p^2-m]u_A(p)&=&  0\;,\\ \nn
 [-\sigma_3+i\sigma^1p^1+i\sigma_2p^2-m]u_B(p)&=&  0\;.
\eea
Once more, if we decompose the spinors into their components
\be
u_A(p)=\left(\begin{array}{c} u_{A_1}(p)\\
u_{A_2}(p)\end{array}\right)\;,
\hspace{1cm}
u_B(p)=\left(\begin{array}{c} u_{B_1}(p)\\  u_{B_2}(p)\end{array}
\right)\;,
\ee
we arrive to the system of equations~:
\bea
 (E-m)u_{A_1}(p)-(p^2+i p^1)u_{A_2}(p)&=&  0\;,\\ \nn
  (p^2-i p^1)u_{A_1}(p)-(E+m)u_{A_2}(p)&=&  0\;,\\ \nn
 -(E+m)u_{B_1}(p)-(p^2+i p^1)u_{B_2}(p)&=&  0\;,\\ \nn
 -(p^2-i p^1)u_{B_1}(p)-(E-m)u_{B_2}(p)&=&  0\;.
\eea
To solve such system, we take advantage of the consequences of choosing a particular normalization for the spinors components, as displayed below~:
\begin{itemize}
\item $ u_{A_1}(p)=1\Rightarrow u_{A_2}(p)=\frac{p^2-i p^1}{E+m}\;,$\\
\item $ u_{A_2}(p)=1\Rightarrow u_{A_1}(p)=\frac{p^2+i p^1}{E-m}\;,$\\
\item $ u_{B_1}(p)=1\Rightarrow u_{B_2}(p)=\frac{p^2-i p^1}{E-m}\;,$\\
\item $ u_{B_2}(p)=1\Rightarrow u_{B_1}(p)=\frac{p^2+i p^1}{E+m}\;.$
\end{itemize}
Therefore, we have the independent solutions
\bea
 u_1(p)&=& \left(\begin{array}{c}
  1\\  \frac{p^2-i p^1}{E+m}\\
 0\\  0\end{array}\right)\;,\hspace{1cm}
 u_2(p)\;=\; \left(\begin{array}{c}  \frac{p^2-i p^1}{E+m}\\  1\\
 0\\  0\end{array}\right) \;,\\ \nn
 u_3(p)&=& \left(\begin{array}{c} 0\\  0\\  1\\
 \frac{p^2-i p^1}{E-m}\end{array}\right) \;,\hspace{1cm}
 u_4(p)\;=\; \left(\begin{array}{c} 0\\  0\\
 \frac{p^2+i p^1}{E+m}\\  1\end{array}\right) \;.
\eea
Since the solutions are decoupled, we can merge the corresponding solutions for positive and negative energy as
\bea
 u_P(p)&=& \left(\begin{array}{c} 1\\  \frac{p^2-i p^1}{E+m}\\
   \frac{p^2+i p}{E+m}\\  1\end{array}\right)\\ \nn
 u_N(p)&=& \left(\begin{array}{c} \frac{p^2+i p^1}{E-m}\\ 1\\ 1\\
 \frac{p^2-i p}{E-m}\end{array}\right) \;,
\eea
where we understand
\be
u_P(p)=\left(\begin{array}{c}u_P^1\\u_P^2\end{array}
\right)\;,\hspace{1cm}
u_N(p)=\left(\begin{array}{c}u_N^1\\u_N^2\end{array}
\right)\;.
\ee
In this way we represent the solutions to the Dirac equation.

\section{Discrete Symmetries}
\noindent
We are now in position to discuss the discrete symmetries of Dirac equation. For that purpose, we will follow the work of~\cite{japon}, where we use only the $\tbyt$ representation for the Dirac matrices. We should first introduce two Dirac fields $\psi$ and $\phi$ which satisfy, respectively,
\bea
  (i\gamma^\alpha\partial_\alpha-\gamma\partial_2-m)\psi\x&=& 0\;,\\ \nn
  (-i\gamma^\alpha\partial_\alpha+\gamma\partial_2-m)\phi\x&=& 0\;,
\eea
where the index  $\alpha$ runs from 0 to 1 and
$\gamma=\gamma^0\gamma^1\gamma^2$ .  In order to study the charge conjugation operation $C$, we must write the Dirac equation when the field $\psi$ interacts with an external magnetic field~:
\be
\{i\gamma^\alpha\partial_\alpha-\gamma\partial_2
-e(\gamma^\alpha A_\alpha+i\gamma A_2)-m\}\psi\x=0\;.
\ee
It follows that, under charge conjugation,
\be
\{i\gamma^\alpha\partial_\alpha-\gamma\partial_2
+e(\gamma^\alpha A_\alpha+i\gamma A_2)-m\}\psi^C\x=0\;.
\ee
On the other hand, we have
\be
\{i\gamma^\alpha\partial_\alpha-\gamma\partial_2
+e(\gamma^\alpha A_\alpha+i\gamma A_2)-m\}C\bar{\psi}^T\x=0\;,
\ee
where $C$ in the charge conjugation matrix and $T$ denotes the matrix transpose operation. Comparing these equation, we obtain that
\be
\psi^C\x=C\bar{\psi}^T\x\;,
\ee
which is the usual charge conjugation operation.

Parity $P$ in three dimensions corresponds to the inversion of one axis, say $x$, because the inversion of both the axis can be obtained from a rotation of $\pi$ of the plane. Dirac equation modified by a Parity transformation then reads~:
\be
(i\gamma^0\partial_0-i\gamma^1\partial_1
-\gamma\partial_2-m)\psi^P\xp=0\;.
\ee
For the field $\phi$, we rewrite the corresponding Dirac equation without performing the Parity operation as follows~:
\be
(i\gamma^0\partial_0+i\gamma^1\partial_1+\gamma\partial_2-m)
\phi\xnp=0\;.
\ee
Multiplying by $1=\gamma^0\gamma^0$ those terms with
$\gamma^1$ and $\gamma$, our equation then reads~:
\bea
  (i\gamma^0\partial_0+i\gamma^1\gamma^0\gamma^0\partial_1+
\gamma\gamma^0\gamma^0\partial_2-m)\phi\xnp&=& 0\;,\\ \nn
  (i\gamma^0\partial_0-i\gamma^0\gamma^1\gamma^0\partial_1-
\gamma^0\gamma\gamma^0\partial_2-m)\phi\xnp&=& 0\;,\\ \nn
  (i\gamma^0\gamma^0\partial_0-i\gamma^0\gamma^1\partial_1-
\gamma^0\gamma\partial_2-m\gamma^0)\gamma^0\phi\xnp&=& 0\;,\\ \nn
  (i\gamma^0\partial_0-i\gamma^1\partial_1-
\gamma\partial_2-m)\gamma^0\phi\xnp&=& 0\;,\\
\eea
such that, under a comparison,
\be
\psi^P\xp=\gamma^0\phi\xnp\;. \label{parity}
\ee
It is convenient at this stage to take a closed look at the mass terms of QED3. If we write the four-dimensional spinors as
\be
\psi=\left(\begin{array}{c} \psi_1\\
\psi_2\end{array}\right)\;,
\ee
a Parity transformation acts on them in the following way~:
\be
\psi_1\rightarrow \sigma_1\psi_2\;,
\hspace{1cm} \psi_2\rightarrow \sigma_1\psi_1\;.
\ee
Obviously, the ordinary mass term is invariant under this transformation,
\be
m\bar{\psi}\psi=m(\psi_1^\dagger\psi_2+\psi_2^\dagger\psi_1)\;,
\ee
since Pauli matrices are unitary and hermitian. The other mass term
\be
\tilde{m}\bar{\psi}\frac{1}{2}[\gamma^3,\gamma^5]\psi=\tilde{m}
(\psi_1^\dagger\sigma_3\psi_1+\psi_2^\dagger\sigma_3\psi_2)\;,
\ee
is invariant under chiral transformations~(\ref{chitrans}), but not under Parity,~(\ref{parity}).

For the Time Reversal $\tau$, we will need to relate somehow the fields $\psi$ and $\phi$. Dirac equation under a Time Revarsal operation for the field $\psi$ reads~:
\be
(-i\gamma^0\partial_0+i\gamma^1\partial_1-\gamma\partial_2-m)
\psi^\tau\xti=0\;.
\ee
Using the same reasoning as before, the field $\phi$ satisfies~:
\bea
(i\gamma^0\partial_0+i\gamma^1\partial_1-\gamma\partial_2-m)
C\bar{\phi}^T\xnt&=& 0\;,\\ \nn
(-i\gamma\gamma\gamma^0\partial_0-i\gamma\gamma\gamma^1\partial_1
-\gamma\partial_2-m)C\bar{\phi}^T\xnt&=& 0\;,\\ \nn
(i\gamma\gamma^0\gamma\partial_0
+i\gamma\gamma^1\gamma\partial_1-\gamma\partial_2-m)
C\bar{\phi}^T\xnt&=& 0\;,\\ \nn
(i\gamma\gamma^0\partial_0+i\gamma\gamma^1\partial_1
-\gamma\gamma\partial_2+m\gamma
)\gamma C\bar{\phi}^T\xnt&=& 0\;,\\ \nn
(-i\gamma^0\partial_0-i\gamma^1\partial_1+\gamma\partial_2
-m)\gamma C\bar{\phi}^T\xnt&=& 0\;,\\ \nn
(-i\gamma^0\partial_0-i\gamma^1\gamma^0\gamma^0\partial_1
+\gamma\gamma^0\gamma^0\partial_2-m)\gamma
C\bar{\phi}^T\xnt&=& 0\;,\\ \nn
  (-i\gamma^0\partial_0+
i\gamma^0\gamma^1\gamma^0\partial_1
-\gamma^0\gamma\gamma^0\partial_2-m)\gamma
C\bar{\phi}^T\xnt&=& 0\;,\\ \nn
  (-i\gamma^0\gamma^0\partial_0+
i\gamma^0\gamma^1\partial_1-\gamma^0\gamma\partial_2
-m\gamma^0)\gamma^0\gamma
C\bar{\phi}^T\xnt&=& 0\;,\\ \nn
  (-i\gamma^0\partial_0+
i\gamma^1\partial_1-\gamma\partial_2-m)
\gamma^0\gamma C\bar{\phi}^T\xnt&=& 0\;,
\eea
from where we conclude that
\be
\psi^\tau\xti=\gamma^0\gamma C\bar{\phi}^T\xnt\;.
\ee
We see that if we require the massive spin-1/2 fields to be invariant under $C$, $P$ and $\tau$ respectively, we need to introduce the field $\phi$ which satisfies Dirac equation with the second representation for $\gamma$. It is worth to mention that if originally the lagrangian does not include such field, as in our case, the usual mass term breaks $P$ and $\tau$, but retains $CP\tau$.

Finally, if we consider the transformation
\be
\phi\rightarrow\phi'=i\gamma\phi\;,
\ee
we observe that $\phi'$ satisfies the same equation as $\phi$, except for the sign for the mass term. The boost along the $y$ axis and the rotation around it for the field $\phi$ correspond to the same transformations for the field $\psi$, but in the opposite direction, due to the sign of
$\gamma$. On the other hand, $\phi'$ behaves as $\psi$ under Lorentz transformations, although with the opposite sign for the mass term. This corresponds to the fact that both the representations of the Clifford's algebra are the same (under transformations of $\gamma$) as the representation of the Lorentz Group. Therefore, we can conclude that if we require invariance under $C$, $P$ and $\tau$, there must exist two fields whith the opposite sign for their mass terms, and under $C$, $P$ and $\tau$, such field are interchanged.
\\
The invariant lagrangian under discrete transformations is given by
\be
{\cal L} =i\bar{\psi}(\gamma^\alpha\partial_\alpha
+i\gamma\partial_2)\psi+m\bar{\psi}\psi
+i\bar{\phi}'(\gamma^\alpha\partial_\alpha+i\gamma\partial_2)\phi'
-m\bar{\phi}'\phi'\;.
\ee

\section{Chiral Symmetry}
\noindent
As we mentioned before, we can choose the $\tbyt$ representation for the $\gamma^\mu$ matrices as the Pauli matrices, and we then use two-dimensional spinors. However, there is no other $\tbyt$ matrix which anticommutes with the $\sigma$ matrices, therefore, the massive theory does not posses a greater symmetry than the massles one~\cite{applequist}. This is an obstacle to define chirality. That is why we use four-dimensional spinors and also those $\gamma^\mu$ matrices from the four-dimesnional space-time. The massless theory in this case is invariat under two chiral-like tranformations~:
\bea
 \psi\rightarrow \sexp^{i\alpha\gamma^3}\psi\;,\\
 \psi\rightarrow \sexp^{i\alpha\gamma^5}\psi\;.\label{chitrans}
\eea
and, therefore, the lagrangian is invariant under a global $U(2)$ symmetry with the generators
\be
1\;,\gamma^3\;,\gamma^5\;,\left[\gamma^3\;,\;\gamma^5\right]\;.
\ee

This symmetry, however, is broken with a mass term of the form
 $m\bar{\psi}\psi$.  Also, a dynamically generated mass will break this symmetry, as in four-dimensions. In QED3 there is another posible mass term which is invariant under Chiral transformations, but not under Parity (as we saw in the last section). Such term has the form
\begin{eqnarray*}
\frac{1}{2}\tilde{m}\bp[\gamma^3\;,\;\gamma^5]\psi\;.
\end{eqnarray*}
We are only considering the typical mass term, since it has been shown, by analyzing the effective potential, that solutions to the SDE with this mass term are energetically preferred
~\cite{applequist,unquenched1}.
Another reason to do so is the conservation of parity in QED3.
As a result, the Lagrangian that we shall be considering in this
thesis is the one we are familiar with in QED4, i.e.,
\be
 {\cal L} =\bar{\psi}( i \gamma^{\mu} \partial_{\mu}
-m  ) \psi - \frac{1}{4} F_{\mu \nu} F^{\mu \nu} - \frac{1}{2 \xi}
(\partial_{\mu} A^{\mu})^2 \;.
\ee
where the notations carry the usual meaning.

These are the main features of Quantum Electrodynamics on a
plane.  To study the phenomenon of Dynamical Mass Generation, the
one we are concerning with, we must firstly study the gauge
structure of QED3, in particular we must know the Schwinger-Dyson
equations in this context, and also we must explore two of the
consequences of gauge covariance of the theory:  the
Ward-Green-Takahashi Identities, which relate Green's functions
among them, and the Landau-Khalatnikov-Fradkin transformations of
these functions, which realize their gauge behavior under a
variation of gauge.  As the Lagrangian does not change its form,
the derivation of the Schwinger-Dyson equations etc. does not
really differ from the one in 3 spatial dimensions.  We take up
these derivations in the next chapter.

\chapter{Schwinger-Dyson Equations and Gauge Invariance Constraints}
\pagestyle{myheadings}
\markboth{Schwinger-Dyson Equations and Gauge Invariance Constraints}
{Schwinger-Dyson Equations and Gauge Invariance Constraints}

%{SDE and Gauge Invariance Constraints}{SDE and Gauge Invariance Constraints}

\section{Introduction}
As in any other theory, the electromagnetic dynamics can be obtained, in arbitrary dimensions, from the lagrangian and its corresponding action
\be
S=\int d^dx {\cal L}(\phi (x),\partial_\mu\phi (x))\;.
\ee
The equation of motion for the field $\phi$ is obtained after impossing the staticity condition on the action
\be
\delta S=0\;.
\ee
From this condition, we obtain the Euler-Lagrange equations for this field
\be
\partial_\mu\frac{\delta{\cal L}}{\delta (\partial_\mu\phi)}-
\frac{\delta {\cal L}}{\delta\phi}=0\;.
\ee
In this chapter we start from the QED action. Functional derivatives of this action lead us to the Schwinger-Dyson Equations (SDE) for the Green's functions. We also repeat the derivation of the Ward-Green-Takahashi Identity (WGTI) and the Landau-Khalatnikov-Fradkin (LKF) transformations, two of the gauge identities of QED necessary in order to ensure that solutions to the SDE in the study of the Dynamical Generation of Masses with a Non Perturbative Vertex reproduce gauge parameter independent physical observables, as they must be.

\section{Electromagnetic Action}
\noindent
Electromagnetic quantization can be obtained either from the canonical formulation or by means of the Feynman's Path Integral. We prefer the later scheme, since the integral formulation allows us to neatly obtain the SDE. Richard's idea was that in order to know the transition amplitude between two quantum states, we must sum over all possible histories in which such transition can take place, i. e.,
\be
U(t_f, x_f\leftarrow t_i,x_i)=\int {\cal D}x{\cal D}p \,\sexp^{i
S}\;,
\ee
where ${\cal D}x$ and ${\cal D}p$ are the measures.

We saw before that the electromagnetic dynamics is obtained from Dirac equation (\ref{dirac})
\begin{eqnarray*}
\left(i \gamma^\mu\partial_\mu-m \right)\psi=0\;.
\end{eqnarray*}
This equation can be obtained from the free lagrangian density
\be
{\cal L}_{\emph{free}}=
\bar{\psi}(i\partial_\mu\gamma^\mu-m)\psi\:.
\ee
We must consider the interaction between fermions and the electromagnetic field, given by the term $e\gamma^\mu A_\mu$. To make sure that the lagrangian density being invariant under global gauge transformations, we must replace the ordinary derivatives by covarian derivatives by using the minimal substitution principle~:
\be
\partial_\mu\to D_\mu=\partial_\mu-i eA_\mu\;.
\ee
In this way, the lagrangian density is expressed as~:
\be
{\cal L}_{\emph{int}}=\bar{\psi}(i \gamma^\mu D_\mu-m)\psi\;.
\ee
We still need to consider the interaction of the magnetic field with itself. This is obtained from the term
\be
-\frac{1}{4}F_{\mu\nu}F^{\mu\nu}\;.
\ee
Gathering terms, QED action is
\be
S\fgp = \int d^dx \left[\sum_{f=1}^N \bp^f (i \gamma^\mu
D_\mu^f-m^f)\psi^f-\frac{1}{4}F_{\mu\nu}F^{\mu\nu}\right]\;,
\ee
where $f$ is a flavor label and $N$ is the number of different fermion flavors. Although we are considering bare quantities, the results we obtain are also valid for renormalizad quantities, provided this process is performed properly. As for the Green's functions, its bareness will be shown explicitly in order to distinguish them from the corresponding complete functions.

Let us turn our atention to the gererating functional
\bea
\hspace{-8mm}
Z[\bar{\eta},\eta,J_\mu]&=&\int d\mu(\bar{\psi},\psi,A)
%\nn\\
%&&\hspace{-19mm}
\sexp^{\left(
i S[\bar{\psi},\psi,A_\mu]\!+\!i\int d^dx\left[\sum_f\left(
\bar{\psi}^f\eta^f\!+\!\bar{\eta}^f\psi^f\right)\!+\!A_\mu J^\mu\right]
\right)}\;,\label{fungen}
\eea
where $\bar{\eta}^f$, $\eta^f$ and $J_\mu$ are the sources for the fermions, antifermions and photons, respectively, and where it is defined, as in~\cite{review}
\be
d\mu (\bar{\psi},\psi, A)=\Pi_f {\cal D}\bar{\psi}^f{\cal D}\psi^f
\Pi_\mu{\cal D}A_\mu\;.
\ee
To complete the operational definition of QED, let us note that the action is invariant under the abelian local transformations
\bea
 \psi (x)\to \psi^\lambda (x)&=& \sexp^{-i e\lambda
 (x)}\psi\;,\nn\\
 \bar{\psi} (x)\to \bar{\psi}^\lambda (x)&=&
\sexp^{i e\lambda (x)}\bar{\psi}\\ \nn
 A_\mu (x)\to A_\mu^\lambda (x)&=&  A_\mu (x)-\partial_\mu
\lambda (x)\label{localgauget}
\eea
where $\lambda(x)$ is an arbitrary scalar function and, for the time being, we have supressed flavor labels. Under such circumstances, the generating functional is senseless, since for every single fields configuration $\{ \bp(x),\psi(x),A_\mu(x)\}$, due to the gauge inavriance, there exists an infinite number of related configurations
$\{\bp^\lambda(x),\psi^\lambda(x), A_\mu^\lambda(x)\}$,  which have the same action
\be
S[\bar{\psi},\psi,A_\mu]=S[\bar{\psi}^\lambda,\psi^\lambda,
A_\mu^\lambda]\;.
\ee
The Grassman integration over $\bp$ and $\psi$ yield the same result, independently of $\lambda(x)$, since the corresponding Jacobian is unity. Then, there is a divergence in the functional integration over the field $A_\mu$. The correct definition of the measure must ensure that the integration over tha gauge field is extended only to inequivalent configurations under gauge transformations.

This problem can be solved by introducing the Fadeev-Popov determinant. The net effect of this procedure in QED is simply to introduce a gauge fixing term in the action. A commonly used choice for this term is
\be
S[\bar{\psi},\psi,A_\mu]\rightarrow S_\xi [\bar{\psi},\psi,A_\mu]=
S[\bar{\psi},\psi,A_\mu]-\frac{1}{2\xi}\int d^dx (\partial_\mu
A^\mu)^2\;,\label{QEDdaction}
\ee
where $\xi$ is the gauge fixing parameter.

\section{Schwinger-Dyson Equations (SDE)}
\noindent
It is known, from some time ago, that from the field equations for a Quantum Field Theory, it can be derived a system of coupled integral equations which relates the Green's functions of such theory among them. This infinite tower of equations is known as the Schwinger-Dyson Equations~\cite{esd}. We will use the integral functionals formulation to derive the SDE following the works on~\cite{review} and~\cite{cdrnotes}.

We start from eq.~(\ref{QEDdaction}) and the gererating functional~(\ref{fungen}). Let us take into account that the fermionic fields $\{\bp,\psi\}$ and their sources $\{ \bar{\eta},\eta\}$
are elements if the Grassman algebra, that is, all of these field anticommute among themselves; and that $A_\mu$ and its source $J_\mu$ are $c$-numbers. We also take the standard notations and conventions, where
\begin{eqnarray*}
{\not \!\! A}=A_\mu\gamma^\mu=g^{\mu\nu}A_\mu\gamma_\nu\;,\quad
\{\gamma^\mu,\gamma^\nu\}=2g^{\mu\nu}\;,
\quad
\mbox{etc}\;.
\end{eqnarray*}
Let us note that for an electron, the physical charge must be $e^{phys}=-e$,
where, by definition, $e=\vert e \vert$ is the magnitude of the charge of the electron.

\subsection{SDE for the Photon Propagator}
\noindent
Let us consider the generating functional~(\ref{fungen}). The generating functional for connected Green's functions ${\cal G}\fge$ is given by
\be
Z[\bar{\eta},\eta,J_\mu]=\sexp^{{\cal G}[\bar{\eta},\eta,J_\mu]}\;.
\label{tLeg}
\ee
To obtain the SDE corresponding to the Photon Propagator we simply use the fact that the functional integral of a total functional derivative vanishes with the apropriate boundary conditions. For example,
\bea
 0&=& \int d\mu(\bar{\psi},\psi,J_\mu)\frac{\delta}{\delta A_\mu (x)}
\sexp^{\left\{i\left(S_\xi [\bar{\psi},\psi, A_\mu]\!+\!\int d^dx\left[
\bar{\psi}^f\eta^f\!+\!\bar{\eta}^f\psi^f\!+\!A_\mu J^\mu\right]\right)
\right\}} \nn\\
&=& \int d\mu(\bar{\psi},\psi,J_\mu)\left\{\frac{\delta S_\xi}
{\delta A_\mu (x)}+J^\mu (x)\right\}
\nn\\
&&\hspace{2mm}
\sexp^{\left\{i\left(S_\xi [\bar{\psi},\psi, A_\mu]+\int d^dx\left[
\bar{\psi}^f\eta^f+\bar{\eta}^f\psi^f+A_\mu J^\mu\right]\right)
\right\}}\\ \nn
&=& \left\{\frac{\delta S_\xi}{\delta A_\mu (x)}\left[-\frac{\delta}
{i\delta \eta},\frac{\delta}{i\delta\bar{\eta}},\frac{\delta}
{i\delta J^\mu}\right]+J^\mu (x)\right\}Z[\bar{\eta},\eta,J_\mu]
\;.\label{zerofoton}
\eea
Differentiating the action~(\ref{QEDdaction}), we immediately
obtain
\be
\frac{\delta S_\xi}{\delta A_\mu (x)}=\left[\partial_\rho\partial^\rho
g^{\mu\nu}-\left(1-\frac{1}{\xi}\right)\partial^\mu\partial^\nu
\right] A_\nu+\sum_fe^f\bar{\psi}^f\gamma^\mu\psi^f\;,
\ee
from where it follows that, after we divide by $Z$, we can write eq.~(\ref{zerofoton}) as~:
\bea
\left[\partial_\rho\partial^\rho
g^{\mu\nu}-\left(1-\frac{1}{\xi}\right)\partial^\mu\partial^\nu
\right]\frac{\delta {\cal G}}{i\delta J^\nu (x)}\;+&&\nn\\
&&\hspace{-5.8cm}
\;\sum_f e^f
\left(\frac{\delta {\cal G}}{\delta\eta^f(x)}\gamma^\mu
\frac{\delta {\cal G}}{\delta\bar{\eta}^f(x)}+\frac{\delta}
{\delta\eta^f(x)}\left[\gamma^\mu\frac{\delta {\cal G}}
{\delta\bar{\eta}^f(x)}\right]\right)
=-J^\mu (x)\;.
\label{maxwellcomp}
\eea
This equation represents a compact form of the non perturbative equivalent to the Maxwell's equations. This is useful for us to obtain an expression for the photon vacuum polarization. Now we can take the Legendre transformation and introduce the generating functional for one-particle irreducible (1PI) Green's functions, $\Gamma[\bp,\psi,A_\mu]$~:
\be
{\cal G}[\bar{\eta},\eta,A_\mu]\equiv i\Gamma [\bar{\psi},
\psi,A_\mu]+i\int d^dx\left[\bar{\psi}^f\eta^f+\bar{\eta}^f
\psi^f+A_\mu J^\mu\right]\;.\label{gtogamma}
\ee
From the Grassman integration, it follows that $Z\fgp$ and consequently
${\cal G}\fge$ depend only on even powers of $\bar{\eta}$ and $\eta$, which in turn implies that setting $\bar{\eta}=\eta=0$ after taking the derivative of ${\cal G}$ (or $Z$), we will have nonvanishing results only for the same number of derivatives with respect to $\bar{\eta}$ and $\eta$. Similarily, in the absence of derivatives with respect to the fermionic fields, it can be seen that only an even number of derivatives of $Z$ and ${\cal G}$ with respect to $J_\mu$ survive when we set $J=0$. From the last expresion, we have
\bea
 A_\mu (x)&=& \frac{\delta {\cal G}}{i\delta J^\mu
 (x)}\;,\hspace{.3cm}
\psi^f(x)=\frac{\delta {\cal G}}{i\delta\bar{\eta}^f(x)}\;,
\hspace{.3cm}\bar{\psi}^f(x)=-\frac{\delta {\cal G}}{i\delta\eta
^f(x)}\:,\\ \nn
 J_\mu (x)&=&  -\frac{\delta\Gamma}{\delta A^\mu (x)}\;,\hspace{.3cm}
\eta^f(x)=-\frac{\delta\Gamma}{\delta\bar{\psi}^f(x)}\;,
\hspace{.3cm}\bar{\eta}^f(x)=\frac{\delta\Gamma}{\delta\psi
^f(x)}\;.\label{relsourfield}
\eea
From here we obtain expressions for  $\bp$, $\psi$ and $A_\mu$ in terms of $\bar{\eta}$, $\eta$ and $J_\mu$ and viceversa; for example,
\begin{eqnarray*}
\bp_\alpha^f(x)=\bp_\alpha^f\fge=i\frac{\delta{\cal
G}\fge}{\delta\eta_\alpha^f}\;,
\end{eqnarray*}
with the spinorial indices explicitly shown. It is now easy to see that, setting $J=0$ after we differentiate $\Gamma$, we will have nonvanishing results only when we have the same number of derivatives of $\bp$ and $\psi$, in analogy with the case of ${\cal G}$. Making use of the expresions~(\ref{relsourfield}), let us take a look at the following term~:
\bea
i \int d^dz\left.\frac{\delta^2{\cal G}}{\delta\eta_\alpha^f(x)
\delta\bar{\eta}_\gamma^h(z)}
\frac{\delta^2\Gamma}{\delta \psi_\gamma^h(z)\delta\bar{\psi}
_\beta^g(y)}\right|_
{\eta=\bar{\eta}=\psi=\bar{\psi}=0}&&\nn\\
&&\hspace{-38mm}
=\int d^dz\left.
\frac{\delta\psi_\gamma^h(z)}{\delta\eta_\alpha^f(x)}
 \frac{\delta\eta_\beta^g(y)}{\delta\psi_\gamma^h(z)}\right|_
 {\eta=\bar{\eta}=\psi=\bar{\psi}=0}\nn\\
&&\hspace{-38mm}
=\left.\frac{\delta\eta_\beta^g(y)}{\delta\eta_\alpha^f(x)}\right|_
{\psi=\bar{\psi}=0}=
\delta_{\alpha\beta}\delta_{fg}\delta^d(x-y)\;.\label{deltasfot}
\eea
Therefore, when the fermionic sources $(\bar{\eta},\eta)$ are null, we can write eq.~(\ref{maxwellcomp}) as~:
\bea
\left.\frac{\delta\Gamma}{\delta A^\mu
(x)}\right|_{\psi=\bar{\psi}=0}&=&\left[
\partial_\rho\partial^\rho g_{\mu\nu}-\left( 1-\frac{1}{\xi }\right)
\partial_\mu\partial\nu\right] A^\nu (x)\nn\\
&&-i\sum_f e ^f
Tr[\gamma_\mu S_F^f(x,x,[A_\mu])]\;,\label{gamaA}
\eea
where we have identified the term
\be
S_F^f(x,y,[A_\mu])=i\frac{\delta{\cal G}}{\delta\eta^f(y)\delta\bar{\eta}^f(x)}
=-i\frac{\delta{\cal
G}}{\delta\bar{\eta(x)}\delta\eta^f(y)}\label{propferrob}
\ee
as the Fermion Propagator of flavor $f$ in an extermal magnetic field $A_\mu$. One of the consequences of eq.~(\ref{deltasfot}) is that the inverse of this Green's function is given by~:
\be
S_F^f(x,y,[A_\mu])^{-1}= \left.\frac{\delta^2\Gamma}
{\delta\psi^f(x)\delta
\bp^f(y)}\right|_{\psi=\bar{\psi}=0}\;.\label{sfDelta}
\ee
Obviously, the complete Green's function for the fermion $S_F(x,y)$ is obtained after setting $A_\mu=0$ in eq.~(\ref{sfDelta}).

To obtain the corresponding SDE for the photon polarization tensor, we only need to act with $\delta/\delta A_\nu(y)$ on eq.~(\ref{gamaA}) and set $J_\mu(x)=0$.  Let us see~:
\bea
\left.\frac{\delta^2\Gamma}{\delta A^\mu(x)\delta
A^\nu(y)}\right|_{A_\mu=\psi=\bp=0}&&\nn\\
&&\hspace{-2.8cm}
=\left[\partial_\rho\partial^\rho g_{\mu\nu}-
\left( 1-\frac{1}{\xi }\right)\partial_\mu\partial\nu\right]
\delta^d(x-y)\nn\\
&&\hspace{-2.8cm}
-i\sum_f e ^f
Tr\left[\gamma_\mu \frac{\delta}{\delta A_\nu(y)}
\left( \left.\frac{\delta^2\Gamma}{\delta\psi^f(x)\delta
\bp^f(x)}\right|_{\psi=\bar{\psi}=0}\right)^{-1}
\right]\;. \label{fpizq}
\eea
The right hand side of this equation can be interpreted in a better way by observing that~:
\bea
\frac{\delta}{\delta A_\nu(y)}
\left( \left.\frac{\delta^2\Gamma}{\delta\psi^f(x)\delta
\bp^f(x)}\right|_{\psi=\bar{\psi}=0}\right)^{-1}&&\nn\\
&&\hspace{-4.8cm}
=-\int d^du\; d^dw
\left( \left.\frac{\delta^2\Gamma}{\delta\psi^f(x)\delta
\bp^f(w)}\right|_{\psi=\bar{\psi}=0}\right)^{-1}\nn\\
&&\hspace{-4.8cm}
\times\frac{\delta}{\delta A_\nu(y)}
\frac{\delta^2\Gamma}{\delta\psi^f(u)\delta\bp^f(w)}
\left( \left.\frac{\delta^2\Gamma}{\delta\psi^f(w)\delta
\bp^f(x)}\right|_{\psi=\bar{\psi}=0}\right)^{-1}\;, \label{fpder}
\eea
an analogous result to
\bea
\frac{d}{dx}[A(x)A^{-1}(x)=I]=&0&=\frac{dA(x)}{dx}A^{-1}(x)
+A(x)\frac{dA^{-1}(x)}{dx}\nn\\
&\Rightarrow&\frac{dA^{-1}(x)}{dx}=-A^{-1}(x) \frac{dA(x)}{dx}
A^{-1}(x)\;,
\eea
which hold for finite-dimensional matrices. Equation~(\ref{fpder}) involves the fermion-boson Vertex of [1-PI]
\be
e ^f\Gamma_\mu^f(x;y,z)=\frac{\delta}{\delta A^\mu (x)}\left.
\frac{\delta^2\Gamma}
{\delta\psi^f(x)\delta\bp^f(y)}\right|_{0=A_\mu=\psi
=\bar{\psi}}\;,\label{vp1pi}
\ee
which should not be confused with the generating functional $\Gamma$. Similarily to the fermionic case, the second derivative of $\Gamma$ with respect to $A_\mu$ generates the inverse of the photon propagator $(\Delta^{-1})^{\mu\nu}(x,y)$ (left hand side of eq.~(\ref{fpizq})). Therefore, from eqs.~(\ref{gamaA}),~(\ref{fpizq}) and~(\ref{fpder}), we obtain the SDE for the inverse of the photon propagator~:
\bea
(\Delta^{-1})^{\mu\nu}(x,y)&=&\left.\frac{\delta^2\Gamma}
{\delta A^\mu (x)\delta A^\nu
(y)}\right|_{A_\mu=\psi=\bar{\psi}=0}\nn\\
&&\hspace{-1.5cm}
=\left[\partial_\rho
\partial^\rho
g_{\mu\nu}-\left( 1-\frac{1}{\xi }\right)\partial_\mu\partial
\nu\right]\delta^d(x-y)+\Pi_{\mu\nu}(x,y)\;,
\eea
where we have identified the photon polarization tensor
$\Pi_{\mu\nu}$~:
\be
\Pi_{\mu\nu}(x,y)=i\sum_f (e ^f)^2\int d^dz_1d^dz_2
Tr[\gamma_\mu S_F^f(x,z_1)\Gamma_\nu^f(y;z_1,z_2)S_F^f(z_2,x)]\;.
\label{Pimunu}
\ee
Making use of the traslational invariance, we can write the photon propagator in momentum space
\be
\Delta_{\mu\nu}(q)=\frac{-g_{\mu\nu}+(q_\mu q_\nu
/(q^2+i\epsilon))}{q^2+i\epsilon}\frac{1}{1+\Pi (q)}-\xi
\frac{q_\mu q_\nu}{(q^2+i\epsilon)^2}\;,
\ee
where, as usual, we define the scalar polarization $\Pi(q)$ as~:
\begin{eqnarray*}
\Pi_{\mu\nu}(q)\equiv
(-g_{\mu\nu}q^2+q_\mu q_\nu)\Pi(q)\;.
\end{eqnarray*}
Let us note that $\Pi(q)$ is independent of the gauge parametre $\xi$ in QED, as a result of current conservation. At the lowest order in Perturbation Theory, we have that $\Pi(q)=0$, as well as
\be
\Gamma_\nu^f(y;z_1,z_2)=\gamma_\nu\delta^d(y-z_1)\delta^d(y-z_2)
\hspace{0.25cm}
 \mbox{y}\hspace{0.25cm}(i \slsh{\partial}-m ^f)S_F^f(x,y)=
 \delta^d(x-y)\;.
\ee
Once we factorized $e^f$, there is no explicit flavor dependence for the proper vertex. We have seen that from the second derivative of the generating functional $\Gamma\fgp$ we obtain the photon and fermion propagators, and from the third one, the proper vertex of the fermion-boson interaction. Generally, higher order derivatives of $\Gamma\fgp$ yield the corresponding proper Green's functions, where the number and type of derivatives yield the number and type of legs in the proper Green's functions. The SDE for the Photon Propagator is shown in Diagram~(1)\footnote{Diagrams were generated with AXODRAW~\cite{axodraw}}.
%ESD para el propagador del fot\'on
\vspace{-1.5cm}
\begin{center}
\SetScale{0.7}
\begin{picture}(500,100)(0,0)
%Propagador completo
\Photon(50,50)(150,50){4}{10}
\PText(145,60)(0)[]{-1}
\CCirc(100,50){5}{}{}
%propagador desnudo
\Photon(200,50)(300,50){4}{10}
\PText(295,60)(0)[]{-1}
%SDE piece
\Vertex(370,50){1}
\ArrowArcn(400,50)(30,180,0)
\ArrowArcn(400,50)(30,0,180)
\Photon(350,50)(370,50){4}{2}
\Photon(430,50)(450,50){4}{2}
\CCirc(430,50){4}{}{}
\CCirc(400,80){4}{}{}
\CCirc(400,20){4}{}{}
%Signos algebraicos
\PText(175,52)(0)[]{=}
\PText(325,52)(0)[]{-}
\end{picture}\\
{\sl Diagram~(1)~: SDE for the Photon Propagator.}
\end{center}
The representation of the SDE in momentum space is immediately obtained by taking the Fourier transform of the expression in coordinate space, or, more easily, using the usual Feynman rules for the diagrams based in the lowest order perturbative contribution to the non perturbative quantities. For example, for the photon polarization tensor we obtain
\bea
\hspace{-1cm}
i\Pi_{\mu\nu}(q)&=&
(-1)\sum_f (e ^f)^2\int\frac{d^dk}{(2\pi )^d}
\nn\\
&&
Tr[(i\gamma_\mu)(i S_F^f(k))(i\Gamma_\nu^f(k,k+q))(i S_F^f(k+q))]\;,
\eea
where the factor $(-1)$ arises from the fermion loop as usual.

\subsection{SDE for the Fermion Propagator}
\noindent
Following a similar procedure, we can derive the integral equation for the Fermion Propagator starting from
\bea
 0&=& \int d\mu
 (\bar{\psi},\psi,A)\frac{\delta}{\delta\bar{\psi}(x)}
 %\nn\\
 %&&\hspace{2mm}
 \sexp^{\left\{
 i\left( S_\xi [\bar{\psi},\psi,A_\mu]+\int
 d^dx[\bar{\psi}^f\eta^f+\bar{\eta}^f\psi^f+A_\mu
 J^\mu]\right)\right\} }\\ \nn
&=& \left\{ \frac{\delta
S_\xi}{\delta\bar{\psi}(x)}\left[-\frac{\delta} {i\delta
\eta},\frac{\delta}{i\delta\bar{\eta}},\frac{\delta}
{i\delta J}\right]+\eta^f (x)\right\}Z[\bar{\eta},\eta,J_\mu]\nn\\
&=&\left[\eta^f(x)+\left(i\slsh{\partial}-m^f+e^f\gamma^\mu
\frac{\delta}{i\delta
J^\mu(x)}\right)\frac{\delta}{i\delta\bar{\eta}(x)} \right]Z\fge\;.
\label{zerofp}
\eea
The last line of this expresion in the functional non perturbative equivalent to the Dirac equation.

As before, we act with $\delta/\delta\eta^f(y)$ on this expresion to obtain
\bea
\left.\delta^d(x-y)Z\fge\right|_{\eta=\bar{\eta}=0}&&\nn\\
&&\hspace{-4.8cm}
 -\left.\left(
i\slsh{\partial}-m^f+e^f\gamma^\mu\frac{\delta}{i\delta J^\mu(x)}
\right) Z\fge\right|_{\eta=\bar{\eta}=0}S_F^f(x,y;[A_\mu])=0\;,
\eea
with obvious notation. Now, using eqs.~(\ref{tLeg}) and~(\ref{relsourfield}), we can rewrite~:
\bea
\delta^d(x-y)&&\nn\\
&&\hspace{-1.8cm}
-\left(i\slsh{\partial}-m^f+e^f{\not \!\! A}(x;[J])
+e^f\gamma^\mu \frac{\delta}{i\delta
J^\mu(x)}\right)S_F^f(x,y;[A_\mu])=0\;,\label{2pcgf}
\eea
which defines the non perturbative connected  two-points Green's function.

The electromagnetic potential vanishes in the absence of an
 external source, that is, $A_\mu(x;[J=0])=0$, in such a way that
 it is only written  to exhibit the content of the remaining
 functional derivation for eq.~(\ref{2pcgf}), which can be done
 exploiting the identity~(\ref{fpder})~:
\bea
\frac{\delta}{i\delta J^\mu(x)}S_F^f(x,y;[A_\mu])&&\nn\\
&&\hspace{-3.8cm}
=\int d^dz\frac{\delta A_\nu(z)}{i\delta J^\mu(x)}
\frac{\delta}{\delta A_\nu(z)}\left(\left.
\frac{\delta^2\Gamma}{\delta\psi^f(x)
\delta\bp^f(y)}\right|_{\psi=\bp=0} \right)^{-1}\nn\\
&&\hspace{-3.8cm}
=-e^f\int d^dz \;d^du\;d^dw \frac{\delta A_\nu(z)}{i\delta
J^\mu(x)} S_F^f(x,u)\Gamma_\nu(u,w;z)S_F^f(w,y)\nn\\
&&\hspace{-3.8cm}
=-e^f\int d^dz \;d^du\;d^dw \;i\Delta_{\mu\nu}(x,z)S_F^f(x,u)
\Gamma_\nu(u,w;z)S_F^f(w,y)\;,
\eea
where in the last line we take $J=0$. From here, in the absence of external sources, eq.~(\ref{2pcgf}) is equivalent to the expression~:
\bea
\hspace{-8mm}
\delta^d(x-y)&-&
(i \slsh{\partial}-m ^f)S_F^f(x,y)\nn\\
&&\hspace{-1.8cm}
=-i (e ^f)^2\int
d^dz\;d^du\;d^dw
\Delta^{\mu\nu}(x,z)\gamma_\mu S_F^f(x,u)\Gamma_\nu(u,w;z)S_F^f(w,y)
\;.
\eea
It is usual to write the SDE corresponding to the inverse Fermion Propagator. Therefore, multiplying by $S_F^{f\,-1}(y,y')$, integrating with respect to $y$ and relabeling $y'=y$~:
\bea
S_F^{f\,-1}(x,y)
&-&(i \slsh{\partial}-m^f)\delta^d(x-y)\nn\\
&&\hspace{-1cm}
=-i (e^f)^2\int d^dz\;d^du
\;\Delta^{\mu\nu}(x,z)\gamma_\mu S_F^f(x,u)\Gamma_\nu(u,y;z)\;.\label{deltaFP}
\eea
The photon porpagator couples eqs.~(\ref{deltaFP}) and~(\ref{Pimunu}).  In this way, one observes that the equations for the two-points functions couple to each other, and both depend on the three-points Green's function $\Gamma^{f\mu}$. This is the first indication of a general rule which says that the SDE for an $n$-points function is coupled to others of the same order or lower orders, and to functions of order $(n+1)$.

Diagram~(2) shows the SDE corresponding to the Fermion Propagator
%SDE for the fermion propagator
\vspace{-1.5cm}
\begin{center}
\SetScale{0.7}
\begin{picture}(500,100)(0,0)
%Propagador Completo
\ArrowLine(50,50)(150,50)
\CCirc(100,50){3}{}{}
\PText(145,60)(0)[]{-1}
%Propagador desnudo
\ArrowLine(200,50)(300,50)
\PText(295,60)(0)[]{-1}
%SDE piece
\Vertex(430,50){1} \Vertex(370,50){1}
\ArrowLine(350,50)(450,50)
\PhotonArc(400,50)(30,0,180){4}{8.5}
\CCirc(400,80){3}{}{}
\CCirc(430,50){3}{}{}
\CCirc(400,50){3}{}{}
%Signos algebraicos
\PText(175,52)(0)[]{=}
\PText(325,52)(0)[]{-}
\end{picture}\\
\vspace{-30pt}
{\sl Diagram~(2)~: SDE for the Fermion Propagator.}
\end{center}

\subsection{SDE for the Fermion-Boson Vertex}
\noindent
The corresponding equation for the three-point vertex can be obtained in a similar fashion. For completeness, we present it in momentum space, where it is written more concisely~:
\bea
 i\Gamma_\mu^f(p',p)&=& i\gamma_\mu +\sum_g \int\frac{d^dl}
{(2\pi )^d} (i S_F^g(p'+l))\nn\\
&\times& (i\Gamma^g_\mu (p'+l,p+l))(i
S_F^g(p+l))K^{gf}(p+l,p'+l,l)\;.
\eea
$K$ is the fermion-antifermion scattering kernel. The diagramatic representation of this equation is shown in Diagram~(3)~:
%SDE para el V\'ertice
\vspace{-1.5cm}
\begin{center}
\SetScale{0.65}
\begin{picture}(500,100)(0,0)
%V\'ertice completo
\Line(75,50)(125,75)
\Line(75,50)(125,25)
\Vertex(75,50){1}
\Photon(25,50)(75,50){-3}{4}
\CCirc(75,50){3}{}{}
%V\'ertice desnudo
\Line(225,50)(275,75)
\Line(225,50)(275,25)
\Vertex(225,50){1}
\Photon(175,50)(225,50){-3}{4}
%SDE piece
\Line(425,50)(475,75)
\Line(425,50)(475,25)
\Vertex(375,50){1}
\Photon(325,50)(375,50){-3}{4}
\CArc(400,50)(25,0,360)
\CCirc(425,50){3}{}{}
\CCirc(400,75){3}{}{}
\CCirc(400,25){3}{}{}
\CCirc(375,50){3}{}{}
%Signos algebraicos
\PText(150,52)(0)[]{=}
\PText(300,52)(0)[]{-}
\end{picture}\\
{\sl Diagram~(3)~: SDE for the Vertex.}
\end{center}
Clearly, $\Gamma_\mu$ couples to the two-point function for the fermion
 $S_F$, and to the fermion-antifermion scattering amplitude $M$, a four-point function, which again illustrates the general rule.

To solve the SDE for the Fermion Propagator, the simplest approximation for the Vertex that has been used is the so-called rainbow or ladder approximation, which consists in approximate $M$ by iterating the lowest order perturbative contribution to the kernel $K$, along with the substitution of the fermionic propagators by their bare couterparts, $S_F^{0f}(p)=1/[\slsh{p}-m^{f}]$. This and other approximation will be discussed below, avoiding flavor labels.

\subsection{Solving the SDE}
\noindent
\subsubsection{Quenched Approximation}
In massless QED3 in the quenched approximation \cite{BR1,quenched1,Dong}, which corresponds to neglect fermion-loop contributions to the vacuum polarization, that is, to take
\be
\Pi(q)=0
\ee
in the Photon Propagator, we face infrared divergences with the
ordinary Perturbation Theory.  A commonly used remedy for this
situation is to soften the infrared behavior of the Photon
Propagator by including fermion-loop contributions to the vacuum
polarization.  At the lowest order for a fermion of mass $m$,
this contribution to the Polarization Scalar is~:
\be
\Pi(k)=\frac{\alpha}{k^2}\left[2m+\frac{k^2-m^2}{k}
\arcsin{\left(\frac{k}{\sqrt{k^2+4m^2}}\right)}\right]\;.
\ee
In a theory with $N$ massless fermions, the Polarization Scalar is then~:
\be
\Pi(k)=\frac{\tilde{\alpha}}{k}\;,
\ee
where
\begin{eqnarray*}
\tilde{\alpha}=\frac{Ne^2}{8}\;,
\end{eqnarray*}
such that the photon propagator behaves as $1/q$ for $q^2\to 0$, that is, the infrared divergence has been softened without altering the ultraviolet properties of the propagator.

The quenched approximation in the SDE for the Fermion Propagator
co\-rres\-ponds to Diagram~(4)~:
%Diagrama quenched
\vspace{-1.5cm}
\begin{center}
\SetScale{0.7}
\begin{picture}(500,100)(0,0)
%Propagador Completo
\ArrowLine(50,50)(150,50)
\CCirc(100,50){3}{}{}
\PText(145,60)(0)[]{-1}
%Propagador desnudo
\ArrowLine(200,50)(300,50)
\PText(295,60)(0)[]{-1}
%SDE piece
\Vertex(430,50){1} \Vertex(370,50){1}
\ArrowLine(350,50)(450,50)
\PhotonArc(400,50)(30,0,180){4}{8.5}
\CCirc(430,50){3}{}{}
\CCirc(400,50){3}{}{}
%Signos algebraicos
\PText(175,52)(0)[]{=}
\PText(325,52)(0)[]{-}
\end{picture}\\
\vspace{-20pt}
{\sl Diagram~(4)~: SDE for the Fermion Propagator (quenched approximation)}
\end{center}
and at the one-loop level in the ordinary Perturbation Theory there is no distiction between the quenched and the unquenched approximations.

\subsubsection{Rainbow or Ladder Approximation }
\noindent
In the study of the Dynamical Generation of Masses, one commonly
used approximation is to set \cite{review}
\be
\Gamma^\mu(k,p)=\gamma^\mu\;,
\ee
which is known as the rainbow or ladder approximation. If this approximation is added to the quenched one, the SDE for the Fermion Propagator decouples for the corresponding equations for the Photon Propagator and for the Vertex. The following Diagram~:
%diagrama quenched+rainbow
\begin{center}
\vspace{-1.5cm}
\SetScale{0.7}
\begin{picture}(500,100)(0,0)
%Propagador Completo
\ArrowLine(50,50)(150,50)
\CCirc(100,50){3}{}{}
\PText(145,60)(0)[]{-1}
%\PText(100,45)(0)[]{p}
%Propagador desnudo
\ArrowLine(200,50)(300,50)
\PText(295,60)(0)[]{-1}
%\PText(250,45)(0)[]{p}
%SDE piece
\Vertex(430,50){1} \Vertex(370,50){1}
\ArrowLine(350,50)(450,50)
\PhotonArc(400,50)(30,0,180){4}{8.5}
\LongArrowArc(400,50)(20,60,120)
%\PText(400,45)(0)[]{k}
%\PText(400,95)(0)[]{q}
\CCirc(400,50){3}{}{}
%Signos algebraicos
\PText(175,52)(0)[]{=}
\PText(325,52)(0)[]{-}
\end{picture}\\
\vspace{-20pt}
{\sl Diagram~(5)~:  SDE for the Fermion Propagator (quenched approximation and bare vertex)}
\end{center}
describes this conjunction of approximations.

\subsubsection{Beyond the Rainbow Approximation}
One of the problems with the rainbow approximation is the
violation of gauge covariance, particularly of the
Ward-Green-Takahashi Identity.  The correct form for the
fermion-boson vertex is crucial to restore the gauge covariance
of the SDE and should be such that the above mentioned identity
is fulfilled, among other requirements.  Ball and Chiu \cite{BC}
have studied the structure of this vertex and have propossed
their now famous ansatz, which we will discus afterwards.  The
restoration of the gauge covariance for the physical observables
is one of the main motivations for the construction of the
fermion-boson vertex, which we will carry out in Chapter{5.

\subsubsection{$1/N$ Expanssion}
Ordinary Perturbation Theory in terms of the coupling seems to break down due to the infrared divergences of the Green's functions. These divergences can be avoided making use of another expanssion parameter. We can take $N$ fermion flavors and expand in $1/N$ \cite{maris}. It has been shown that massless QED3 is finite order by order in this approximation, and besides, the infrared behavior of the Photon Propagator softenes, as we pointed out before. Therefore, we have a theory with $N$ massless fermions and we take the large-$N$ limit, keeping $Ne^2$ fixed in the weak coupling regime. If we write
\be
e^2=\frac{8}{N}\;,
\ee
the new perturbative expanssion for the Photon Propagator is shown in Diagram~(6)~:
%diagrama SDE \pf en 1/N
\begin{center}
\vspace{-1.5cm} %\hfil\\
\SetScale{0.6}
\begin{picture}(500,100)(0,0)
%Propagador completo
\Photon(50,50)(150,50){4}{10}
\PText(145,60)(0)[]{-1}
\CCirc(100,50){5}{}{}
%propagador desnudo
\Photon(200,50)(300,50){4}{10}
\PText(295,60)(0)[]{-1}
%SDE piece
\Vertex(370,50){1}
\ArrowArcn(400,50)(30,180,0)
\ArrowArcn(400,50)(30,0,180)
\Photon(350,50)(370,50){4}{2}
\Photon(430,50)(450,50){4}{2}
%\CCirc(430,50){4}{}{}
%\CCirc(400,80){4}{}{}
%\CCirc(400,20){4}{}{}
%Signos algebraicos
\Text(110,30)[]{=}
\Text(190,30)[]{+}
\Text(200,30)[]{N}
\Text(300,30)[]{+ O(1/N)}
\end{picture}\\
%\vspace{-10pt}
{\sl Diagram~(6)~: SDE for the Fermion Propagator ($1/N$ expanssion).}
\end{center}
and after its evaluation, it yield a Polarization Scalar for the Photon
\be
\Pi(q)=q\;,
\ee
which leads to a $1/q$ behavior for the Photon Propagator in the infrared domain.

The SDE for the Fermion Propagator in this scheme is given by
\be
\Sf{p}^{-1}=\slsh{p}-\frac{8i}{N}\int\frac{d^3k}{(2\pi)^3}\gamma^\mu S_F(p)
\Gamma^\nu(p,k)\Delta_{\mu\nu}(p-k)\;.
\ee
There exists a controversy on whether in the study of this equation, see
for example Pennington
\etal~\cite{pen1enn} and Atkinson \etal~\cite{atk1enn},
DCSB takes place for arbitrary number of flavours or there exists
a critical number of such flavours separating the chirally
symmetric and asymmetric phases of unquenched QED. However,
we do not take up these matters in this thesis.

\subsubsection{Gauge Technique}
There exist a scheme to solve the SDE which differs substantially in the method with the previously mentioned studies~: The Gauge Technique~\cite{Salam1,SD1,S1,DW1,DW2,D1,YH,Keck1,D2}.
This scheme, based in Minkowski space, assumes that the elements of the SDE (Propagators and Vertices) have a spectral representation, in term of which the SDE are reformulated and directly solved. For example, it assumes that there exists a spectral function $\rho_\psi$ such that the Fermion Propagator can be written as~:
\be
\Sf{p}=\int_{-\infty}^\infty d\omega\frac{\rho_\psi (\omega)}
{\slsh{p}-\omega}\;,
\ee
and the fermion-boson Vertex has a similar form. In fact, an ansatz in this scheme is
\be
S_F(p)\Gamma_\mu^{GT}(p,q)S_F(q)=\int_{-\infty}^\infty d\omega
\rho_\psi (\omega)
\frac{1}{\slsh{p}-\omega}\gamma_\mu\frac{1}{\slsh{q}-\omega},\;,
\ee
Inserting these expressions into the SDE, we obtain a linear equation for the spectral density. This is an important feature of the Gauge Technique~: It reduces the SDE to linear equations.
Another advantage is that the
Ward-Green-Takahashi
identity is automatically taken into account.

\section{ Ward-Green-Takahashi Identity}

One of the consequences of gauge covariance is that Green's functions obey certain identites which relate one of these functions to the others. Thes are called Ward-Geen-Takahashi identities (WGTI), \cite{W1,G1,T1},  and they come out from the Becci-Rouet-Stora-Tyutin (BRST) symmetry. They play a crucial role in the proof for the renormalizability of the theory. One of them, simply known as the WGTI, relates the [1-PI] Vertex to the propagators, and it has been widely implemented in SDE studies based either on the Gauge Technique, for intance,\cite{Salam1,SD1,S1,DW1,DW2,D1,Keck1,D2}, and on making an ansatz for the fermion-boson vertex,
\cite{BC,CP1,CP2,CP3,ABGPR1,AGM1,BP1,BP2}.

We derive this identity following the textbook~\cite{ryder},
starting from the ge\-ne\-ra\-ting functional~(\ref{fungen}) with the
action~(\ref{QEDdaction}), which includes the gauge fixing term.
Let us recall that without such term (and the source terms), the
lagrangian corresponding to this action is gauge invariant.  This
makes $Z$ to be infinite and spoils the search for the photon
propagator.  In order to find a finite propagator, we are forced
to introduce a gauge fixing term (and a ghost term, which in the
abelian case, we can absorbed into the normalization).  This means
that the lagrangian related to the action~(\ref{QEDdaction}) is
no longer gauge invariat.  The physical consequences of the
theory, expressed in terms of Green's functions, should not
depend upon the gauge, in such a way that $Z$ \emph{must be}
gauge in\-va\-riant.  This is a nontrivial requirement, and leads us
to a differential equation for $Z$, which we will find below.

Let us take the transformations~(\ref{localgauget}) to be infinitesimal, that is,
\bea
A_\mu&\to&A_\mu+\partial_\mu\lambda(x)\nn\\
\psi&\to&\psi-ie\lambda(x)\psi\nn\\
\bp&\to&\bp+ie\lambda(x)\bp\;.
\eea
Under these transformations, neither the gauge fixing term, nor the source terms are gauge invariant, in such a way that the integrand of $Z$ acquires a factor
\be
\sexp^{{\left\{i\int dx\left[-\frac{1}{\xi}(\partial^\mu
A_\mu)\partial_\rho\partial^\rho\lambda +J^\mu\partial_\mu\lambda+
ie\lambda(\bar{\eta}\psi-\bp\eta)\right]\right\}}}\;,
\ee
which, being $\lambda$ infinitesimal, can be rewritten as
\be
1+i\int dx\left[ -\frac{1}{\xi}(\partial^\mu
A_\mu)\partial_\rho\partial^\rho
+ \partial^\mu J_\mu\lambda-
ie(\bar{\eta}\psi-\bp\eta)\right]\lambda\;,\label{inflamb}
\ee
where we have integrated by parts to remove the derivative operator from $\lambda$. Gauge invariance of $Z$ implies that the operator~(\ref{inflamb}), when acting on $Z$, is merely the identity. Since $\lambda$ is an arbitrary function, this implies that
\be
\left[-\frac{1}{\xi}\partial_\rho\partial^\rho (\partial^\mu A_\mu) +
\partial^\mu J_\mu\lambda-
ie(\bar{\eta}\psi-\bp\eta)\right]Z=0\;.
\ee
Substituting the fields by derivatives with respect to their sources,
\be
\psi\to\frac{1}{i}\frac{\delta}{\delta\bar{\eta}}\;,\hspace{5mm}
\bp\to\frac{1}{i}\frac{\delta}{\delta\eta}\;,\hspace{5mm}
A_\mu\to\frac{1}{i}\frac{\delta}{\delta J^\mu}\;,
\ee
we find the following functional differential equation for $Z$~:
\be
\left[ \frac{i}{\xi}\partial_\rho\partial^\rho\partial^\mu
 \frac{\delta}{\delta J^\mu}-\partial^\mu J_\mu
-e\left(\bar{\eta}\frac{\delta}{\delta\bar{\eta}}
-\eta\frac{\delta}{\delta\eta}\right)\right]Z\fge=0\;.
\ee
Taking the transformation~(\ref{tLeg}), the last expression can be written as an equation for  ${\cal G}$~:
\be
\frac{i}{\xi}\partial_\rho\partial^\rho\partial^\mu\frac{\delta
{\cal G}}{\delta J^\mu}-\partial^\mu J_\mu -e\left(\bar{\eta}
\frac{\delta {\cal G}}{\delta\bar{\eta}} -\eta\frac{\delta {\cal
G}}{\delta\eta}\right)=0\;,\label{ztow}
\ee
where ${\cal G}={\cal G}\fge$.  Finally, let us turn this expression into an equation for the vertex function $\Gamma$, given by the transformation~(\ref{gtogamma}). Making use of the identities~(\ref{relsourfield}), eq.~(\ref{ztow}) becomes~:
\be
-\frac{1}{\xi}\partial_\rho\partial^\rho\partial^\mu A_\mu(x)+\partial_\mu
\frac{\delta\Gamma}{\delta A_\mu(x)}
-ie\psi\frac{\delta\Gamma}{\delta\psi(x)}
+ie\bp\frac{\delta\Gamma}{\delta\bp(x)}=0\;.
\ee
Now, taking the functional derivative with respect to $\bp(x_1)$ and $\psi(y_1)$, and setting $\bp=\psi=A=0$, the first term vanishes, and therefore,
\bea
\left.-\partial_x^\mu\frac{\delta^3\Gamma}{\delta\bp (x_1)
\delta\psi(y_1)\delta A^\mu(x)}\right|_{\bp=\psi=A=0}
&=& \left.ie\delta(x-x_1)
\frac{\delta^2\Gamma}{\delta\bp (x_1)\delta\psi(y_1)}
\right|_{\psi=\bp=0}\nn\\
&&\hspace{-1.3cm}
\left.-ie\delta(x-y_1)
\frac{\delta^2\Gamma}{\delta\bp (x_1)\delta\psi(y_1)}\right|_{\psi=\bp=0}\;.
\label{iwtc}
\eea
The left hand side of this equation is the derivative of the [1-PI] fermion-boson vertex~(\ref{vp1pi}), and the next two terms are the inverses of the exact Fermion Propagators~(\ref{propferrob}).  The content of eq.~(\ref{iwtc}) becomes clear if we expres it in momentum space. For such purpose, we define the proper vertex function
$\Gamma_\mu(k,p,q)$ as
\bea
\int dxdx_1dy_1\sexp^{i(qx_1-ky_1-px)}
\left.\frac{\delta^3\Gamma}{\delta\bp (x_1)\delta\psi(y_1)\delta
A^\mu(x)}\right|_{\bp=\psi=A=0}&&\nn\\
&&\hspace{-5cm}
=ie(2\pi)^4\delta(q-k-p)\Gamma_\mu(k,p,q)\;.
\eea
On the other hand, we define the Fermion Propagator in momentum space as~:
\be
\int dx_1dy_1 \sexp^{i(qx_1-ky_1)} \left.\frac{\delta^2\Gamma}{\delta\bp
(x_1)\delta\psi(y_1)}\right|_{\psi=\bp=0}=(2\pi)^4\delta(q-p)iS_F^{-1}(p)\;.
\ee
Therefore, multiplying  eq.~(\ref{iwtc}) by $\sexp^{i(qx_1-ky_1-px)}$ and integrating over
$x$, $x_1$ and $y_1$, we have
\be
q^\mu\Gamma_\mu(k,p,q)=S_F^{-1}(k)-S_F^{-1}(p)\;.
\ee
or, in the limit $k\to p$,
\be
\frac{\partial S_F^{-1}}{\partial p^\mu}=\Gamma_\mu(p,p)\;.
\ee
WGTI is one of the requirements for the restoration of the gauge covariance of the physical observables that have been employed in SDE studies. Its implementation is pretty much simple, and it has been widely used.

There exist also a WGTI for the Photon Propagator, which is given by the expression~:
\be
q_\mu\Pi^{\mu\nu}(q)=0\;.
\ee
This expression is useful, because it allows us to define the Polarization Scalar in the traditional way. The fact that $\Pi^{\mu\nu}$ is transverse, leads us to the masslessness for the photon. We no longer take into account this identity.

\section[Landau-Khalatnikov-Fradkin Transformations]{Landau-Khalatnikov-Fradkin\\ Transformations}

In a gauge field theory, Green's functions transfrom in a specific manner under a variation of gauge. In Quantum Electrodynamics, and in honor to Lev Davidovich and his collegues who firs obatained them, these transformations carry the name of  Landau-Khalatnikov-Fradkin (LKF) transformations, \cite{LK1,LK2,F1}.  These were also derived by Johnson and Zumino through functional methods,\cite{JZ1,Z1}
\footnote{Fukuda, Kubo and Yokoyama have looked for a possible formalism where renormalization constants of the wave function are in fact gauge invariant~\cite{Fukuda}}.  LKF transformations are non perturbative in nature, and therefore, they have the potential to play an important role to address the problems of gauge invariance which plague the strong coupling SDE studies. In general, the rules governning these transformation are far from simple. The fact that they better describe their essence in coordinate space, make them even more complex. As a result, these transformations have played a less significant and practical role than desired in SDE studies.

We display its derivation below, following the work of Zumino~\cite{Z1}. We start by noticing that in Landau gauge, the Photon Propagator can be written as
\be
\Delta_{\mu,\nu}(x;0)=\left[
g_{\mu\nu}-\frac{\partial\mu\partial\nu} {\partial^2}\right]
\Delta_c(x)\;,
\ee
where $\Delta_c(x)$ is the so-called Feynman function
\be
\Delta_c(x)=-\frac{\delta(x)}{\partial^2}\;.
\ee
In an arbitrary covariant gauge, the Photon Propagator can be parametrized by an arbitrary function $\Delta_d$ in the form~:
\be
\Delta_{\mu\nu}(x;\Delta_d)=\Delta_{\mu\nu}(x;0)+\partial_\mu
\partial_\nu\Delta_d(x)\;.\label{lkffot}
\ee
Let us recall now that the generating functional is expressed as
\be
Z\fge=\bra{0} T\sexp^{i\int dx (\bar{\eta}\psi
+\bar{\psi}\eta+A_\mu J^\mu)}\ket{0}\;.
\ee
Just to obtain the LKF transformations, for the moment we are not assuming that
\be
\partial_\mu J^\mu=0\;.
\ee
Now, under a gauge transformation,
\be
Z_\lambda\fge = Z_0\left[\bar{\eta}\sexp^{ie\lambda},
\eta\sexp^{-ie\lambda},J^\mu\right]\sexp^{ie\int dx
J_\mu\partial^\mu \lambda}\;, \label{gtZ1}
\ee
which can be written in a differential form with the expression
\be
i\frac{\delta Z}{\delta\lambda}=\left( \partial^\mu J_\mu
+e\eta\frac{\delta}{\delta\eta}-e\bar{\eta}\frac{\delta}{\delta
\bar{\eta}} \right)Z\;.\label{gtZ2}
\ee
With these definitions, it is easy to verify that the generating functional satisfies the following set of differential equations~:
\bea
\hspace{-8mm}
\left\{ \partial^\sigma \left(\partial_\mu \frac{\delta}{i\delta
J^\sigma}\! -\!\partial_\sigma \frac{\delta}{i\delta J^\mu}\right)
\!+\!(\delta_\mu^\sigma\!+\!a_\mu\partial^\sigma)\left( ie
\frac{\delta}{i\delta\eta}\gamma_\sigma \frac{\delta}{i\delta
\bar{\eta}}\!-\!J_\sigma\right)\right\} Z&=&0\label{edZ1}\\
\hspace{-8mm}
\left\{ \left[\gamma^\mu \left(ie \frac{\delta}{i\delta
J^\mu}\right) +m \right]\frac{\delta}{i\delta\bar{\eta}}
-\eta\right\}Z&=&0\label{edZ2}\\
\hspace{-8mm}
\left\{ -\frac{\delta}{i\delta\eta} \left[-\gamma^\mu \left(
\partial_\mu+ie\frac{\delta}{i\delta J^\mu}\right)+m \right]
-\bar{\eta}\right\}Z&=&0\label{edZ3}\\
\hspace{-8mm}
\left( a^\mu\frac{\delta}{i\delta J^\mu}+\lambda\right)Z&=&0\;.
\label{edZ4}
\eea
The vector operator $a_\mu$ is introduced firstly for the sake of
consistency of the notation.  It satisfies
\be
\partial^\mu a_\mu = -1\;,
\ee
and a convenient choice for it defines the different gauges as well. For instance,
\be
a_\mu=\partial_\mu (-\partial^2-i\epsilon)^{-1}
\ee
defines the Landau gauge, and for a Lorentz frame, characterized by a time-like unitary vector $n_\mu$,
\be
a_\mu=\frac{\partial_\mu+n_\mu (n\cdot
\partial)}{\partial^2+(n\cdot \partial)^2}
\ee
corresponds to the Coulomb gauge.  Turning our attention back to
the ge\-ne\-ra\-ting functional, if an object ${\cal F}$, constructed
as a functional derivative of the generating functional, is gauge
invariant in the sense that it does not change with the choice of
$\lambda$, then we have that
\be
\frac{\delta{\cal F}}{\delta\lambda}=0\;.
\ee
In particular, if we take the Fermion Propagator as
\be
S_F(x,y)=\frac{1}{iZ}\frac{\delta^2Z}{\delta\eta(y)\delta \bar{\eta}(x)}\;,
\label{fpZZ}
\ee
we obtain from eq.~(\ref{gtZ1}) or~(\ref{gtZ2}), after setting $\eta=\bar{\eta}=0$,
\be
Z_\lambda[0,0,J^\mu] S_{F\,\lambda}(x,y) =
\sexp^{\left[ie(\lambda(x)-\lambda(y))+i\int \partial_\mu
J^\mu\lambda\right]} Z_0[0,0,J^\mu] S_{F\,0}(x,y)\;,
\ee
or,
\bea
i\frac{\delta}{\delta\lambda(z)}[Z[0,0,J^\mu]S_F(x,y)]& = &\nn\\
&&\hspace{-3cm}
[\partial_\mu
J^\mu(z)-e\delta(x-z)+e\delta(y-z)]Z[0,0,J^\mu]S_F(x,y) \;.\label{ZSF}
\eea
To obtain a convenient expression for the generating functional which allows us to deduce the LKF transformations, let us consider first the Fermion Propagator in an external magnetic field $B_\mu$ given by
\be
\tilde{S}_F[x,y;B_\mu]\equiv\tilde{S}_F(B) =
\frac{\delta(x-y)}{\gamma^\mu (\partial_\mu-ieB_\mu)+m}\;.
\ee
We write then the vaccum Polarization Scalar, in obvious notation, as~:
\be
\Pi(B)=\sexp^{-Tr (\ln{\tilde{S}_F(B)\tilde{S}_F^{-1}(0)})}\;.
\ee
By direct verification we have the identity
\be
\left(\frac{\delta}{\delta B_\mu}-e\frac{\delta}{i\delta\eta}
\gamma^\muð\frac{\delta}{i\delta\bar{\eta}} \right)[\sexp^{i
\bar{\eta}\tilde{S}_F(B)\eta}\Pi(B)]=0\;.\label{idZZ}
\ee
The generating functional can then be written in the following way~:
\bea
Z\fge& =& \sexp^{i
\bar{\eta}\tilde{S}_F\left[\frac{\delta}{i\delta J_\mu}\right]
\eta} \Pi\left[\frac{\delta}{i\delta J_\mu}\right]\nn\\
&&\hspace{-5mm} \sexp^{\left[
\frac{i}{2}(J_\mu+a_\mu\partial^\rho J_\rho) \Delta_c (J^\mu
+a^\mu\partial^\rho J_\rho )-i\partial^\rho J_\rho \lambda
-\frac{i}{2}\partial^\rho J_\rho \Delta_d \partial^\sigma
J_\sigma \right]}\;.\label{ZZ}
\eea
This expression satisfies eqs.~(\ref{edZ1}) to~(\ref{edZ3}) automatically, and eq.(\ref{edZ4}) is satisfied when we take $\Delta_d=0$. We can verify directly this sentence, except in the case of eq.~(\ref{edZ1}), where we should observe firstly that, on acting with the operator
\bea
\partial^\sigma \left(\partial_\mu\frac{\delta}{i\delta
J^\sigma}-\partial_\sigma\frac{\delta}{i\delta J^\mu} \right)
\nn
\eea
in the last exponential of~(\ref{ZZ}), we obtain a factor
\be
J_\mu+a_\mu\partial^\rho J_\rho\;.
\ee
Those factors with $\lambda$ and $\Delta_d$ do not contibute.  We
must now move this factor to the left of the terms
of~(\ref{ZZ}) which contain derivatives with respect to $J_\mu$.
The net effect of this operation is to replace
\be
J_\mu\to J_\mu-ie\frac{\delta}{i\delta\eta} \gamma_\mu
\frac{\delta}{i\delta\bar{\eta}}\;,
\ee
by virtue of~(\ref{idZZ}). This verifies eq.~(\ref{edZ1}). We also take advantage of
\be
\Pi\left[\frac{\delta}{i\delta J_\mu}\right]\partial^\rho J_\rho
= \partial^\rho J_\rho \Pi\left[\frac{\delta}{i\delta
J_\mu}\right]
\ee
and
\bea
\tilde{S}_F\left[\frac{\delta}{i\delta J_\mu};x,y \right]
\partial^\rho J_\rho(z) &=&\nn\\
&&\hspace{-2.5cm}
[\partial^\rho J_\rho(z)-e\delta(x-z)+e\delta(y-z)]
\tilde{S}_F\left[\frac{\delta}{i\delta J_\mu};x,y \right]\;,
\eea
to find the relation
\be
Z[0,0,J^\mu]S_F[x,y;J^\mu] =
\tilde{S}_F\left[\frac{\delta}{i\delta J_\mu};x,y \right]
Z[0,0,J^\mu]\;,
\ee
which involves the propagator~(\ref{fpZZ}).  This expression is very useful to obtain in a direct way the LKF transformation for the Fermion Propagator. We satart from considering an infinitesimal transformation of the function
$\Delta_d$.  Such variation induces a transfrormation on the generating functional given by
\be
\delta Z =\frac{i}{2}\int\!\!\!\int \frac{\delta}{\delta\lambda}
(\delta\Delta_d) \frac{\delta}{\delta\lambda}Z\;.
\ee
Now, setting $\eta=\bar{\eta}=0$, the induced change in the generating functional without fermionic sources is
\be
\delta Z[0,0,J^\mu]=-\frac{i}{2}\int\!\!\!\int \partial^\mu
J_\mu (\delta\Delta_d)\partial^\rho J_\rho Z[0,0,J^\mu]\;,
\ee
or, in finite form
\be
Z'[0,0,J^\mu]=\sexp^{-\frac{i}{2}\int\!\!\!\int \partial^\mu
J_\mu (\delta\Delta_d)\partial^\rho J_\rho}Z[0,0,J^\mu]\;.
\ee
Therefore, for the Fermion Propagator we have
\be
\delta(Z[0,0,J^\mu] S_F) = \frac{i}{2}\int\!\!\!\int
\frac{\delta}{\delta\lambda}(\delta\Delta_d)\frac{\delta}
{\delta\lambda} (Z[0,0,J^\mu]S_F)\;,
\ee
which, along with eq.~(\ref{ZSF}), implies
\bea
S'_F(x,y;J^\mu)&=&\sexp^{ie^2[\delta\Delta_D(x-y)-\delta\Delta_d(0)]
+ie\int [\delta\Delta_d(x-z)-\delta\Delta_d(y-z)]\partial^\rho
J_\rho(z)dz} \nn\\
&&\hspace{5mm}
S_F(x,y;J^\mu)\;.
\eea
For $J=0$ we find
\be
S'_F(x,y;0)=\sexp^{ie^2[\delta\Delta_d(x-y)-\delta\Delta_d(0)]}S_F(x,y;0)
\ee
Then, for a finite change of  $\Delta_d$, the transformation law for the Fermion Propagator reads~:
\be
S_F(x;\xi)=S_F(x;0)e^{-i[\Delta_d(0)-\Delta_d(x)]}\;.
\label{lkfrev}
\ee
For the Vertex, we have
\bea
B_\mu(z;x,y\vert\Delta)&=&B_\mu(z;x,y\vert 0)e^{-i[\Delta_d(0)
-\Delta_d(x-y)]}\nn\\
&&\hspace{-2.5cm}
+S_F(x-y;0)e^{-i[\Delta_d(0)-\Delta_d(x-y)]}
\frac{\partial}{\partial z_\mu}[\Delta_d(x-z)-\Delta_d(z-y)]\;,
\label{lkfver}
\eea
where $B_\mu$ is the non-amputated vertex, defined in momentum space in terms of the amputated vertex $\Gamma_\mu$ as~:
\be
B_\mu(k,p)=S_F(k)\Gamma_\nu(k,p)S_F(p)\Delta_{\mu\nu}(q)\;.
\ee
The transformation rule for the partially amputated Vertex
\be
\Lambda_\mu(k,p)=S_F(k)\Gamma_\mu(k,p)S_F(p)
\ee
follows from eqs.~(\ref{lkffot}), (\ref{lkfrev}) and~(\ref{lkfver}), and it is simply given by~:
\be
\Lambda_\mu(z;x,y\vert\Delta_d)=\Lambda_\mu(z;x,y\vert 0)
e^{-i[\Delta_d(0) -\Delta_d(x-y)]} \;.
\ee
In the usual covariant way for gauge fixing, the Photon Propagator takes the form~:
\be
\Delta_{\mu\nu}(q,\xi)=\frac{1}{q^2[1+\Pi(q)]} \left(g_{\mu\nu}
-\frac{q_\mu q_\nu}{q^2}\right)+\xi\frac{q_\mu q_\nu}{q^4}\;,
\ee
which is obtained by taking $\Delta_d$ in eq.~(\ref{lkffot}) as~:
\be
\Delta_d(x)=-i\xi e^2\mu^{4-d}\int\frac{d^dq}{(2\pi)^d} \frac{e^{-iq\cdot
x}}{q^4}\;, \label{deltadlkf}
\ee
where $e^2$ is the dimensionless electromagnetic coupling, and $\mu$ is the 't Hooft mass scale in Dimensional Regularization.

These LKF transformations are ruled by very complex laws.  Being
written in coordinate space adds to their complexity, and they
have been less used in the context of SDE, as compared to the
WGTI, particularly in the study of the phenomenon of Dynamical
Mass Generation.

To have a better understanding of the role that either the WGTI
and the LKF transformations play in the restoration of gauge
independence for the phy\-si\-cal observables, it is necessary to
first know the phenomenon we are dealing with, and those
assumptions that simplify the most the SDE, to our knowledge, to
make use of the bare vertex.  This is the scenario we will
develop in the next chapter.

\chapter{Gauge Dependence of Physical Observables}
\pagestyle{myheadings}
\markboth{Gauge Dependence of Physical Observables}
{Gauge Dependence of Physical Observables}

\noindent
In a gauge theory, at the level of physical observables, gauge
symmetry reflects as the fact that they be independent of the
gauge parameter.  Perturbation theory respects these requirements
and besides the Ward-Green-Takahashi identities (WGTI) and the
Landau-Khalatnikov-Fradkin (LKF) transformations remain valid at
every level of approximation.  However, this has not been
achieved in general in the non perturbative study of gauge field
theories through Schwinger-Dyson equations (SDE) carried out so
far although significant progress has been made.  The gauge
technique of Salam, Delbourgo and later collaborators,
\cite{Salam1,SD1,S1,DW1,DW2,D1,YH}, was developed to incorporate
the constraint imposed by WGTI.  However, as pointed out in
\cite{D2}, gauge technique can become completely reliable only
after incorporating transverse Green functions with correct
analytic and gauge-covariance properties.  Another method widely
used to explore the non-perturbative structure of the SDE is to
make an {\em ansatz} for t he full fermion-boson vertex and then
study the gauge dependence of the physical observables related to
the phenomenon of Dynamical Chiral Symmetry Breaking (DCSB).
This method has been quite popular in four dimensional Quantum
Electrodynamics (QED).  For example, the vertex {\em ansatz}
proposed by Curtis and Pennington, \cite{CP1}, has been
extensively used to study the gauge dependence of the fermion
propagator and the dynamical generation of fermion mass in
Quenched QED, e.g., \cite{CP2,CP3,ABGPR1,AGM1}.  Later on, in the
work of Bashir and Pennington, \cite{BP1,BP2},  an
improved vertex which achieves complete gauge independence of the
critical coupling above which mass is dynamically generated is proposed.
These methods use the cut-off regularization to study the gauge
dependence of the physical observables.  As the cut-off method in
general does not respect gauge symmetry, a criticism of these
works has been raised recently, \cite{GSSW,SSW,KSW}.  They
suggest dimensional regularization scheme to study t he chirally
asymmetric phase of QED so that the possible gauge dependence
coming from the inappropriate regulator could be filtered out.

 Three dimensional Quantum Electrodynamics (QED3) provides us with a neat  laboratory to study DCSB as it is ultraviolet  well-behaved and hence the source of gauge non-invariance finds its roots only in the simplifying assumptions employed and {\em not} in the choice of the regulator.
Burden and Roberts, \cite{BR1}, studied the gauge dependence of the chiral condensate in quenched QED3 and proposed a vertex which  appreciably reduces this gauge dependence in the range $0-1$ of the covariant gauge parameter $\xi$. Unfortunately, the choice of their  vertex does not transform correctly under the operation of charge  conjugation. Moreover, the selected range of values for $\xi$ is very narrow, close to the vicinity of the Landau gauge.  In this chapter, we  undertake the calculation of the Euclidean mass of the fermion (referred to as {\em mass} from now onwards)  and the condensate for a wide range of values of $\xi$ in the bare vertex approximation in the followig schemes~:
\begin{itemize}
\item Setting $F(p)=1.$

\item Including the equation for $F(p)$, and

\item Making a partial use of the WGTI.
\end{itemize}

We have not achieved a complete gauge independence neither for the mass nor for the chiral condensate, which is a sing for the necesity of the construction and use of the full vertex.

\section{The Fermion Propagator}

In quenched QED3, the SDE for the fermion propagator in the Minkowski space
can be written as~:
\bea
S_F^{-1}(p) &=& S_F^{0 -1}(p) -
ie^2 \int \frac{d^3k}{(2\pi)^3} \Gamma^{\nu}(k,p)
S_F(k) \gamma^{\mu} \Delta^0_{\mu \nu}(q) \label{prop} \;,
\eea
where $q=k-p$, $e$ is the electromagnetic coupling, $\Gamma^{\nu}(k,p)$ is the full fermion-photon vertex, $S_F^{0}(p)$ and $\Delta^0_{\mu \nu}(q)$ are the bare fermion and photon propagators defined as
\be
  S_F^{0}(p)=1/\not \! p\;, \hspace{10mm} \Delta^0_{\mu \nu}(q) =
-\frac{g_{\mu \nu}}{q^2} + (1-\xi)\frac{ q_{\mu}q_{\nu} }{q^4}
\label{bareprop} \;,
\ee
and $S_F(p)$ is the full fermion propagator, which we prefer to write in
the following most general form~:
\be
S_F(p)=\frac{F(p)}{\not \! p- {\cal M}(p)} \label{fullprop} \;.
\ee
$F(p)$ is referred to as the wavefunction renormalization and ${\cal M}(p)$
as the mass function and $\xi$ is the usual covariant gauge parameter.

Eq.~(\ref{prop}) is a matrix equation.  It consists of two
independent equations, which can be decoupled by taking its trace
after multiplying it with $1$ and $\slsh{p}$, respectively.
Making use of Eqs.~(\ref{bareprop},\ref{fullprop}) and replacing
the full vertex by its bare counterpart, these equations can be
written as~:
\bea
\frac{1}{F(p)}&=& 1 + \frac{\alpha}{2 \pi^2 p^2} \int d^3k
\, \frac{F(k)}{k^2+{\cal M}^2(k)} \, \frac{1}{q^4} \;\nn\\
&&\hspace{5mm}
\left[\,-2 (k \cdot p)^2 + (2-\xi) (k^2+p^2) k \cdot p
- 2 (1-\xi) k^2 p^2\,
\right] \nn\;,\\
\frac{{\cal M}(p)}{F(p)}&=&   \frac{\alpha(2+\xi)}{2 \pi^2} \int d^3k
\, \frac{F(k)\,{\cal M}(k)}{k^2+{\cal M}^2(k)} \, \frac{1}{q^2}
\;,
\eea
where $\alpha=e^2/(4\pi)$ as usual. Carrying out angular integration after the Wick rotation to the Euclidean space, the above equations acquire the form~:
\bea
\frac{1}{F(p)} &=& 1 - \frac{\alpha \xi}{\pi p^2} \int_0^\infty dk \,
\frac{k^2 F(k)}{k^2+ {\cal M}^2(k)}
\left[\; 1 - \frac{k^2+p^2}{2kp} \, {\rm ln}\left| \frac{k+p}{k-p} \, \right|
\;  \right]  \;, \label{FNoWT} \\  \nn  \\
\frac{{\cal M}(p)}{F(p)} &=& \frac{\alpha( \xi +2 )}{\pi p}
\int_0^\infty dk \,
\frac{k F(k)  {\cal M}(k) }{k^2+ {\cal M}^2(k)} \,
{\rm ln}\left| \frac{k+p}{k-p} \, \right|
\; .     \label{MNoWT}
\eea

A trivial solution to Eq.~(\ref{MNoWT}) is ${\cal M}(p)=0$, which
 corresponds to the usual perturbative solution.  We are
 interested in a non-trivial solution by solving
 Eqs.~(\ref{FNoWT}) and (\ref{MNoWT}) simultaneously.  Such a
 solution for ${\cal M}(p)$ is related to the mass $m$ and the
 chiral condensate $\langle \bar{\psi} \psi \rangle $.  Assuming
 a simple analytic continuation from Minkowski to the Euclidean
 space, neglecting the rotation of the integration contour, we
 define $m={\cal M}(m)$.  It is true that this is not the
 physical mass for the fermion, and we do not expect it to be
 exactly gauge invariant.  However, since $m\sim \M (0)$, we can
 consider it as an effective mass.  At most we can expect this
 Euclidean mass to be approximately gauge invariant, in the sense
 that it is close to the physical mass~\cite{ABGPR1}.  On the
 other hand, in reference \cite{BR1}, Burden and Roberts
 demonstrated that the standard definition of the fermion
 condensate (Eq. (3.9) in \cite{BR1}) in terms of an integral
 over the mass function is in excellent numerical agreement with
 the prediction of the operator product expansion \cite{Politzer}
 which allows us to write $\langle \bar{\psi} \psi\rangle=4 p^2
 {\cal M}(p)/(2+\xi)$ (in units of $e^4$) in the limit when $p^2
 \to \infty$.  Such an expansion is valid only for values of the
 gauge parameter in the range $[0,1]$.  As expected, we find that
 in this limit, ${\cal M}(p)$ falls as $1/p^2$ so that the
 condensate does not depend upon the momentum variable $p$.  We
 shall study the gauge dependence of the mass and the condensate
 in the next section.

\section{Effect of the Wavefunction Renormalization}

In studying DCSB, it has been a common practice to make the  approximation $F(p)=1$ so that we only have to solve Eq.~(\ref{MNoWT}). The justification for this approximation stems from the fact that perturbatively $F(p)=1+{\cal O}(\alpha \xi/\pi)$. If $\alpha$ is small and we are sufficiently close to the Landau gauge, one would naturally expect that $F(p) \approx 1$. Although, it has been quite customary to employ this  approximation, there exist several works which include  both the equations. We study the effects of neglecting the wavefunction renormalization quantitatively. Fig.~(\ref{fnoWTIF1}) depicts the mass function ${\cal M}(p)$ for $F(p)=1$ in various gauges. As expected, the mass function is roughly a constant for low values of $p$ and  falls as $1/p^2$ for large values of $p$. The integration region chosen is from $10^{-3}$ to $10^3$ and we select 26 points per decade.  The mass  probes low momentum region of this graph, whereas,  the condensate is extracted from its asymptotic behaviour.  Obviou
sly, the mass seems to vary in more or less  equally spaced steps with the variation of the gauge parameter. In order to obtain a quantitative value of the mass, we select neighbouring points $p_a$ and $p_b$ ($p_a>p_b$), such that ${\cal M}(p_a)<p_a$ and ${\cal M}(p_b)>p_b$. We  then approximate the mass by the following relation~:
\bea
m &=& \frac{{\cal M}(p_b)-{\cal M}(p_a)}{p_b-p_a}(m-p_a)+{\cal M}(p_a) \;.
\eea
%%%%%%%%%%%%%%%%%%%%%%%%%%%%%%%%%%%%%%%%%%%%%%%%%%%%%%%%%%%%%%%%%%%%%%%%%%%%%%
%figura 3.1
\begin{figure}[t]
\ponfigura{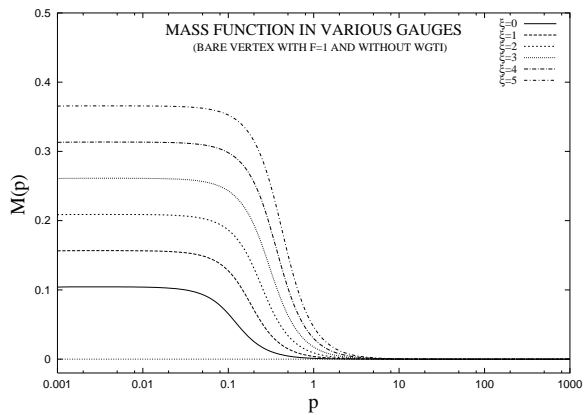}
\vspace*{5mm}
\caption{Mass Function ${\cal M}(p)$ in the $F(p)=1$ approxmation.}
\label{fnoWTIF1}
\end{figure}
%%%%%%%%%%%%%%%%%%%%%%%%%%%%%%%%%%%%%%%%%%%%%%%%%%%%%%%%%%%%%%%%%%%%%%%%%%%%%%
As for the condensate, the figure does not distinguish between the results for various  gauges. Therefore, we have to look at the numbers explicitly. Table~(1)  shows the value of the condensate for $\xi$ ranging from $1-5$. Momentum $p$ is displayed in units of $e^2$ and the condensate in units of $10^{-3} e^4$. The point $p=1000$ was chosen to calculate the condensate. This number seems  sufficiently large as the $1/p^2$ behaviour seems to set in much earlier ($p \approx 300$), as noted also in \cite{BR1}.  In Figs.~(\ref{fnoWTIF1condensate1}) and  (\ref{fmassnoWTIF1}) we display the gauge dependence of the chiral condensate and the mass for $F(p)=1$ in a wide range of values of the gauge parameter. The condensate varies heavily with the  change of gauge, roughly twice per unit change in the value of $\xi$.  Gauge dependence of the mass is not too different either.
%%%%%%%%%%%%%%%%%%%%%%%%%%%%%%%%%%%%%%%%%%%%%%%%%%%%%%%%%%%%%%%%%%%%%%%%%%%%
%figura 3.2
\begin{center}
\begin{figure}[t]
\ponfigura{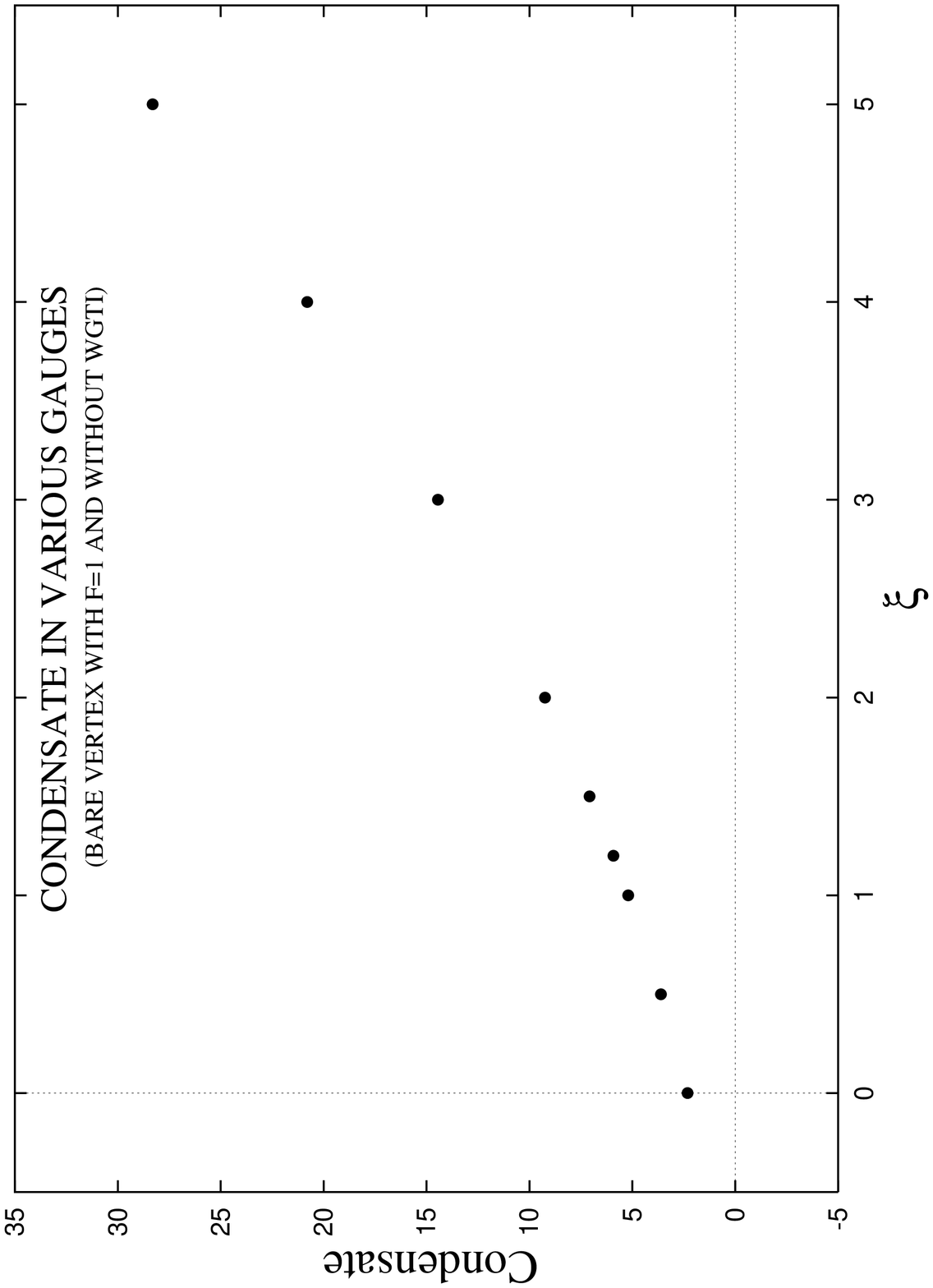}
\vspace{5mm}
\caption{Condensate $<\bar{\psi} {\psi}>$ in the $F(p)=1$ approxmation.}
\label{fnoWTIF1condensate1}
\end{figure}
\end{center}
%%%%%%%%%%%%%%%%%%%%%%%%%%%%%%%%%%%%%%%%%%%%%%%%%%%%%%%%%%%%%%%%%%%%%%%%%%%%%
%%%%%%%%%%%%%%%%%%%%%%%%%%%%%%%%%%%%%%%%%%%%%%%%%%%%%%%%%%%%%%%%%%%%%%%%%%%%%
%figura 3.3
\begin{center}
\begin{figure}[t]
\ponfigura{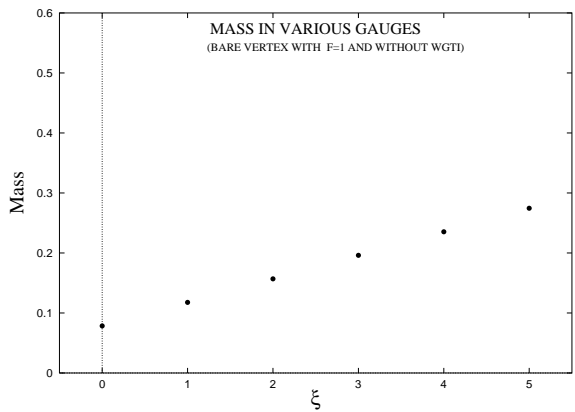}
\vspace{5mm}
\caption{Mass in the $F(p)=1$ approxmation.}
\label{fmassnoWTIF1}
\end{figure}
\end{center}
%%%%%%%%%%%%%%%%%%%%%%%%%%%%%%%%%%%%%%%%%%%%%%%%%%%%%%%%%%%%%%%%%%%%%%%%%%%%

Repeating the exercise by taking both the equations, namely Eqs.~(\ref{FNoWT}) and (\ref{MNoWT}), into account, we see similar qualitative behaviour of the mass function. It is roughly a constant  for low values of $p$ and falls as $1/p^2$ for large values of $p$,  Fig.~(\ref{fnoWTI}). The large $p$ behaviour is also evident from the entries in Table~(2). As for the wavefunction renormalization, it also is constant for small values of $p$. As $p$ becomes large it goes to $1$,  Fig.~(\ref{fnoWTIwave}). In Table~(2) we also give a comparison with the work of Burden and Roberts,  \cite{BR1}. As mentioned earlier, they restrict themselves to the close  vicinity of the Landau gauge, where our results are in excellent agreement. We investigate the gauge dependence of the condensate as well as the  mass far beyond the Landau gauge. A graphical description can be found in Figs.~(\ref{fnoWTIcondcomp}) and (\ref{feucmass1}). The following points are important to note~:
%%%%%%%%%%%%%%%%%%%%%%%%%%%%%%%%%%%%%%%%%%%%%%%%%%%%%%%%%%%%%%%%%%%%%%%%%
%figura 3.4
\begin{center}
\begin{figure}[t]
\ponfigura{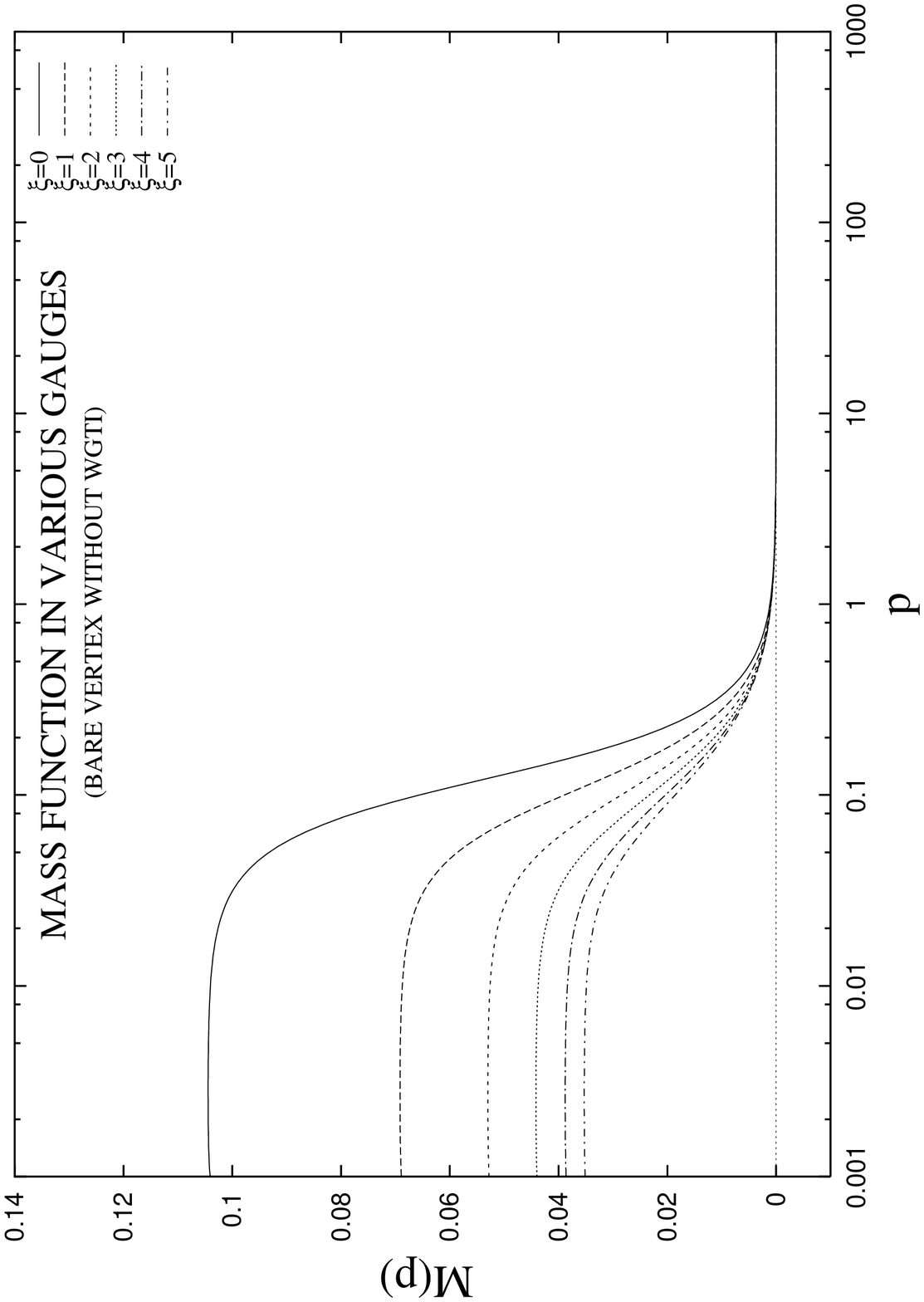}
\vspace{5mm}
\caption{Mass Function incliding the equation for the
Wavefunction Renormalization.}
\label{fnoWTI}
\end{figure}
\end{center}
%%%%%%%%%%%%%%%%%%%%%%%%%%%%%%%%%%%%%%%%%%%%%%%%%%%%%%%%%%%%%%%%%%%%%%%%
%%%%%%%%%%%%%%%%%%%%%%%%%%%%%%%%%%%%%%%%%%%%%%%%%%%%%%%%%%%%%%%%%%%%%%%%
%figura 3.5
\begin{center}
\begin{figure}[t]
\ponfigura{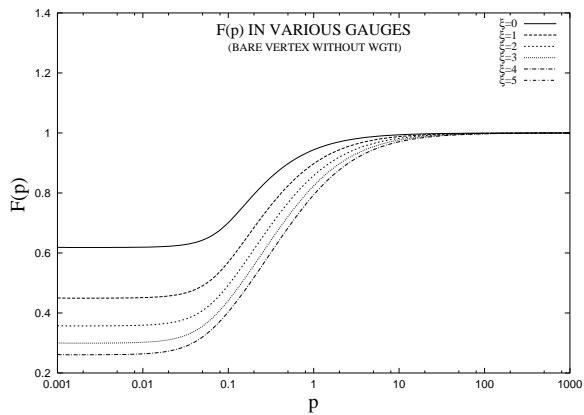}
\vspace{5mm}
\caption{Wavefunction Renormalization.}
\label{fnoWTIwave}
\end{figure}
\end{center}
%%%%%%%%%%%%%%%%%%%%%%%%%%%%%%%%%%%%%%%%%%%%%%%%%%%%%%%%%%%%%%%%%%%%%%%%
%%%%%%%%%%%%%%%%%%%%%%%%%%%%%%%%%%%%%%%%%%%%%%%%%%%%%%%%%%%%%%%%%%%%%%%%
%fugura 3.6
\begin{center}
\begin{figure}[t]
\ponfigura{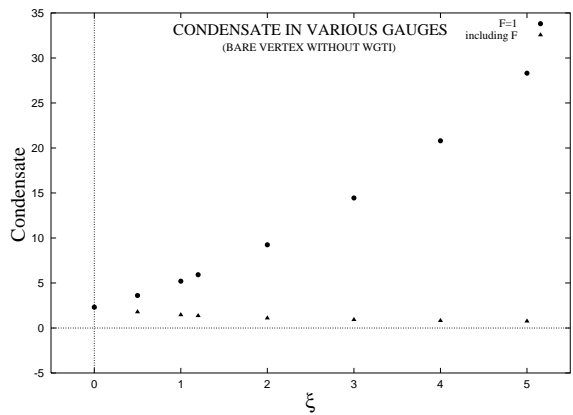}
\vspace{5mm}
\caption{Condensate:  Comparison between the cases with and
without the usage of $F(p)=1$.}
\label{fnoWTIcondcomp}
\end{figure}
\end{center}
%%%%%%%%%%%%%%%%%%%%%%%%%%%%%%%%%%%%%%%%%%%%%%%%%%%%%%%%%%%%%%%%%%%%%%%%
%%%%%%%%%%%%%%%%%%%%%%%%%%%%%%%%%%%%%%%%%%%%%%%%%%%%%%%%%%%%%%%%%%%%%%%%
%figura 3.7
\begin{center}
\begin{figure}[t]
\ponfigura{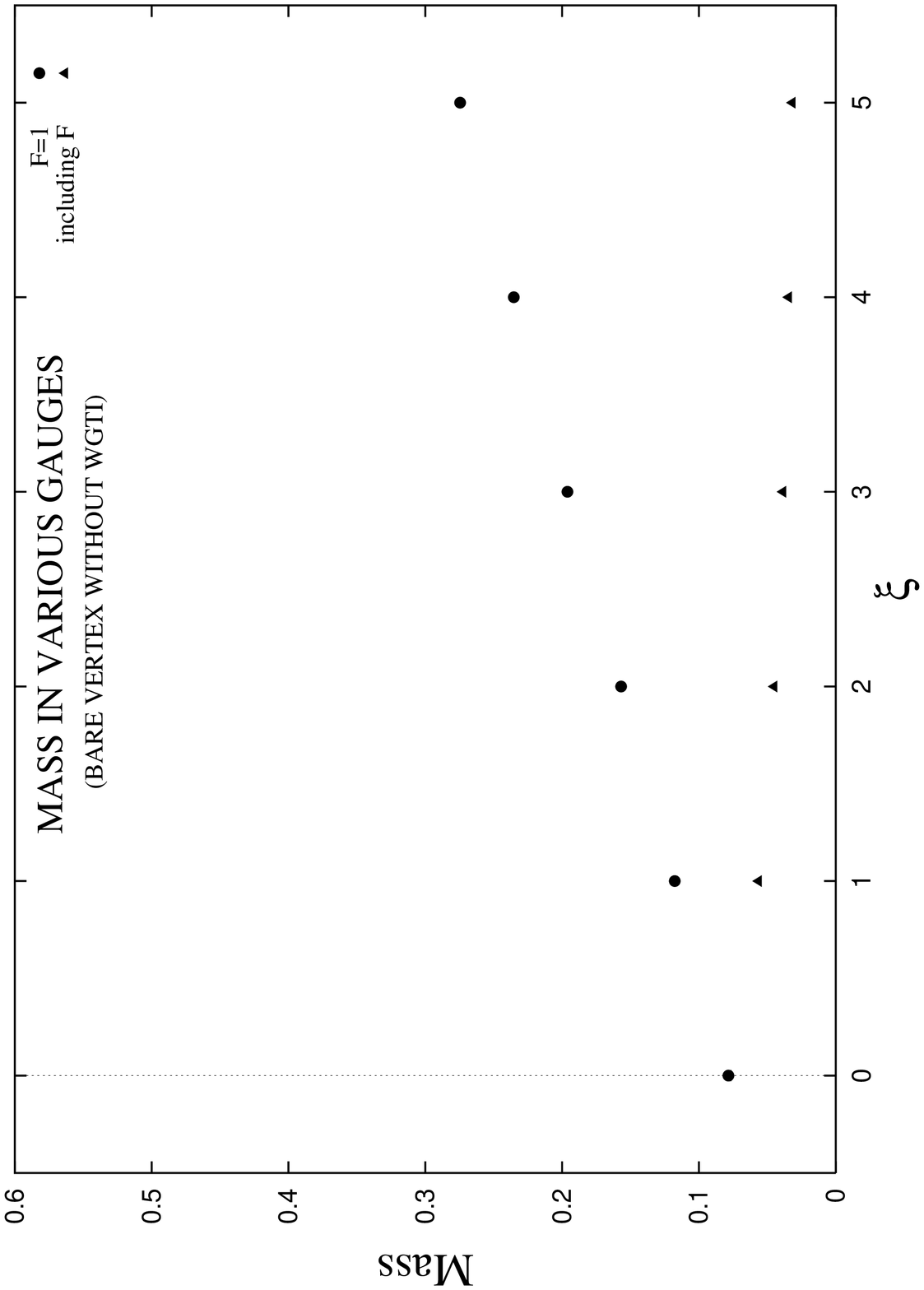}
\vspace{5mm}
\caption{Mass:  Comparison between the cases with and
without the usage of $F(p)=1$.}
\label{feucmass1}
\end{figure}
\end{center}
%%%%%%%%%%%%%%%%%%%%%%%%%%%%%%%%%%%%%%%%%%%%%%%%%%%%%%%%%%%%%%%%%%%%%%%%
\begin{itemize}

\item The wavefunction renormalization plays an extremely important role in
restoring the gauge invariance of the chiral condensate as well as the
mass of the fermion. Although the qualitative behaviour of the
mass function in various regimes of momenta remains largely unchanged,
whether or not we employ the approximation $F(p)=1$, quantitative dependence
of the physical observables mentioned above on the covariant gauge parameter
$\xi$ reduces a great deal by including the wavefunction renormalization.

\item As we move away from the Landau gauge towards large positive values
of $\xi$, the gauge dependence of the condensate as well as the mass
keeps di\-mi\-ni\-shing, without resorting to any sophisticated {\em ansatze}
for the fermion-boson interaction.

\end{itemize}

\section{Effect of the Ward-Green-Takahashi Identity}

\noindent
The bare photon propagator which appears in Eq.~(\ref{prop}) can be split
up in longitudinal and transverse parts as follows~:
\be
\Delta^0_{\mu \nu}(q) =  \Delta^{0 T}_{\mu \nu}(q)  -\xi \,
\frac{ q_{\mu}q_{\nu} }{q^4} \:,
\ee
where
\begin{eqnarray*}
\Delta_{\mu\nu}^{0   T}(q)=-g_{\mu\nu}/q^2+q_\mu   q_\nu/q^4\;.
\end{eqnarray*}
Employing this decomposition, we can rewrite Eq.~(\ref{prop}) as~:
\bea
S_F^{-1}(p) &=& S_F^{0 -1}(p) -
ie^2\int \frac{d^3k}{(2\pi)^3} \Gamma^{\nu}(k,p)  S_F(k)
\gamma^{\mu}\Delta^{0 T}_{\mu \nu}(q)  \nn\\
&&+ ie^2 \xi
\int \frac{d^3k}{(2\pi)^3} \Gamma^{\nu}(k,p) S_F(k)
\gamma^{\mu} \frac{q_{\mu}q_{\nu}}{q^4} \label{propWTI} \;.
\eea

It is well known that the use of the WGTI, in the equivalent of the  last term of Eq.~(\ref {propWTI}) in QED4, filters out a spurious term which is an artifact of using the gauge dependent cut-off regulator. Therefore, one is naturally motivated to use this decomposition in dimensions other than four. Now employing the bare vertex ansatz  $\Gamma^{\mu}(k,p)=\gamma^{\mu}$, multiplying Eq.~(\ref{propWTI}) by $1$ and $\not \! p$ respectively and Wick-rotating to the Euclidean space, we obtain the following equations~:
\bea
\frac{1}{F(p)}&=& 1 + \frac{\alpha}{2{\pi}^2p^2} \int d^3k
\, \frac{F(k)}{k^2+{\cal M}^2(k)} \, \frac{1}{q^4}\nn\\
&&\left[2(q\cdot p)(q\cdot k)
\!+\!  \frac{\xi}{F(p)}[p^2(q \cdot k)\!+\!{\cal M}(k){\cal M}(p)
(q\cdot p)]
 \right]  \;,  \\
\frac{{\cal M}(p)}{F(p)}&=&  \frac{ \alpha}{2\pi^2} \int d^3k
\, \frac{F(k)}{k^2+{\cal M}^2(k)} \, \frac{1}{q^2} \nn\\
&&\left[\,2 {\cal M}(k)\,
- \frac{\xi}{q^2}\, \frac{1}{F(p)}\,[{\cal M}(k)\,
(p\cdot q) - {\cal M}(p)\,(k\cdot q)\,]
\; \right] \;. \label{Mangle}
\label{Fangle}
\eea
On carrying out angular integration,
\bea
\frac{1}{F(p)}&=& 1 + \frac{\alpha \xi}{\pi p^2}\int_0^{\infty}dk
\, \frac{k^2  F(k)/F(p)}{k^2+{\cal M}^2(k)} \nn\\
&&\hspace{-5mm}
\Bigg[
\frac{p^2}{k^2-p^2}+\frac{p}{2k} {\rm ln}\left| \frac{k+p}{k-p}  \right|
\nn\\
&&\hspace{-3mm}
+{\cal M}(k){\cal M}(p)\left\{
\frac{1}{k^2-p^2}-\frac{1}{2kp} {\rm ln}\left| \frac{k+p}{k-p}  \right|
 \right\}
 \Bigg]  \;, \label{3 1/F}   \\
\frac{{\cal M}(p)}{F(p)}&=& \frac{\alpha}{\pi}\int_0^{\infty}dk
\, \frac{k^2 F(k)}{k^2+{\cal M}^2(k)}
\Bigg[
 \frac{2 {\cal M}(k)}{kp}{\rm ln}\left|
\frac{k+p}{k-p} \, \right|
\nn\\
&&\hspace{-10mm}
- \frac{\xi}{F(p)} \left\{ \frac{{\cal M}(k) -
{\cal M}(p)}{k^2-p^2}
%\right.\left.
-
\frac{{\cal M}(k)+{\cal M}(p)}{2kp}  {\rm ln}
\left|  \frac{k+p}{k-p} \right|  \right\}
 \Bigg] \;.   \label{3 M/F}
\eea
As the terms of the type $1/(k^2-p^2)$ are harder to deal with numerically, we use the approximation $F(p)=1$ to analyze the effect of the WGTI. Under this simplification, we only have to solve
\bea
{\cal M}(p)&=& \frac{\alpha}{\pi}\int_0^{\infty}dk
\, \frac{k^2 }{k^2+{\cal M}^2(k)} \Bigg[
 \frac{2 {\cal M}(k)}{kp}{\rm ln}\left|
\frac{k+p}{k-p} \, \right|\nn\\
&&- \xi \left\{ \frac{{\cal M}(k) -
{\cal M}(p)}{k^2-p^2} -
\frac{{\cal M}(k)+{\cal M}(p)}{2kp}  {\rm ln}
\left|  \frac{k+p}{k-p} \right|  \right\}
 \Bigg] \;.   \label{3M}
\eea
The mass function obtained on solving Eq.~(\ref{3M}) is depicted in Fig.~(\ref{fWTIF1}), which, along with Table~(3), reveals that its qualitative behaviour remains unchanged both for small and large values  of $p$. In Figs.~(\ref{fcondF1WTIcompnoWTI}) and~(\ref{feucmass2}), we  compare the gauge dependence of the condensate and the mass with and  without the usage of the WGTI. We find that the gauge dependence of these quantities seems to increase by incorporating the said identity. Similar behaviour was observed in QED4 by Gusynin  {\em et. al.} \cite{GSSW} in studying the gauge dependence of the critical coupling above which mass is generated. They carried out a numerical  analysis of the  criticism raised by Dong et. al. \cite{Dong} on the work of Atkinson et. al. \cite{Atkinson} who did not employ the WGTI as suggested in Eq.~(\ref{propWTI}), resulting in the appearance of a spurious cut-off dependent term~\footnote{To trace back the origin of this error, consult the footnote on page 7680 of the referenc
e \cite{BP1}}.  Gusynin {\em et. al.} found that if one employs the WGTI, the critical coupling is more steeply gauge dependent. We find similar behaviour for the mass and the condensate  in QED3 in the approximation $F(p)=1$ in this section.
%%%%%%%%%%%%%%%%%%%%%%%%%%%%%%%%%%%%%%%%%%%%%%%%%%%%%%%%%%%%%%%%%%%%%%%%%%%%%
%figura 3.8
\begin{center}
\begin{figure}[t]
\ponfigura{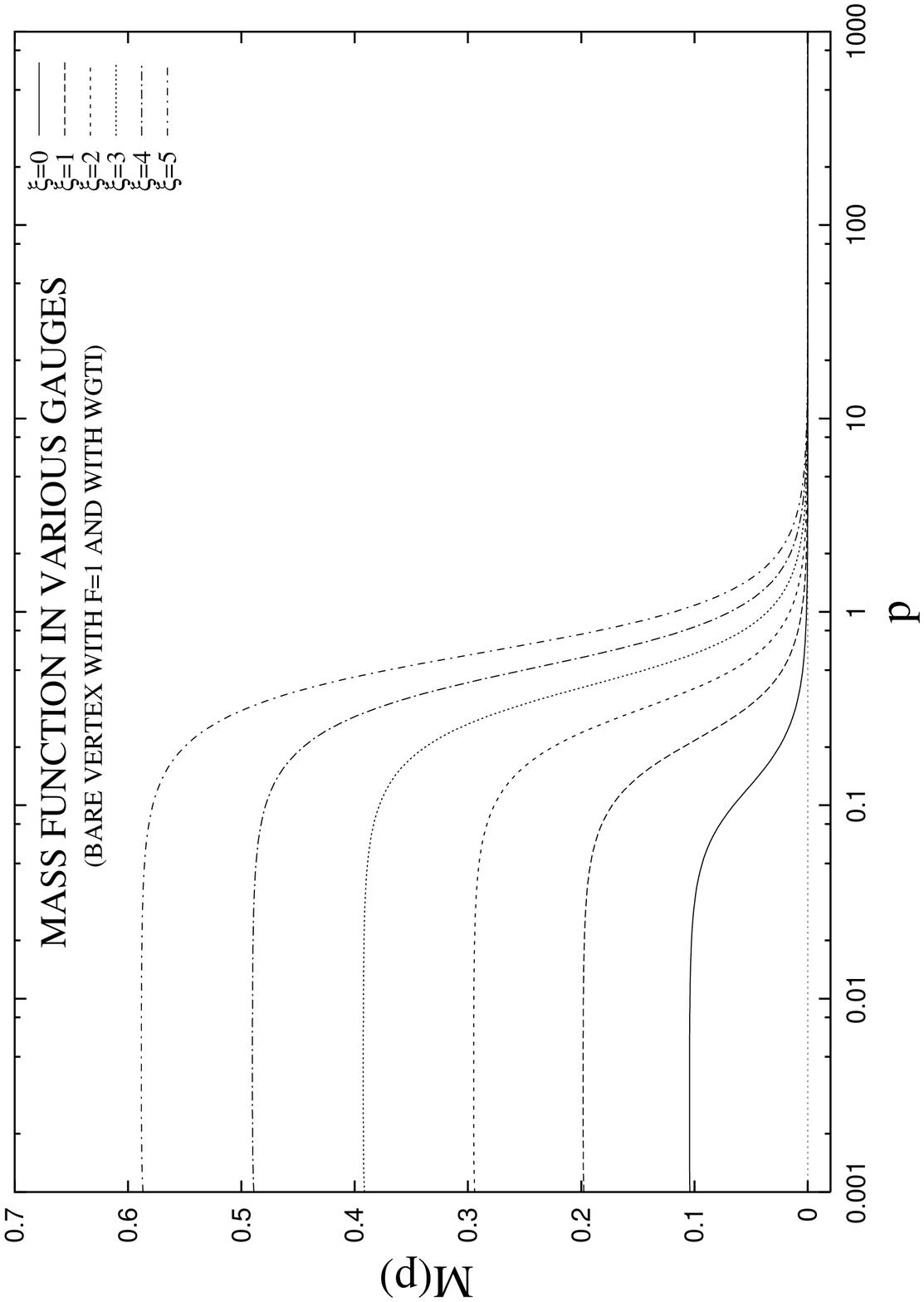}
\vspace{5mm}
\caption{Mass Function incliding the effect of the WGTI for
 $F(p)=1$.}
\label{fWTIF1}
\end{figure}
\end{center}
%%%%%%%%%%%%%%%%%%%%%%%%%%%%%%%%%%%%%%%%%%%%%%%%%%%%%%%%%%%%%%%%%%%%%%%%%%%%%%%
%%%%%%%%%%%%%%%%%%%%%%%%%%%%%%%%%%%%%%%%%%%%%%%%%%%%%%%%%%%%%%%%%%%%%%%%%%%%%%%
%figura 3.9
\begin{center}
\begin{figure}[t]
\ponfigura{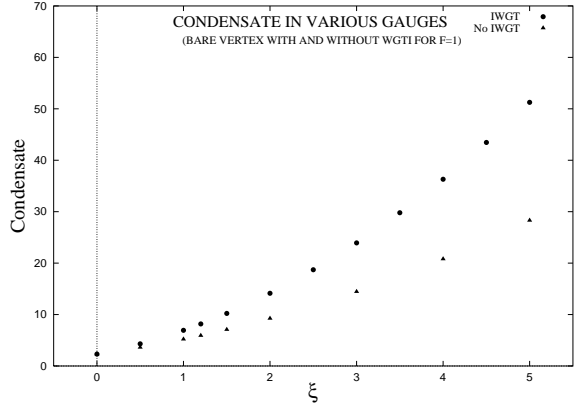}
\vspace{5mm}
\caption{Effect of the WGTI on the gauge dependence of the
condensate for $F(p)=1$.}
\label{fcondF1WTIcompnoWTI}
\end{figure}
\end{center}
%%%%%%%%%%%%%%%%%%%%%%%%%%%%%%%%%%%%%%%%%%%%%%%%%%%%%%%%%%%%%%%%%%%%%%%%%%%%%%%
%%%%%%%%%%%%%%%%%%%%%%%%%%%%%%%%%%%%%%%%%%%%%%%%%%%%%%%%%%%%%%%%%%%%%%%%%%%%%%%
%figura 3.10
\begin{center}
\begin{figure}[t]
\ponfigura{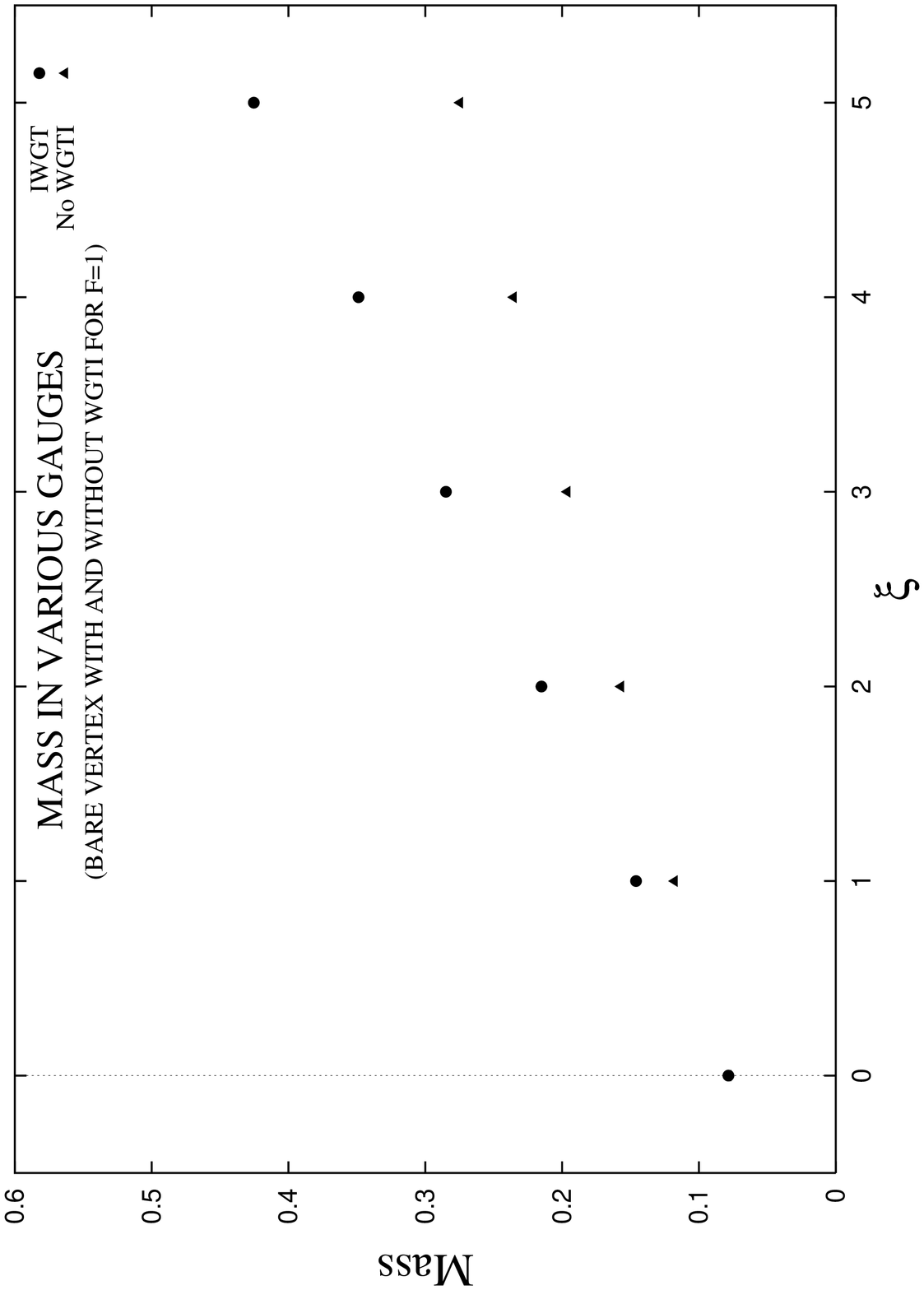}
\vspace{5mm}
\caption{Effect of the WGTI on the gauge dependence of the
mass for $F(p)=1$.}
\label{feucmass2}
\end{figure}
\end{center}
%%%%%%%%%%%%%%%%%%%%%%%%%%%%%%%%%%%%%%%%%%%%%%%%%%%%%%%%%%%%%%%%%%%%%%%%%%%%%%%

\section{Dimensional Regularization Method}
In this section we compare our numerical results with those obtained
by employing the dimensional regularization scheme, \cite{GSSW,SSW,KSW}. For
simplicity, we restrict ourselves only to the Landau gauge
without incorporating the WGTI. In this case, the
equation for the mass function acquires the following form in
Euclidean space in arbitrary dimensions~:
\be
{\cal M}(p)=4\pi\alpha (d-1)\int\frac{d^dk}{(2\pi)^d}
\frac{{\cal M}(k)}{k^2+{\cal M}^2(k)}\frac{1}{q^2},
\ee
where $\alpha$ is a dimensionful coupling except in four dimensions. We define
$d=4- 2 \epsilon$ and
\bea
\alpha &=& \alpha_d \; {\mu}^{2 \epsilon}  \;,
\eea
$\alpha_d$ being dimensionless. We now use the volume element
$d^dk=k^{d-1} \,dk \, d \Omega_d$, where $d \Omega_d$ is the $d$-dimensional
solid angle defined as
\begin{eqnarray*}
d\Omega_d=\prod_{l=1}^{d-1}\sin^{d-1-l}\theta_l d\theta_l\;.
\end{eqnarray*}
The angle $\theta_{d-1}$ varies from 0 to
$2 \pi$, whereas all other angles vary from 0 to $\pi$. Choosing
$\theta_1$ to be the angle between $k$ and $p$, we can easily carry out
the remaining angular integrations to arrive at~:
\bea
     {\cal M}(p) &=& \frac{2(d-1) \alpha}{(4 \pi)^{\frac{d-1}{2}}
\Gamma(\frac{d\!-\!1}{2})}   \int_0^\infty dk^2 \;
\frac{k^{d-2} \, {\cal M}(p)}{k^2 \!+\! {\cal M}^2(p)}
\int_0^{\pi} \, d \theta_1 \; \frac{\sin^{d-2} \theta_1}{q^2} \;.
\eea
Using the standard formula, \cite{GR},
\begin{eqnarray*}
   \int_0^{\pi} dx
\frac{\sin^{2 \sigma -1}x}{\left[1+2 a {\rm cos}x +a^2 \right]^{\lambda}}
&=& B(\sigma,1/2) _2F_1(\lambda,\lambda-\sigma+1/2,\mu+1/2;a^2) \\
&&\hspace{50mm}  |a|<1 \;,
\end{eqnarray*}
integration over $\theta_1$ yields~:
\bea
{\cal M}(p)&=& \frac{ (3-2 \epsilon) \alpha}{(4 \pi)^{1- \epsilon}
\Gamma(2- \epsilon)}  \int_0^\infty dk^2
\frac{(k^2)^{1-\epsilon}{\cal M}(k)}{k^2+{\cal M}^2(k)}\nn\\
&&\hspace{-15mm}
\left[ \frac{1}{k^2} \,_2F_1\left( 1,\epsilon;2\!-\!\epsilon;\frac{p^2}{k^2}
\right) \theta (k^2-p^2)
\!+\!\frac{1}{p^2} \,_2F_1 \left(1,\epsilon;2\!-\!\epsilon;\frac{k^2}{p^2}\right)
\theta (k^2\!-\!p^2)\right]\;. \nn \\
\eea

This equation was studied in detail in \cite{GSSW} in four dimensions, taking $\epsilon$ to be a small positive number. The  factor  $(k^2)^{- \epsilon}$ in the numerator regulates the otherwise divergent behaviour of the integrand for large momenta. As noted in \cite{GSSW}, the hypergeometric function does not play any  role in regularization and hence can simply be replaced by  $F(1,0;2,z)=1$. In case of three dimensions, the hypergeometric function develops a pole for $k^2=p^2$, as is obvious from the following identity~:
\be
_2F_1(1,\epsilon;2-\epsilon;1)=\frac{1-\epsilon}{1-2\epsilon} \;. \label{pole}
\ee

As was pointed out earlier, such terms are hard to deal with numerically. Due to increasing computational time  and memory, we go only up to $\epsilon =0.48$, starting from  $\epsilon=0.4$. To obtain satisfying results, we need to use increasingly more points per decade as we approach closer to $\epsilon=0.5$. For instance, we use $100$ points per decade for $\epsilon =0.46$. The problems of ever increasing computational time and memory limited us to use  $140$ points per decade for the case of $\epsilon=0.48$. Despite this  large number, we belive that the corresponding result falls short of the  desired accuracy. As a result, there is a slight rise at the end of the flat region of the mass function, and the final descent begins rather late, Fig. (\ref{fdimreg}). The problematic  pole  for $\epsilon=0.5$ in Eq.~(\ref{pole}) corresponds to the relatively well-controlled singularity in the following expression
\bea
_2F_1\left( 1,\frac{1}{2};\frac{3}{2};z^2 \right) &=& \frac{1}{2z}
\ln{\frac{1+z}{1-z}}
\eea
for $z\rightarrow 1$.  A comparison between the mass function obtained from techniques based upon the dimensional regularization scheme and the one  computed in Section 3.3 is also depicted in Fig ~(\ref{fdimreg}). Taking the numerical limitation  for $\epsilon=0.48$ into account, we note that as  $\epsilon$ approaches the value of $0.5$, we get  closer and closer to the result obtained  previously, where we work in 3-dimensions to start with.
%%%%%%%%%%%%%%%%%%%%%%%%%%%%%%%%%%%%%%%%%%%%%%%%%%%%%%%%%%%%%%%%%%%%%%%%%%%%%%%%%%%
%figura 3.11
\begin{center}
\begin{figure}[t]
\ponfigura{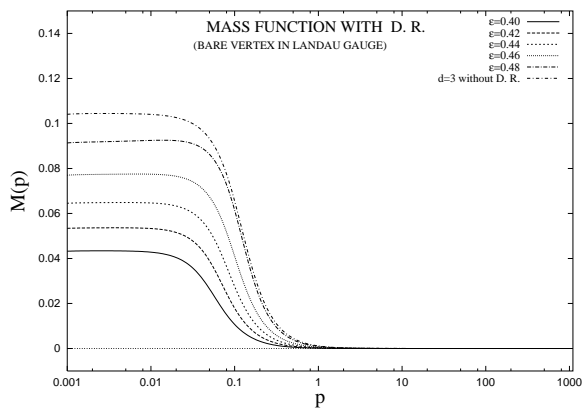}
\vspace{5mm}
\caption{Mass Function for various values of $\epsilon$.  Notice
that $\epsilon=0.5$ co\-rres\-ponds to 3 dimensions.  The result for
$\epsilon=0.48$ could not be achived with the desired accuracy
due to the computing time and memory increase.}
\label{fdimreg}
\end{figure}
\end{center}
%%%%%%%%%%%%%%%%%%%%%%%%%%%%%%%%%%%%%%%%%%%%%%%%%%%%%%%%%%%%%%%%%%%%%%%%%%%%%%%%%%%%

\section{Towards the Full Vertex}

WGTI with the bare vertex is valid only if bare fermion propagators are involved. When we use full propagators, physical observables calculated from the corresponding SDE exhibit  gauge dependence. Temptation emerges then to move towards the full vertex, and the first requirement we must think of is that it should satisfy the WGTI. Hence we face the construction of the Ball-Chiu vertex detailed displayed below. The full vertex satisfies WGTI
\be
q_{\mu}\Gamma^{\mu}(k,p)={\it S}^{-1}_{F}(k)-{\it S}^{-1}_{F}(p)  \; .
\label{WTI}
\ee
This relation allows us to decompose the complete vertex into its longitudinal $\Gamma_L^\mu(k,p)$  and transverse $\Gamma_T^\mu(k,p)$ parts~:
\be
\Gamma^{\mu}(k,p)=\Gamma^{\mu}_{L}(k,p)+\Gamma^{\mu}_{T}(k,p) \;,
\label{vertex}
\ee
where the transverse part satisfies
\be
q_{\mu}\Gamma^{\mu}_{T}(k,p)=0\;\;\;\;\;\mbox{y} \;\;\;\;
\Gamma^{\mu}_{T}(p,p)=0 \;,\label{defofTvertex}
\ee
and therefore, remains undetermined by the WGTI. Following the work of Ball and Chiu, we define the longitudinal part of the vertex in terms of the Fermion Propagator. In the limit $k\to p$, WGTI is written as~:
\be
\Gamma^\mu(p,p)= \frac{\partial}{\partial p_\mu} S_F^{-1}(p)\;.
\ee
Substituting the expresion for the full Fermion Propagator, we observe that
\bea
\Gamma^\mu(p,p)&=& \frac{\partial}{\partial p_\mu} S_F^{-1}(p)\nn\\
&=& \frac{\partial}{\partial p_\mu}\frac{\slsh{p}-{\cal M}(p)}{F(p)}
\nn\\
&=&\frac{\gamma^\mu}{F(p)}+2p^\mu\slsh{p}\frac{\partial}{\partial p^2}
-2p^\mu\frac{\partial}{\partial p^2}\frac{{\cal M}(p)}{F(p)}\;.
\eea
After we make the substitutions
\bea
\frac{1}{F(p)}&\to&\frac{1}{2}\left[\frac{1}{F(k)}+\frac{1}{F(p)}\right]\; ,\nn\\
p^\mu &\to& \frac{1}{2}(k^\mu+p^\mu) \; ,\nn\\
\slsh{p}&\to&\frac{1}{2}(\slsh{k}+\slsh{p})\;, \nn\\
\frac{\partial}{\partial p^2}\frac{1}{F(p)}&\to&
\frac{1}{k^2-p^2}\left[ \frac{1}{F(k)}-\frac{1}{F(p)}\right] \; ,\nn\\
\frac{\partial}{\partial p^2}\frac{{\cal M}(p)}{F(p)}&\to&
\frac{1}{k^2-p^2}\left[ \frac{{\cal M}(k)}{F(k)}
-\frac{{\cal M}(p)}{F(p)}\right]\;,
\eea
we can define the longitudanal or Ball-Chiu (BC) vertex as~:
\bea
\Gamma^{\mu}_{BC}&=&\frac{\gamma^{\mu}}{2}
\left[ \frac{1}{F(k)}+\frac{1}{F(p)} \right] \; + \;
\frac{1}{2} \, \frac{(\slsh{k}+\slsh{p})(k+p)^{\mu}}
{(k^2-p^2)}\left[ \frac{1}{F(k)}-\frac{1}{F(p)} \right] \nn\\
&+&   \frac{(k+p)^{\mu}}
{(k^2-p^2)}\left[ \frac{\M(k)}{F(k)}-
\frac{\M(p)}{F(p)}\right] \; . \label{Lvertex}
\eea
Without loss of generality, the transverse vertex can be expressed as~:
\be
\Gamma^{\mu}_{T}(k,p)=\sum_{i=1}^{8} \tau_{i}(k^2,p^2,q^2)T^{\mu}_{i}(k,p)
\;,\label{VT}
\ee
with the apropriate $\{T^\mu\}$ basis.  Functions $\tau_1$,
$\tau_4$, $\tau_5$ and $\tau_7$ are proportional to the mass $m$,
and therefore in massless studies they do not appear.  During the
last few years, a programme has been started towards the
construction of vertex ans\"atze which impose constraints on the
transverse part of the vertex.  Some of the most famous attempts
are discussed below.

\subsection{Curtis-Penington Vertex}
\noindent
In four dimensions, the question of why cannot we set all of the
$\tau$'s as zero was answered by Curtis and Pennington, arguing
to Perturbation Theory and the multiplicative renormalizability
properties of the Fermion Propagator.

Their vertex (CP)
\be
\Gamma^\mu_{CP}=\Gamma^\mu_{BC}+\frac{1}{2}\left[\frac{1}{F(k)}-\frac{1}{F(p)}\right]\frac{\gamma^\mu(k^2-p^2)-(k+p)^\mu\slsh{q}}{d(k,p)}\;,
\ee
where
\be
d(k,p)=\frac{(k^2-p^2)^2+[ {\cal M}^2(k)+ {\cal
M}^2(p)]^2}{k^2+p^2}\;,
\ee
comes out from the following assumptions~:
\begin{itemize}
\item Transverse vertex must agree with perturbative results at the one-loop level in the relevant kinematic regime $k>>p$.

\item Fermion Propagator obtained from this vertex should be multiplicatively renormalizable.

\item This Propagator should transform correctly under its LKF transformation.

\item Vertex should have correct charge conjugation properties.

\item Transverse vertex should vanish in Landau gauge.

\item Transverse Vertex should not depend upon the angle between fermion momenta.

\item The only contributing coefficient is $\tau_6$.

\item Transverse vertex as well as the longitudinal one are writen without explicit dependence on the gauge parameter $\xi$~\footnote{This, as we will show, is not possible.}.
\end{itemize}

Transverse vertex is written in such a way that, in the massless case, their
\be
\tau_6=\frac{1}{2}\frac{k^2+p^2}{(k^2-p^2)^2}
\left[\frac{1}{F(k)} -\frac{1}{F(p)}\right]
\ee
is antisymmetric under the interchange of $k$ and $p$.

This vertex, however, does not lead to a completelly gauge
independent critical coupling.  In the massless case, it exhibits
a kinematic singularity when $k^2\to p^2$.  In comparison with
Perturbation Theory at one-loop, CP vertex agrees with
perturbative results only in the kinematic regime above
mentioned.  Multiplicative renormalizability for the Fermion
Propagator is achieved only in the leading term of the power law
for the wavefunction renormalization.  Finally, the LKF
transformation of this propagator is valid only up to the leading
logarithmic term.

\subsection{Burden-Roberts Vertex}
\noindent
Burden and Roberts have parametrized a slight modification to the BC vertex in the following way~\cite{BR1}~:
\bea
\Gamma^\mu_{BR}&=&\left[ a\frac{1}{F(k)}+(1-a)\frac{1}{F(p)}\right]
\gamma^\mu \nn\\
&&+ \frac{(k+p)^\mu((1-a)\slsh{k}-a\slsh{p})}{k^2-p^2}
\left[\frac{1}{F(k)}-\frac{1}{F(p)}\right]\nn\\
&&-\frac{(k+p)^\mu}{k^2-p^2}\left[ \frac{{\cal M}(k)}{F(k)}
-\frac{{\cal M}(p)}{F(p)}\right]\;,\label{BRvertex}
\eea
where
\be
a=\frac{1}{2}+\delta\;.
\ee
With the assumptions that the vertex should satisfy the WGTI and that it should be free from kinematical singularities to all orders in Perturbation Theory, their parametrization allows to optimize $\delta$ such that the most gauge independence of the chiral condensate is achieved. They report $\delta=0.03$ as the best value. However, we observe that the BR vertex can be written as~:
\bea
\Gamma^\mu_{BR}&=&\Gamma^\mu_{BC}+\delta\left[\frac{1}{F(k)}-\frac{1}{F(p)}
\right][k^2-p^2]\left\{\gamma^\mu(p^2-k^2)+(k+p)^\mu\slsh{q}\right\}\nn\\
&=&\Gamma^\mu_{BC}+\delta\tau_6 T^\mu_6\;,
\eea
with
\be
\tau_6=(k^2-p^2)\left[\frac{1}{F(k)}-\frac{1}{F(p)}\right]\;.
\ee
Charge conjugation symmetry for the vertex requires $\tau_6$ to be antisymmetric under the interchage of $k$ and $p$, but the BR vertex is symmetric. Besides, for $\delta=0.03$ this vertex does not agree with perturbative results, and therefore we neglect it as a good choice.

\subsection{Dong-Munczek-Roberts Vertex}
\noindent
Dong \etal \cite{Dong} have proposed a vertex (DMR) that satisfies WGTI and ensures gauge covariance of the Fermion Propagator in the massless case by means of the so-called Transversality Condition~:
\be
\int\frac{d^dk}{(2\pi)^d}\Delta_{\mu\nu}^T(p-k)\gamma^\mu
S_F(k)\Gamma^\nu(k,p)=0\; .
\ee
DMR vertex is constructed under the assumption that the coefficionts of the transverse part are independent of the angle between fermion momenta. In it, the following constraints are set on the transverse coefficients~:
\bea
f^i&=&0\hspace{5mm}\mbox{for}\hspace{5mm}i\ne3,8\nn\\
f^3&=&\frac{1}{2}\left(\frac{d}{2}-1\right)f^8\nn\\
f^8&=&\frac{1}{\frac{d}{2}-1}\frac{(d-1)I_3}{I_1-I_3}\;,
\eea
$d$ is the number of space-time dimensions and
\bea
I_1&=&k^2p^2{\cal I}_1\nn\\
I_2&=&\frac{1}{2}\left((k^2+p^2) {\cal I}_1-1 \right)\nn\\
I_3&=&\frac{1}{2}[k^2+p^2]I_2\; ,
\eea
where they define
\bea
{\cal I}_n&=&\int d\Omega_d\frac{1}{(k-p)^{2n}}\; ,
\eea
being the solid angle
\bea
\hspace{-3mm}
\int d\Omega_d&\equiv& \frac{1}{{\cal N}}\left[\int_0^\pi
d\theta_2\sin^{d-2}\theta_2 \int_0^\pi d\theta_3
\sin^{d-3}\theta_3\ldots \int_0^{2\pi}d\theta_{d-1}\right]\;,
\eea
and
\bea
{\cal N}&=&\frac{2\pi^{d/2}}{\Gamma(d/2)}\;.
\eea
Functions $f^i$ are related to the $\tau_i$ in the form~:
\be
\tau_i = \frac{1}{k^2-p^2}\left[\frac{1}{F(k)}-\frac{1}{F(p)}\right]f^i\;.
\label{rel_tau_f}
\ee
In this construction it is shown that the effect of including only
$\tau_6$ as in the CP vertex can also be reproduced by
considering $\tau_3$ and $\tau_8$.  This vertex, however, exhibit
logarithmic kinematic singularities.  It has the correct charge
conjugation properties and leads to a multiplicatively
renormalizable Fermion Propagator.

\subsection{Burden-Tjiang Vertex}

\noindent
Using a similar reasoning, Burden and Tjiang~\cite{BT1} have
deconstructed a one-parameter ans\"atze family for massless QED3.
Constraints on the transverse coefficients in the BT vertex are
the following~:
\bea
\bar{f}&=&-2(1+\beta)\frac{I(k,p)}{J(k,p)}\nn\\
f^3 &=&-\beta\frac{I(k,p)}{J(k,p)}\nn\\
f^6&=&0\;,
\eea
where
\bea
I(k,p)&=&\frac{(k^2+p^2)^2}{16kp}\ln{\left(\frac{(k+p)^2}{(k-p)^2}\right)}-\frac{1}{4}(k^2+p^2)\nn\\
J(k,p)&=& \frac{(k^2-p^2)^2}{16kp}\ln{\left(\frac{(k+p)^2}{(k-p)^2}\right)}-\frac{1}{4}(k^2+p^2)\;,
\eea
They have used the relation~(\ref{rel_tau_f}) and they define
\be
\bar{\tau}=\tau_8+(k^2+p^2)\tau_2\;.
\ee
Choosing $\beta=1$ leads to the DMR vertex. This deconstruction is based upon the assumptions that the transverse vertex vanishes in Landau gauge and takes $\tau_6=0$. Transversality Condition is the strongest argument to ensure the gauge covariance of the Fermion Propagator. A possible drawback mentioned by the authors is that for any value of $\beta$, for $k^2=p^2$ but $k_\mu\ne p_\mu$, the BT vertex exhibits a logarithmic singularity.

A criticism to ths vertex has been exposed by Bashir \etal \cite{BKP2}. They assure that $\tau_6$ cannot be taken as zero and that the transverse coefficients should depend upon the angle between fermion momenta. One important observation is that the parameter $\beta$, introduced explicitly gauge parameter independent by the authors, should have such dependence as pointed out by the corresponding perturbative calculation. Finally, in this work they point out  that the Transversality Condition is not valid beyond the one-loop level.

\subsection{Bashir-Pennington Vertex}
With the assumptions that the transverse vertex vanishes in Landau gauge and thet it has no dependence upon the angle between fermion momenta, Bashir and Pennington \cite{BP1}. have constructed a vertex that satisfies~:
\begin{itemize}
\item WGTI,

\item that the Fermion Propagator is multiplicatively renormalizable,

\item that with the correct choice of the functions that define
this vertex, it agrees with Perturbation Theory in the weak
coupling regime is warrantied,

\item offers a strictly gauge independent critical coupling.
\end{itemize}

BP Vertex imposes integral and diferential contraints on the functions that define its transverse part. For the massless case,  K{\i}z{\i}lers{\"u}, Bashir and Pennington~\cite{BKP1} removed the initial assumptions, including all the corrections in the exponent of the power law for the wavefunction renormalization. This is done by introducing new functions and constraints on them.

Continuing with the programme, we should start an ambitious quest towards the construction  of a vertex for the fermion-boson interaction in massive QED3, taking into account all the nice features of the above mentioned vertices, but getting rid of their drawbacks. It is clear that the inclusion of the WGTI is common in all the vertex ans\"atze. This is due to the fact that its implementation is straghtforward. Demanding correct gauge behavior of Green's functions is a nontrivial requirement. We will show thos in the next chapter, where we study the LKF transformation for the Fermion Propagator at the tree level.
\newpage
\section*{Appendix: Tables}
Below we display the Tables that contain the numeric results of our calculation~:
\vspace{1cm
\hspace{5mm
\begin{tabular}{|crr|}  \hline %\hline
%&     &    \\
\multicolumn{1}{|c}{$\xi$}  &
\multicolumn{1}{c}{$p$} &
\multicolumn{1}{c|}{$\frac{4}{2+\xi}\,p^2\,M(p)$}  \\
%&     &    \\
                               \hline  \hline
%&     &    \\
0.0 &  1000      &  2.31109     \\
    &  642.233   &  2.31117     \\
    &  316.228   &  2.31119     \\
%&     &    \\
                        \hline
%&     &    \\
0.5 &  1000      &  3.61103     \\
    &  642.233   &  3.61119     \\
    &  316.228   &  3.61124     \\
%&     &    \\
                        \hline
%&     &    \\
1.0 &  1000      &  5.19982     \\
    &  642.233   &  5.20009     \\
    &  316.228   &  5.20017     \\
%&     &    \\
                       \hline
%&     &    \\
1.2 &  1000      &  5.91622     \\
    &  642.233   &  5.91654     \\
    &  316.228   &  5.91664     \\
%&     &    \\
                        \hline
%&     &    \\
1.5 &  1000      &  7.07745     \\
    &  642.233   &  7.07787     \\
    &  316.228   &  7.07800     \\
%&     &    \\
                        \hline
%&     &    \\
2.0 &  1000      &  9.24390     \\
    &  642.233   &  9.24452     \\
    &  316.228   &  9.24473     \\
%  &     &    \\
                                 \hline   %\hline
\end{tabular}
\begin{tabular}{|crr|}  \hline %\hline
%&     &    \\
\multicolumn{1}{|c}{$\xi$}  &
\multicolumn{1}{c}{$p$} &
\multicolumn{1}{c|}{$\frac{4}{2+\xi}\,p^2\,M(p)$}  \\
%&     &    \\
                                 \hline  \hline
%&     &    \\
2.5 &  1000      &  11.6992     \\
    &  642.233   &  11.7001     \\
    &  316.228   &  11.7003     \\
%&     &    \\
                        \hline
%&     &    \\
3.0 &  1000      &  14.4432     \\
    &  642.233   &  14.4444     \\
    &  316.228   &  14.4448     \\
%&     &    \\
                        \hline
%&     &    \\
3.5 &  1000      &  17.4761     \\
    &  642.233   &  17.4777     \\
    &  316.228   &  17.4782     \\
%&     &    \\
                       \hline
%&     &    \\
4.0 &  1000      &  20.7977     \\
    &  642.233   &  20.7998     \\
    &  316.228   &  20.8005     \\
%&     &    \\
                        \hline
%&     &    \\
4.5 &  1000      &  24.4081     \\
    &  642.233   &  24.4108     \\
    &  316.228   &  24.4117     \\
%&     &    \\
                       \hline
%&     &    \\
5.0 &  1000      &  28.3073     \\
    &  642.233   &  28.3106     \\
    &  316.228   &  28.3117     \\
%  &     &    \\
                                 \hline   %\hline
\end{tabular}} } \label{table1} \\ \\ \\
{\centerline {TABLE~1. Condensate in Various Gauges for $F(p)=1$}}
%\vspace{5mm}
%
\newpage
\vspace{1cm
\hspace{15mm
\begin{tabular}{|crrr|}  \hline
&     &     &   \\
\multicolumn{1}{|c}{$\xi$}  &
\multicolumn{1}{c}{$p$} &
\multicolumn{1}{c}{$\frac{4}{2+\xi}\,p^2\,M(p)$} &
\multicolumn{1}{c|}{$\frac{4}{2+\xi}\,p^2\,M(p)$} \\
&     &     &   \\
&     & \small{BHR}    &  \small{BR} \\   \hline  \hline
%\hline \hline
&     &     &   \\
0.0   &  1000      &  2.31109   &  2.316 \\
      &  642.233   &  2.31117   &   \\
      &  316.228   &  2.31119   &   \\
&     &     &   \\
  \hline
&     &     &   \\
0.5   &  1000      &  1.77309   &  1.775 \\
      &  642.233   &  1.77313   &   \\
      &  316.228   &  1.77306   &   \\
&     &     &   \\
 \hline
&     &     &   \\
1.0   &  1000      &  1.44791   &  1.447 \\
      &  642.233   &  1.44793   &   \\
      &  316.228   &  1.44780   &   \\
&     &     &   \\
 \hline
&     &     &   \\
1.2   &  1000      &  1.35288   &  1.352 \\
      &  642.233   &  1.35288   &   \\
      &  316.228   &  1.35274   &   \\
&     &     &   \\
 \hline
&     &     &   \\
1.5   &  1000      &  1.23591   &   \\
      &  642.233   &  1.23591   &   \\
      &  316.228   &  1.23574   &   \\
&     &     &   \\
 \hline
&     &     &   \\
2.0   &  1000      &  1.09014   &   \\
      &  642.233   &  1.09012   &   \\
      &  316.228   &  1.08992   &   \\
      &     &     &   \\   \hline
\end{tabular} } }
\newpage
\hspace{15mm
\begin{tabular}{|crrr|}  \hline
&     &     &   \\
\multicolumn{1}{|c}{$\xi$}  &
\multicolumn{1}{c}{$p$} &
\multicolumn{1}{c}{ $\frac{4}{2+\xi}\,p^2 \,M(p)$ } &
\multicolumn{1}{c|}{$\frac{4}{2+\xi}\,p^2\,M(p)$} \\
&     &     &   \\
&     & \small{BHR}    &  \small{BR} \\   \hline  \hline
&     &     &   \\
%\hline
2.5   &  1000      &  0.98597   &   \\
      &  642.233   &  0.98594   &   \\
      &  316.228   &  0.98571   &   \\
&     &     &   \\
  \hline
&     &     &   \\
3.0   &  1000      &  0.90941   &   \\
      &  642.233   &  0.90938   &   \\
      &  316.228   &  0.90912   &   \\
&     &     &   \\
 \hline
&     &     &   \\
3.5   &  1000      &  0.85201   &   \\
      &  642.233   &  0.85197   &   \\
      &  316.228   &  0.85169   &   \\
&     &     &   \\
   \hline
&     &     &   \\
4.0   &  1000      &  0.80839   &   \\
      &  642.233   &  0.80833   &   \\
      &  316.228   &  0.80803   &   \\
&     &     &   \\
 \hline
&     &     &   \\
4.5   &  1000      &  0.77497   &   \\
      &  642.233   &  0.77491   &   \\
      &  316.228   &  0.77458   &   \\
&     &     &   \\
 \hline
&     &     &   \\
5.0   &  1000      &  0.74933   &   \\
      &  642.233   &  0.74927   &   \\
      &  316.228   &  0.74891   &   \\
&     &     &   \\
 \hline
\end{tabular}}
\\ \\ \\
{\centerline {TABLE~2.  Condensate In Various Gauges Including
$F(p)$}}
%\vspace{5mm}
\newpage
%\vspace{5mm
\hspace{0.5cm
\begin{tabular}{|crr|}  \hline %\hline
%&     &    \\
\multicolumn{1}{|c}{$\xi$}  &
\multicolumn{1}{c}{$p$} &
\multicolumn{1}{c|}{$\frac{4}{2+\xi}\,p^2\,M(p)$}  \\
%&     &    \\
                                 \hline  \hline
%&     &    \\
0.0 &  1000      &  2.31109     \\
    &  642.233   &  2.31117     \\
    &  316.228   &  2.31119     \\
%&     &    \\
                      \hline
%&     &    \\
0.5 &  1000      &  4.30196     \\
    &  642.233   &  4.30221     \\
    &  316.228   &  4.30248     \\
%&     &    \\
                        \hline
%&     &    \\
1.0 &  1000      &  6.93473     \\
    &  642.233   &  6.93529     \\
    &  316.228   &  6.93609     \\
%&     &    \\
                       \hline
%&     &    \\
1.2 &  1000      &  8.16826     \\
    &  642.233   &  8.16900     \\
    &  316.228   &  8.17011     \\
%&     &    \\
                       \hline
%&     &    \\
1.5 &  1000      &  10.2122     \\
    &  642.233   &  10.2133     \\
    &  316.228   &  10.2150     \\
%&     &    \\
                       \hline
%&     &    \\
2.0 &  1000      &  14.1356     \\
    &  642.233   &  14.1375     \\
    &  316.228   &  14.1406     \\
%  &     &    \\
                               \hline   %\hline
\end{tabular}
\begin{tabular}{|crr|}  \hline %\hline
%&     &    \\
\multicolumn{1}{|c}{$\xi$}  &
\multicolumn{1}{c}{$p$} &
\multicolumn{1}{c|}{$\frac{4}{2+\xi}\,p^2\,M(p)$}  \\
%&     &    \\
                             \hline  \hline
%&     &    \\
2.5 &  1000      &  18.7057     \\
    &  642.233   &  18.7085     \\
    &  316.228   &  18.7136     \\
%&     &    \\
                     \hline
%&     &    \\
3.0 &  1000      &  23.9226     \\
    &  642.233   &  23.9268     \\
    &  316.228   &  23.9345     \\
%&     &    \\
                      \hline
%&     &    \\
3.5 &  1000      &  29.7865     \\
    &  642.233   &  29.7924     \\
    &  316.228   &  29.8037     \\
%&     &    \\
                      \hline
%&     &    \\
4.0 &  1000      &  36.2975     \\
    &  642.233   &  36.3056     \\
    &  316.228   &  36.3212     \\
%&     &    \\
                      \hline
%&     &    \\
4.5 &  1000      &  43.4556     \\
    &  642.233   &  43.4663     \\
    &  316.228   &  43.4872     \\
%&     &    \\
                      \hline
%&     &    \\
5.0 &  1000      &  51.2608     \\
    &  642.233   &  51.2746     \\
    &  316.228   &  51.3020     \\
 % &     &    \\
                                \hline   %\hline
\end{tabular}}  \\ \\ \\
{\centerline {TABLE~3. Condensate in Various Gauges for $F(p)=1$ using WGTI}}
\vspace{5mm}

\chapter{Landau-Khalatnikov-Fradkin  Transformations and the Fermion Propagator}
\pagestyle{myheadings}
\markboth{Landau-Khalatnikov-Fradkin  Transformations and the Fermion Propagator}
{Landau-Khalatnikov-Fradkin  Transformations and the Fermion Propagator}

\section{Introduction}
\noindent
Looking for the gauge independence of the physical observables, we found that the Ward-Green-Takahashi Identity (WGTI) is a necessary condition for it, but not sufficient. We see, therefore, the need for the incorporation of other gauge invariance constraints for this purpose. We incorporate the Landa-Khalatnikov-Fradnin (LKF) transformations, \cite{LK1,LK2,F1,JZ1,Z1}, which tell us the manner in which Green's functions change under a variation of gauge. Rules governing these transformations are far from simple. In counterpart, WGTI \cite{W1,G1,T1}, are more simple, and therefore, they have been widely implemented. We can make larger this set of identities by transforming also the gauge parameter $\xi$ \cite{Nielsen,Sibold}, and arrive to the Nielsen Identities (NI). One advantage of these identities over the conventional Ward identities is that $\partial /\partial \xi$ becomes part of the new relations involving Green's functions. This fact was exploited in \cite{Steele,Grassi} to prove the gauge independenc
e of some quantities related to two-point Green's functions at the one-loop level and to all orders in Perturbation Theory, respectively. Since it is a difficult task to establish tha gauge independence of the physical observables in the Schwinger-Dyson Equations (SDE) study, NI can play a significant role in addressing this issue, along with the WGTI and the LKF transformations. However, in this chapter, we have focused only in the later.

 The LKF transformation for the three-point vertex is complicated
and hampers direct extraction of analytical restrictions on its
structure.  Burden and Roberts, \cite{BR1}, carried out a
numerical analysis to compare the self-consistency of various
{\em ansatze} for the vertex, \cite{BC,CP1,H1}, by means of its
LKF transformation.  In addition to these numerical constraints,
indirect analytical insight can be obtained on the
nonperturbative structure of the vertex by demanding correct
gauge covariance properties of the fermion propagator.  In the
context of gauge technique, examples are
\cite{Keck2,Keck3,Waites1}.  Concerning the works based upon
choosing a vertex {\em ansatze}, references
\cite{Dong,CP1,BP1,BP2,BT1,BKP1} employ this idea \footnote{A
criticism of the vertex construction in \cite{BT1} was raised in
\cite{BKP2}.}.  However, all the work in the later category has
been carried out for massless QED3 and QED4.  The masslessness of
the fermions implies that the fermion propagator can be written
only in terms of one function, the so called wavefunction
renormalization, $F(p)$.  In order to apply the LKF transform,
one needs to know a Green function at least in one particular
gauge.  This is a formidable task.  However, one can rely on
approximations based on perturbation theory.  It is customary to
take $F(p)=1$ in the Landau gauge, an approximation justified by
one loop calculation of the massless fermion propagator in
arbitrary dimensions, see for example, \cite{davydychev}.  The
LKF transformation then implies a power law for $F(p)$ in QED4
and a simple trigonometric function in QED3.  To improve upon
these results, one can take two paths:  (i) incorporate t he
information contained in higher orders of perturbation theory and
(ii) study the massive theory.  As pointed out in \cite{BKP1}, in
QED4, the power law structure of the wavefunction renormalization
remains intact by increasing order of approximation in
perturbation theory although the exponent of course gets
contribution from next to leading logarithms and so on.  In
\cite{BKP1}, constraint was obtained on the 3-point vertex by
considering a power law where the exponent of this power law was
not restricted only to the one loop fermion propagator.  In QED3,
the two loop fermion propagator was evaluated in
\cite{adnan1,BKP2,BKP3}, where it was explicitly shown that the
the approximation $F(p)=1$ is only valid upto one loop, thus
violating the {\em transversality condition} advocated in
\cite{BT1}.  The result found there was used in
\cite{adnan2} to find the improved LKF transform.

In the present Chapter, we calculate the LKF transformed fermion
propagator in massive QED3 and QED4 \footnote{In the context of
gauge technique, gauge covariance of the spectral functions in
QED was studied in \cite{Keck2,Keck3,Waites1}.}.  We start with
the simplest input which corresponds to the lowest order of
perturbation theory, i.e., $S(p)=1/ i \slsh{p}-m$ in the Landau
gauge.  On LKF transforming, we find the fermion propagator in an
arbitrary covariant gauge.  In the case of QED3, we obtain the
result in terms of basic functions of momenta.  In QED4, the
final expression is in the form of hypergeometric functions.
Coupling $\alpha$ enters as parameter of this transcendental
function.  A comparison with perturbation theory needs the
expansion of the hypergeometric function in terms of its
parameters.  We use the technique de\-ve\-lo\-ped by Moch {\em et.
al.}, \cite{Moch}, for the said expansion.  We compare our
results with the one loop expansion of the fermion propagator in
QED4 and QED3, \cite{Reenders,BashirR1}, and find perfect
agreement upto terms independent of the gauge parameter at one
loop, a difference permitted by the structure of the LKF
transformations.  We believe that the incorporation of LKF
transformations, along with WGTI, in the SDE can play a key role
in addressing the problems of gauge invariance.  For example, in
the study of the SDE of the fermion propagator, only those
assumptions should be permissible which keep intact the correct
behaviour of the Green functions under the LKF transformations,
in addition to ensuring that the WGTI is satisfied.  It makes it
vital to explore how two and three point Green functions
transform in a gauge covariant fashion.  In this Chapter, we
consider only a two point function, namely, the fermion
propagator.

\section{Fermion Propagator and the LKF Transformation}

\noindent
We start by expanding out the fermion propagator, in momentum and
coordinate spaces
respectively, in its most general form as follows~:
\bea
S_F(p;\xi)&=& A(p;\xi) + i  \frac{B(p;\xi)}{{\not \! p}}   \equiv
\frac{F(p;\xi)}{i {\not \! p}-{\cal M}(p;\xi)}
\;,
\label{fpropmoment} \\
\nonumber   \\
S_F(x;\xi)&=&{\not \! x}X(x;\xi)+ Y(x;\xi)
\label{fpropcoord} \;,
\eea
where we explicitly write the gauge in which we specify the Fermion Propagator. Motivated from the
lowest order perturbation theory, we take
\be
F(p;0)=1\hspace{.5cm}\mbox{and}\hspace{.5cm} {\cal M}(p;0)=m \;.
\label{lowestFM}
\ee
Perturbation theory also reveals that this result continues to hold true
to one loop order for the wavefunction renormalization.
Eqs.~(\ref{fpropmoment},\ref{fpropcoord}) are related to each
other through the following Fourier transforms~:
\bea
S_F(p;\xi)&=&\int d^dx \sexp^{i p\cdot x}S_F(x;\xi)   \label{pFourier} \\
S_F(x;\xi)&=&\int\frac{d^dp}{(2\pi )^d}\sexp^{-i p\cdot x}S_F(p;\xi)
\;,
\label{xFourier}
\eea
where $d$ is the dimension of space-time. Let us recall that the LKF transformation
relating the coordinate space fermion propagator in Landau gauge
to the one in an arbitrary covariant gauge reads~(\ref{lkfrev})~:
\begin{eqnarray*}
S_F(x;\xi) = S_F(x;0)\sexp^{-i [\Delta_d(0)-\Delta_d(x)]} \;,
\end{eqnarray*}
and, also that~(\ref{deltadlkf})
\begin{eqnarray*}
\Delta_d(x)=-i \xi e^2\mu^{4-d}\int\frac{d^dp}{(2\pi )^d}
\frac{\sexp^{- i p\cdot x}}{p^4} \; .
\end{eqnarray*}
Taking $\psi$ to be the angle between $x$ and $p$, we can write
\begin{eqnarray*}
d^dp=dp p^{d-1}\sin^{d-2}\psi d\psi \Omega_{d-2}\;,
\end{eqnarray*}
 where
$\Omega_{d-2}=2 \; \pi^{(d-1)/2}/\Gamma\left((d-1)/2 \right)$. Hence
\be
\Delta_d(x)=-i \xi e^2\mu^{4-d}f(d)\int_0^\infty dp p^{d-5}
\int_0^{\pi} d\psi \sin^{d-2}\psi \sexp^{-i p x \cos \psi} \;,
\ee
where $f(d)=\Omega_{d-2}/(2\pi )^d$. Performing angular and radial
integrations, we arrive at the following equation
\be
\Delta_d (x)=-\frac{i \xi e^2}{16 (\pi)^{d/2}}
(\mu x)^{4-d}\Gamma\left(\frac{d}{2}-2\right) \;.\label{deltad}
\ee
With these tools at hand, the procedure now is as follows~:
\begin{itemize}
\item Start with the lowest order fermion propagator and Fourier
transform it to coordinate space.

\item Apply the LKF transformation law.

\item Fourier transform the result back to momentum space.

\end{itemize}

\section{Three Dimensional Case}

\noindent
Employing
Eqs.~(\ref{fpropmoment},\ref{fpropcoord},\ref{lowestFM},\ref{xFourier}),
the lowest order three dimensional fermion propagator in Landau gauge
in the position space is given by
\bea
X(x;0)&=& -\frac{\sexp^{-m x}(1+ m x)}{4\pi x^3} \;,\\ \nonumber
Y(x;0)&=& -\frac{m \sexp^{-m x}}{4\pi x}  \;.
\eea
Once in the coordinate space, we can apply the LKF transformation
law using expression (\ref{deltad}) explicitly in three dimensions~:
\be
\Delta_3(x)=-\frac{i \alpha \xi x}{2} \;,
\ee
where $\alpha=e^2/4 \pi$. The fermion propagator in an arbitrary gauge
is then
\bea
S_F(x;\xi)=S_F(x;0)\sexp^{-(\alpha\xi/2)x} \;.
\eea
For Fourier transforming back to momentum space, we use
\bea
A(p;\xi)&=&-\frac{F(p;\xi ){\cal M}(p;\xi)}{p^2+{\cal M}^2(p;\xi )}
=\int d^3x \; \sexp^{i p\cdot x} \; Y(x;\xi) \, \\ \nonumber
i B(p;\xi)&=&-\frac{i p^2F(p;\xi)}{p^2+{\cal M}^2(p;\xi)}
=\int d^3x \; p \cdot x \; \sexp^{i p\cdot x} \; X(x;\xi) \;.
\eea
Performing the angular integration, we get
\bea
A(p;\xi)&=& - \frac{m}{p} \; \int_0^{\infty} dx \; \sin{px} \,
{\sexp}^{-(m+\alpha \xi/2)x}    \;, \\ \nonumber \\
 B(p;\xi)&=& \frac{1}{p} \; \int_0^{\infty} \frac{dx}{x^2} \, (1+mx) \,
\left[ px \cos px - \sin px \right] {\sexp}^{-(m+\alpha \xi/2)x} \;,
\eea
and the radial integration then yields
\bea
A(p;\xi)&=&-\frac{4 m}{4 p^2 + \left( 2 m+ {\alpha \xi} \right)^2 } \\
B(p;\xi)&=&- \frac{4p^2+\alpha\xi (2m+\alpha\xi)}{4p^2+(2m+\alpha\xi)^2}
+ \frac{\alpha\xi}{2p}\arctan{ \left[ 2 p/(2 m+ {\alpha\xi})
\right]} \;.
\eea
We can now arrive at the following expressions for the wavefunction
re\-nor\-ma\-li\-za\-tion and the mass function, respectively~:
\bea
F(p;\xi)&=&-\frac{\alpha\xi}{2p}
\arctan{\left[2p/(2m+\alpha\xi) \right]}
+ \!\frac{2p(4p^2\!+\!\alpha^2\xi^2)}{ \phi(p;\xi)  }\nn\\
 &&-\frac{\!\alpha\xi (4p^2\!+\!\alpha\xi
(2m\!+\!\alpha\xi))
\arctan{[2p/(2m\!+\!\alpha\xi)]}}{\phi(p;\xi)} \;,  \label{LKFF3} \\
{\cal M}(p;\xi)&=&\frac{8p^3 m}{\phi(p;\xi) } \;,   \label{LKFM3}
\eea
where
\be
\phi(p;\xi)=2p(4p^2\!+\!\alpha\xi(2m\!+\!\alpha\xi))\!-\!
\alpha\xi(4p^2\!+\!(2m\!+\!\alpha\xi)^2)\arctan{\left[2p/(2m\!+\!\alpha\xi)
\right]} \;.
\ee
In the massless limit, we immediately recuperate the well-known results~:
\bea
F_{\rm nm}(p;\xi)&=&1-\frac{\alpha\xi}{2p}\arctan{\frac{2p}{\alpha\xi}}
\;,\\ \nonumber
{\cal M}_{\rm nm}(p;\xi)&=&0 \;.
\eea
In the weak coupling, we can expand out Eqs.(\ref{LKFF3},\ref{LKFM3})
in powers of $\alpha$. To ${\cal O}(\alpha)$, we find
\bea
F(p;\xi)&=&1+\frac{\alpha\xi}{2p^3}\left[(m^2-p^2) \; {\rm
arctan} \; [p/m] - m p\right] \;,   \\
{\cal M}(p;\xi)&=&m\left[ 1+\frac{\alpha\xi}{2p^3}\left\{ (m^2+p^2) \; {\rm
arctan} \; [p/m] - m p \right\} \right] \;. \label{LKFM1lazo}
\eea
Let us compare these results with the ones obtained in ~\cite{BashirR1}~:
\bea
\hspace{-8mm}F_{1-{\rm loop}}(p;\xi)&=&1+\frac{\alpha\xi}{2p^3}\left[(m^2-p^2) \; {\rm
arctan} \; [p/m] - m p\right] \;,   \label{QED3F1lazo} \\
\hspace{-8mm}{\cal M}_{1-{\rm loop}}(p;\xi)&=&m\left[ 1\!+\!\frac{\alpha}{2p^3}
\left\{ [\xi (m^2\!+\!p^2) \!+\! 4 p^2]  {\rm
arctan}  [p/m] \!-\!\xi m p \right\} \right] \;.    \label{QED3M1lazo}
\eea

We of course only expect the results to be in agreement upto a term
proportional to $\alpha$, as allowed by the structure of the LKF
transformations. There is no such term in
Eq.~(\ref{QED3F1lazo}). Therefore, the agreement is exact.
Eq.~(\ref{LKFM1lazo}) and Eq.~(\ref{QED3M1lazo}) become identical only
after we subract out the non-vanishing term in the Landau gauge from
Eq.~(\ref{QED3M1lazo}) to write out the {\em subtracted} mass function
at one loop as~:
\bea
{\cal M}_{1-{\rm loop}}^{S}(p;\xi)&=&m\left[ 1+\frac{\alpha\xi}{2p^3}
\left\{  (m^2+p^2)   {\rm
arctan}  [p/m] - m p \right\} \right] \;.
\eea
One can numerically check that without the above mentioned subtraction,
Eqs.~(\ref{LKFM1lazo},\ref{QED3M1lazo}) approach the same value only in the
large momentum regime.

\section{Four Dimensional Case}

\noindent
Employing
Eqs.~(\ref{fpropmoment},\ref{fpropcoord},\ref{lowestFM},\ref{xFourier}),
the
lowest order four dimensional fermion pro\-pa\-ga\-tor in coordinate space
is given by
\bea
X(x;0)&=&-\frac{m^2}{4\pi^2x^2} \;  K_2(mx)  \;,\\
Y(x;0)&=&-\frac{m^2}{4\pi^2x} \; K_1(mx) \;,
\eea
where $K_1$ and $K_2$ are Bessel functions of the second kind. In
order to apply the LKF transformation in four dimensions, we expand
Eq.~(\ref{deltad}) around $d=4- \epsilon$ and use the following identities
\begin{eqnarray*}
\Gamma\left(-\frac{\epsilon}{2}\right)&=&-\frac{2}{\epsilon}-\gamma
+{\cal O}(\epsilon)  \;, \\ \nonumber
x^{\epsilon}&=&1+\epsilon \ln{x}+{\cal O}(\epsilon^2) \;,
\end{eqnarray*}
to obtain
\be
\Delta_4(x)=i \frac{\xi e^2}{16\pi^{2-\epsilon/2}}
\left[ \frac{2}{\epsilon}+\gamma+2\ln{\mu x}+{\cal O}(\epsilon)\right]
\; .
\ee
Note that we cannot write a similar expression for $\Delta_4(0)$
because of the presence of the term proportional to
$\ln{x}$. Therefore, we introduce a cut-off scale $x_{min}$. Now
\be
\Delta_4(x_{min})-\Delta_4(x)=-i \ln{\left(\frac{x^2}{x_{min}^2}\right)^\nu}\;,
\ee
where $\nu=\alpha\xi/4\pi$. Hence
\be
S_F(x;\xi)=S_F(x;0)\left(\frac{x^2}{x_{min}^2}\right)^{-\nu} .
\ee
For Fourier transforming back to momentum space we use the following
expressions
\bea
A(p;\xi)&=&-\frac{F(p;\xi ){\cal M}(p;\xi)}{p^2+{\cal M}^2(p;\xi )}
=\int d^4x \; \sexp^{i p\cdot x} \; Y(x;\xi)  \\ \nonumber
i B(p;\xi)&=&-\frac{i p^2F(p;\xi)}{p^2+{\cal M}^2(p;\xi)}
=\int d^4x \; p \cdot x \; \sexp^{i p\cdot x}    \; X(x;\xi) \;.
\eea
On carrying out angular integration, we obtain~:
\bea
A(p;\xi)&=& - \frac{m^2}{p} \, x_{min}^{2 \nu} \;
\int_0^{\infty} dx x^{-2 \nu+1} \; K_1(mx) \, J_1(px) \;, \\
B(p;\xi)&=& - m^2 \;
\int_0^{\infty} dx x^{-2 \nu+1} \; K_2(mx) \, J_2(px) \;.
\eea
The radial integration then yields~:
\bea
\hspace{-8mm}A(p;\xi)&=&- \frac{1}{m} \left( \frac{m^2}{\Lambda^2} \right)^{\nu}
\Gamma(1\!-\!\nu)\Gamma(2\!-\!\nu)
{~}_2F_1\left(1\!-\!\nu,2\!-\!\nu;2;-\frac{p^2}{m^2}\right)  \;, \\
\nonumber \\
\hspace{-8mm}B(p;\xi)&=& -\frac{p^2}{2 m^2}
\left( \frac{m^2}{\Lambda^2} \right)^{\nu}
\Gamma(1\!-\!\nu)
\Gamma(3\!-\!\nu) {~}_2F_1\left(1\!-\!\nu,3\!-\!\nu;3;-\frac{p^2}{m^2}\right) \;,
\eea
where we have identified $2/x_{min} \rightarrow \Lambda$. The above
equations imply
\bea
\nonumber \\ F(p;\xi)&=&
\frac{ \Gamma(1-\nu)}{
2 m^2 \; \Gamma(3-\nu) \;  {~}_2F_1\left(1-\nu,3-\nu;3;
-p^2/m^2 \right)} \; \left( \frac{m^2}{\Lambda^2} \right)^{\nu} \nonumber  \\
&&\hspace{-8mm}
\Bigg[4m^2\Gamma^2(2-\nu) \; {~}_2F_1^2\left(1-\nu,2-\nu;2;
-\frac{p^2}{m^2} \right)\nn\\
&&\hspace{8mm}
+p^2 \; \Gamma^2(3-\nu) \;
_2F_1^2\left( 1-\nu,3-\nu;3;-\frac{p^2}{m^2}\right) \Bigg] \;,
\label{ourresultF}  \\ \nonumber \\
{\cal M}(p;\xi)&=&\frac{2m \; {~}_2F_1\left(
1-\nu,2-\nu;2;- p^2/m^2 \right)}
{(2-\nu) \; {~}_2F_1\left(1-\nu,3-\nu;3;- p^2/m^2 \right)} \;.
\label{ourresultM}
\eea
Eqs.~(\ref{ourresultF},\ref{ourresultM}) constitute the LKF transformation
of Eqs.~(\ref{lowestFM}). We shall now see that although
Eqs.~(\ref{lowestFM}) correspond to the lowest order propagator, their
LKF transformation, Eqs.~(\ref{ourresultF},\ref{ourresultM}), is
nonperturbative in nature and contains information of higher orders.

\subsection{Case $\alpha=0$}

\noindent
Let us switch off the coupling and put $\alpha=0$ which implies
$\nu=0$. Now using the identity
\bea
 _2F_1(1,2;2;-p^2/m^2)&=& \, _2F_1(1,3;3;-p^2/m^2)= (1+p^2/m^2)^{-1} \;,
\eea
it is easy to see that
\be
F(p;\xi)=1\hspace{.5cm}\mbox{and}\hspace{.5cm} {\cal M}(p;\xi)=m \;,
\label{lowestFMxi}
\ee
which coincides with the lowest order perturbative result as expected.

\subsection{Case $m >> p$}

\noindent
In the limit $m >> p$, the hypergeometric functions in
Eqs.~(\ref{ourresultF},\ref{ourresultM}) can be easily
expanded in powers of $p^2/m^2$, using the identity
\be
_2F_1\left(\alpha,\beta;\gamma;-\frac{p^2}{m^2} \right)=1-
\frac{\alpha\beta}{\gamma} \frac{p^2}{m^2} +{\cal O}
\left(\frac{p^2}{m^2} \right)^2 \; .
\ee
Retaining only ${\cal O}(p^2/m^2)$ terms, we arrive at~:
\bea
F(p;\xi)&=&
\frac{\Gamma(1\!-\!\nu) \Gamma(2\!-\!\nu)}{(1\!-\!\nu/2)}
\left( \frac{m^2}{\Lambda^2} \right)^{\nu}  \nn  \\
&& \hspace{5mm}
\left[
1 \!+\! \frac{2 \nu}{3} \left( 1 \!-\! \frac{5 \nu}{8} \right)
\frac{p^2}{m^2} \!+\! {\cal O}
\left(\frac{p^2}{m^2} \right)^2 \right] \;,\nn\\
{\cal M}(p;\xi)&=& \frac{m}{(1-\nu/2)} \left[ 1 + \frac{\nu}{6}
(1-\nu) \frac{p^2}{m^2} + {\cal O}
\left(\frac{p^2}{m^2} \right)^2 \right] \;.
\eea
Now carrying out an expansion in $\alpha$ and substituting $\nu =
\alpha \xi/ 4 \pi$, we get the following ${\cal O}(\alpha)$ expressions~:
\bea
F(p;\xi)&=&1+\frac{\alpha\xi}{4\pi}\left[2\gamma
-\frac{1}{2}+\frac{2p^2}{3m^2}+\ln{\frac{m^2}{\Lambda^2}}\right] \;,
\label{largemF}
\\
{\cal M}(p;\xi)&=&m\left\{1+\frac{\alpha\xi}{8\pi}
\left[1+\frac{p^2}{3m^2}\right]\right\}  \;.  \label{largemM}
\eea
Let us now compare these expressions against the one-loop perturbative
evaluation of the massive fermion propagator, see e.g., \cite{Reenders}~:
\bea
\hspace{-5mm}F_{\rm 1-loop}(p;\xi)&=&1-\frac{\alpha\xi}{4\pi}
\left[ C\mu^\epsilon+\left(1-\frac{m^2}{p^2}\right) (1-L)\right] \;,
 \label{KRPF} \\
\hspace{-5mm}{\cal M}_{\rm 1-loop}(p;\xi)&=&m\!+\!\frac{\alpha m}{\pi}
\left[\left(1\!+\!\frac{\xi}{4}\right)\!+\!\frac{3}{4}(C\mu^\epsilon\!-\!L)
\!+\!\frac{\xi}{4}\frac{m^2}{p^2}(1\!-\!L)\right] \label{KRPMC} \;,
\eea
where
\bea
L&=&\left(1+\frac{m^2}{p^2} \right) \ln{\left(1+\frac{p^2}{m^2}\right)}
\;,
\nonumber \\ \nonumber
C&=&-\frac{2}{\epsilon}-\gamma-\ln{\pi}-\ln{\left(\frac{m^2}{\mu^2}\right)}
\;.
\eea
          Knowing the fermion propagator even in one particular gauge is a
prohibitively difficult task. Therefore, Eqs.~(\ref{lowestFM}) have to
be viewed only as an approximation. For the wavefunction
renormalization $F(p;0)$, this approximation is valid upto one loop
order, whereas, for the mass function, it is true only to the lowest
order. Therefore we cannot expect the LKF transform of
Eqs.~(\ref{lowestFM}) to yield correctly each term in the perturbative
expansion of the fermion propagator. However, it should correctly
reproduce all those terms at every order of expansion which vanish in
the Landau gauge at ${\cal O}(\alpha)$ and beyond. Therefore, we
expect Eq.~(\ref{KRPF}) to be exactly reproduced and Eq.~(\ref{KRPMC}) to be
reproduced upto the terms which vanish in the Landau gauge at
${\cal O}(\alpha)$. After subtracting these terms,  the
resulting {\em subtracted} mass function is~:
\be
{\cal M}_{\rm 1-loop}^S(p;\xi)=
m+\frac{\alpha\xi m}{4\pi}\left[1+\frac{m^2}{p^2}(1-L)\right] \; .
\label{KRPMS}
\ee
In the limit $m\to\infty$, the wavefunction renormalization
acquires the form
\be
F(p;\xi)_{\rm 1-loop} =1+\frac{\alpha\xi}{4\pi}
\left[- C\mu^\epsilon-\frac{1}{2}+\frac{2p^2}{3m^2}\right] \;,
\ee
while the {\em subtracted} mass function is
\be
{\cal M}_{\rm 1-loop}^S(p;\xi)=
m\left\{1+\frac{\alpha\xi}{8\pi}\left[1+\frac{p^2}{3m^2}\right]\right\}
\;.
\ee
The last two expressions are in perfect agreement with
Eqs.~(\ref{largemF},\ref{largemM}) after we make the identification~:
\be
- C\mu^\epsilon \rightarrow  2\gamma + \ln{\frac{m^2}{\Lambda^2}} \;.
\label{cutoffdimreg}
\ee

\subsection{Case of Weak Coupling}

\noindent
The case $m >> p$ is relatively easier to handle as we merely have to expand
$_2F_1(\beta, \gamma; \delta; x)$ in powers of $x$ and retain only the
leading terms. If we want to obtain a series in powers of the coupling
alone, we need the expansion of the hypergeometric functions
in terms of its parameters $\beta$ and $\gamma$. We follow the
technique developed in ~\cite{Moch}. One of the mathematical objects
we shall use for such an expansion are the $Z$-sums defined as~:
\be
Z(n;m_1, \ldots ,m_k;x_1, \ldots , x_k)=
\sum_{n\ge i_1>i_2>\ldots >i_k>0} \frac{x_1^{i_1}}{i_1^{m_1}}\ldots
\frac{x_k^{i_k}}{i_k^{m_k}} \;.
\ee
For $x_1=\ldots =x_k=1$ the definition reduces to the Euler-Zagier
sums, \cite{Zsums1,Zsums2}~:
\be
Z(n; m_1, \ldots ,m_k;1, \ldots ,1)=Z_{m_1,\ldots ,m_k}(n) \;.
\ee
Euler-Zagier sums can be used in the expansion of Gamma functions. For
positive integers $n$ we have~\cite{Moch}:
\be
\Gamma(n+\epsilon)=\Gamma (1+\epsilon)\Gamma (n) \left[
1+\epsilon Z_1(n-1)+\ldots +
\epsilon^{n-1}Z_{11\ldots 1}(n-1) \right] \;. \label{gammaexp}
\ee
The first sum $Z_1(n-1)$, e.g., is just the $(n-1)$-th harmonic number,
$H_{n-1}$, of order 1~:
\be
Z_1(n-1)=\sum_{i=1}^{n-1}\frac{1}{i} \equiv H_{n-1}  \;.
\ee
With these definitions in hand, we proceed to expand a hypergeometric
function, $_2F_1(1+\varepsilon,2+\varepsilon;2;x)$, as an example,
assuming $\vert x \vert<1$~:
\bea
\nonumber _2F_1(1+\varepsilon,2+\varepsilon;2;x)&=&1+
\frac{\Gamma (2)}{\Gamma (1+\varepsilon)
\Gamma(2+\varepsilon)}\nn\\
&&\sum_{n=1}^\infty
\frac{\Gamma(1+\varepsilon+n)\Gamma(2+\varepsilon+n)}{\Gamma(2+n)}
\frac{x^n}{n!}\\ \nonumber
&=&1+\frac{1}{(1+\varepsilon)\Gamma^2(1+\varepsilon)}\nn\\
&&\sum_{n=1}^\infty \frac{(1+\varepsilon+n)(\varepsilon+n)^2
\Gamma^2(\epsilon+n)}{\Gamma(2+n)} \frac{x^n}{n!} \;.
\eea
Employing Eq.~(\ref{gammaexp}), we can expand the last expression
in powers of $\varepsilon$ to any desired order of approximation. We
shall be interested only in terms upto ${\cal O}(\alpha)$.
\bea
_2F_1(1+\varepsilon,2+\varepsilon;2;x)
&=&1+\sum_{n=1}^\infty x^n-\varepsilon\sum_{n=1}^\infty
x^n+\varepsilon
\sum_{n=1}^\infty \frac{2+3n}{n(n+1)}x^n\nn\\
&&+2\varepsilon \sum_{n=1}^\infty
H_{n-1}x^n \;.
\eea
Performing the summations, we obtain
\be
_2F_1(1+\varepsilon,2+\varepsilon;2;x)=\frac{1}{1-x}
\left[ 1-\varepsilon\left\{
1+ \frac{1+x}{x} \; \ln{(1-x)}\right\} \right] \;.
\ee
Similarly,
\bea
_2F_1(1+\varepsilon,3+\varepsilon;3;x)&=&\frac{1}{1-x}\nn\\
&&\hspace{-2cm}-
\varepsilon\left\{ \frac{1}{x}+\frac{3}{2}\frac{1}{1-x}+
\left(\frac{1+x}{x^2}+\frac{2}{1-x} \right)
\ln{(1-x)}\right\} \;.
\eea
Substituting back into Eqs.~(\ref{ourresultF},\ref{ourresultM}) and
identifying $\varepsilon=-\nu$, we obtain
\bea
F(p;\xi)&=&1-\frac{\alpha\xi}{4\pi}
\left[-2\gamma-\ln{\frac{m^2}{\Lambda^2}}+
\left(1-\frac{m^2}{p^2}\right)(1-L)\right]  \;, \\ \nonumber
{\cal M}(p;\xi)&=&m+\frac{\alpha\xi m}{4\pi}\left[1+\frac{m^2}{p^2}
(1-L)\right]  \;,
\eea
which matches exactly onto the one loop result of
Eqs.~(\ref{KRPF},\ref{KRPMS}) after the same identification as before,
i.e., (\ref{cutoffdimreg}). Therefore, we have seen that the LKF
transformation of the bare propagator contains important information
of higher orders in perturbation theory.

LKF transformations, being non perturbative in nature, tell us the non perturbative way in which the Fermion Propagator~\cite{BaRa4} and the Vertex transform under a gauge variation. Most of the works on SDE violate these transformations, with the exception of few of them which include the corresponding transformation to the Fermion Propagator at one-loop level. Since in Perturbation Theory all of the gauge invariance constraints are valid order by order, we can exploit this fact to construct a vertex ansatz such that in its perturbative expansion, it automatically satisfies, along with its associated Fermion Propagator, their corresponding LKF transformation,  as we will do in the next Chapter.
\section*{Apendix}
\noindent
Most of the integrals involved in this Chapter are listed below for a
quick re\-fe\-ren\-ce~\cite{GR,moretables}~:
\bea
\hspace{-5mm}\int_0^\pi d\psi \sin^{d\!-\!2}\psi\cos{\psi} \sexp^{-i p x
\cos{\psi}}&=&- i \sqrt{\pi}\left(\frac{px}{2}\right)^{1\!-\!\frac{d}{2}}
\Gamma\left(\frac{d\!-\!1}{2}\right)J_{\frac{d}{2}}(px) \;,  \\  \nonumber \\
\int_0^{\infty} x^{d/2-1} \; J_{d/2}(ax) &=&
\frac{\Gamma(d/2)}{2^{1-d/2} \; a^{d/2}}  \;.
\eea
For the three dimensional case, the needed integrals are~:
\bea
\int_0^\pi d\theta \; \sin{\theta} \; \sexp^{-i px\cos{\theta}}
&=&\frac{2\sin{px}}{px}   \;, \\ \nonumber \\
\int_0^\pi d\theta \; \cos{\theta} \; \sin{\theta} \; \sexp^{- i
px\cos{\theta}}&=&
2 i \left[ \frac{\cos{px}}{px }-\frac{\sin{px}}{(px)^2} \right] \;,\\
\nonumber \\
\int_0^\infty dp  \;
\frac{p^3}{(p^2+m^2)}   \left[ \frac{\cos{px}}{px}
-\frac{\sin{px}}{(px)^2}\right]
&=&-\frac{\pi}{2} \; \frac{(1+mx)}{x^2} \; \sexp^{-mx} \;, \\ \nonumber \\
\int_0^\infty dp\frac{p\sin{px}}{(p^2+m^2)}&=&\frac{\pi}{2} \sexp^{-mx} \;,
\\
\nonumber \\
\frac{1}{p}\int_0^\infty\frac{dx}{x^2}\sexp^{-ax}[px\cos{px}-\sin{px}]&=&-1
+\frac{a}{p}\arctan{\frac{p}{a}} \;, \\
\nonumber \\
\frac{1}{p}\int_0^\infty\frac{dx}{x}\sexp^{-ax}[px\cos{px}-\sin{px}]&=&
\frac{a}{a^2+p^2} - \frac{1}{p}\arctan{\frac{p}{a}} \;, \\
\nonumber \\
\int_0^\infty dx\sin{px}\sexp^{-(m+ \alpha\xi/2)x}&=&
\frac{p}{\left(m+ \alpha\xi/2 \right)^2+p^2}  \;.
\eea
For the four dimensional case, we used the following integrals in particular~:
\bea
\int_0^{\pi} d \theta {\sin}^2 \theta \sexp^{- i px \cos \theta} &=&
\frac{\pi}{p x} \; J_1(px)  \;, \\ \nonumber \\
\int_0^\infty dp \frac{p^{\nu+1}J_\nu(px)}{(p^2+m^2)^{\mu+1}}&=&
\frac{m^{\nu-\mu}x^\mu}{2^\mu\Gamma(\mu+1)}K_{\nu-\mu}(mx) \;, \\ \nonumber \\
\hspace{-3mm}\int_0^\infty dx x^{-\lambda}K_\mu (ax)J_\nu (bx)&=&
\frac{a^{\lambda\!-\!\nu\!-\!1}  b^\nu}{2^{\lambda\!+\!1}\Gamma(1\!+\!\nu)}
  \Gamma\left(\frac{\nu\!-\!\lambda\!+\!\mu\!+\!1}{2}\right)\nn\\
&&\Gamma\left(\frac{\nu\!-\!\lambda\!-\!\mu\!+\!1}{2}\right)  \\
&  \times &{~}_2F_1
\left(\frac{\nu\!-\!\lambda\!+\!\mu\!+\!1}{2},
\frac{\nu\!-\!\lambda\!-\!\mu\!+\!1}{2};\nu\!+\!1;-\frac{b^2}{a^2}\right) \;.\nn
\eea
Some of the series used in our calculation are as follows~:
\bea
\sum_{n=1}^{\infty}  H_{n-1} x^{n} &=& - \frac{x \ln (1-x)}{1-x} \;, \\
\nonumber \\
\sum_{n=1}^{\infty} \frac{n+1}{n (n+2)} \; x^n &=& - \frac{2+x}{4x} -
\frac{(1+x^2) \ln(1-x)}{2 x^2}  \;, \\ \nonumber \\
\sum_{n=1}^{\infty} \frac{1}{(n+1)(n+2)} \; x^{n} &=& \frac{2-x}{2x} +
\frac{(1-x) \ln(1-x)}{x^2} \;.
\eea

\chapter{Constructing the Vertex}
\pagestyle{myheadings}
\markboth{Constructing the Vertex}
{Constructing the Vertex}

\section{Introduction}

\noindent
In studies on the Dynamical Breaking of Chiral Symmetry (DBCS) in
QED3 in the quenched and unquenched approximations
\cite{BR1,applequist,unquenched1,quenched1,Dong,GSSW,BT1,unquenched3,
unquenched4, unquenched5, unquenched6,GHR1,GMS,GHR2}, we look for
gauge independent physical observables.  For that purpose, we
have seen previously the necessity of including the
Ward-Green-Takahashi Identity (WGTI) as well as the
Landau-Khalatnikov-Fradkin (LKF) transformations.  We know that,
in Perturbation Theory, these gauge invariance constraints are
satisfied order by order, thus, the vertex should be modified in
every order of approximation.  This fact has been exploited by,
for example, ~\cite{adnan1,adnan2,BKP2,BKP3}.  In this Chapter we
carry out the constructionn of the non perturbative Vertex based
on its perturbative counterpart, following the work~\cite{raya1}.

Making use of the WGTI, that relates the Vertex to the Fermion
Propagator, one part of the Vertex, called longitudinal, can be
expressed in terms of the said propagator \cite{BC}.  We perform
the evaluation of the Fermion Propagator at the one-loop level
and hence we determine the longitudinal vertex at the same order.
We also calculate the Full vertex at one-loop, and a mere
substraction of the longitudinal part yields the transverse part,
which is not fixed by the WGTI.  According to the choice of Ball
and Chiu, later modified by K{\i}z{\i}lers{\"u} \etal
\cite{Reenders}, the transverse vertex can be expressed in terms
of 8 independent spin structures.  Vertex should be free from
kinematical singularities.  Ball and Chiu choose the basis in
such a way that the coefficient of every element of this basis is
independently free from kinematical singularities in Feynman
gauge.  It was later shown by K{\i}z{\i}lers{\"u} \etal
\cite{Reenders} that a similar calculation to the one of Ball and
Chiu in an arbitrary covariant gauge does not have the same nice
feature.  Consequently, they proposed a modified basis whose
coefficients are free from kinematical singularities in an
arbitrary covariant gauge.  The calculation in the present
Chapter confirms that all the vectors of the modified basis also
retain this feature for massive QED3.  The final result for the
transverse vertex is written in terms of basic functions of the
momenta in a form suitable for its extension to the
non-perturbative domain, following the ideas of Curtis and
Pennington \cite{CP1}.

Using perturbative constraints as a guide, we carry out a
construction of the non-perturbative vertex, which has no
explicit dependence on the coupling $\alpha$.  This vertex has an
explicit dependence on the gauge parameter $\xi$.  For practical
purposes of the numerical study of DCSB, we also construct an
effective vertex which shifts the angular dependence from the
unknown Fermion Propagator functions to the known basic
functions, without changing its per\-tur\-ba\-ti\-ve pro\-per\-ties at the
one-loop level.  We believe that this vertex should lead to a
more realistic study of the dynamically generated masses through
the corresponding SDEs.

\section{Longitudinal and Transverse Vertex to One-Loop}
\subsection{The Fermion Propagator}
\noindent
One-loop Fermion Propagator can be obtained by evaluating the graph in Diagram~(7).
\begin{center}
\vspace{-1.5cm}
\SetScale{0.7}
\begin{picture}(500,100)(0,0)
%Propagador Completo
\ArrowLine(50,50)(150,50)
\CCirc(100,50){3}{}{}
\PText(145,60)(0)[]{-1}
\PText(100,45)(0)[]{p}
%Propagador desnudo
\ArrowLine(200,50)(300,50)
\PText(295,60)(0)[]{-1}
\PText(250,45)(0)[]{p}
%SDE piece
\Vertex(430,50){1} \Vertex(370,50){1}
\ArrowLine(350,50)(450,50)
\PhotonArc(400,50)(30,0,180){4}{8.5}
\LongArrowArc(400,50)(20,60,120)
\PText(400,45)(0)[]{k}
\PText(400,95)(0)[]{q}
%Signos algebraicos
\PText(175,52)(0)[]{=}
\PText(325,52)(0)[]{-}
\end{picture}\\
\vspace{-10pt}
{\sl Diagram~7~: One-loop Correction to the Fermion Propagator.}
\end{center}
This graph corresponds to the following equation~:
\bea
  i \Sf{p}^{-1} &=& i S_F^{0\; -1}(p) + e^2 \; \int \frac{d^3k}{(2 \pi)^3}
  \; \gamma^{\mu} \,  S_F^0(k) \, \gamma^{\nu} \, \Delta_{\mu \nu}^0(q)
\label{propSDE}   \;,
\eea
where $q=k-p$.The bare Fermion and Photon Propagators are, respectively~:
\bea
      S_F^0(p) &=&  \frac{1}{\slsh{p}-m}  \;,  \nn \\
   \Delta^0_{\mu\nu}(q)&=&- \left[q^2
 g_{\mu\nu}+(\xi-1)q_{\mu}q_{\nu}\right]/q^4 \;, \label{barepropc}
\eea
where $m$ is the bare mass of the fermion and $\Sf{p}$ represents the full propagator, defined in eq.~(\ref{fullprop}).

Taking the trace of Eq.~(\ref{propSDE}), having multiplied it with $\slsh{p}$ and  with $1$ respectively, one can obtain two independent equations. On simplifying, these equations can be written as~:
\bea
\hspace{-5mm}
  \frac{1}{F(p)} &=& 1  + \frac{i 4 \pi \alpha \xi }{p^2}
  \int \!\frac{d^3k}{(2 \pi)^3}\!  \frac{1}{q^4(k^2-m^2)}
\left[ (k^2+p^2) k \cdot p - 2 k^2 p^2 \right]  \label{Fangular}
\\ \nn \\
 \frac{\M(p)}{F(p)} &=& m
-i 4\pi\alpha \, (\xi+2) \int \frac{d^3k}{(2 \pi)^3} \;
\frac{m}{q^2(k^2-m^2)}  \;,   \label{Mangular}
\eea
On Wick rotating to the Euclidean space and carrying out angular and radial integrations, we arrive at~:
\bea
\frac{1}{F(p)}&=&1-\frac{\alpha\xi}{2 p^2}  \,
\left[ m - (m^2+p^2) \, I(p)
\right] \;,\nn\\
\frac{\M(p)}{F(p)}&=&m\left[1+
\alpha (\xi+2) \, I(p) \right] \;, \label{FMradial}
\eea
where we have used the simplifying notation $I(p^2)=({1}/{\sqrt{-p^2}}) \arctan \sqrt{{-p^{2}}/{m^{2}}}$. Equations~(\ref{fullprop}) and~(\ref{FMradial}) form the complete Fermion Propagator at one loop.

\subsection{Longitudinal Vertex to One Loop}
\noindent
We take the longitudinal part of the Vertex  as the one that satisfies the WGTI, i. e., the Ball-Chiu Vertex, eq.~(\ref{Lvertex}).
On substituting eq.~(\ref{FMradial}) in the said expression, we obtain~:
\bea
   \Gamma^{\mu}_{L}(=\Gamma^\mu_{BC}) &=& \left[ 1 + \frac{\alpha \xi}{4} \, \sigma_1 \right]
    \, \gamma^{\mu} \; + \;  \frac{\alpha \xi}{4} \, \sigma_2 \, \left[
{k^{\mu}}\slsh{k} \, +  \, {p^{\mu}}\slsh{p} \, + \,
{k^{\mu}}\slsh{p} \, +  \, {p^{\mu}}\slsh{k}  \right]\nn \\
 &+&
\alpha (\xi+2) \sigma_3 \, \left[ k^{\mu} + p^{\mu}  \right]\;,
\label{1loopLvertex}
\eea
where
\bea
\sigma_1 &=&   \frac{m^2+k^2}{k^2}  \, I(k) \; + \;
 \frac{m^2+p^2}{p^2}  \, I(p) \; - \; m \frac{k^2+p^2}{k^2 p^2} \;,
  \nn \\
\sigma_2 &=& \frac{1}{k^2-p^2} \, \left[
 \frac{m^2+k^2}{k^2}  \, I(k) \; - \;
 \frac{m^2+p^2}{p^2}  \, I(p) \; + \; m \frac{k^2-p^2}{k^2 p^2}
  \right] \;,  \nn \\
\sigma_3 &=& m \;  \left[ I(k) \; - \; I(p)  \right]   \;.
\label{sigmas}
\eea
Eqs.~(\ref{1loopLvertex}) and~(\ref{sigmas}) give the longitudinal part of the
fermion-photon vertex to one loop for the massive QED3.

\subsection{Full Vertex to One Loop}
\noindent
Full Vertex can be obtained from Diagram~(8)~:
\begin{center}
\vspace{-1.5cm}
\SetScale{0.7}
\begin{picture}(500,100)(0,0)
%V\'ertice completo
\ArrowLine(125,75)(75,50)
\ArrowLine(75,50)(125,25)
\Vertex(75,50){1}
\Photon(25,50)(75,50){-3}{4}
\CCirc(75,50){3}{}{}
\PText(50,63)(0)[]{q}
\LongArrow(55,65)(45,65)
\PText(100,72)(0)[]{k}
\PText(100,35)(0)[]{p}
%V\'ertice desnudo
\ArrowLine(275,75)(225,50)
\ArrowLine(225,50)(275,25)
\Vertex(225,50){1}
\Photon(175,50)(225,50){-3}{4}
\PText(200,63)(0)[]{q}
\LongArrow(205,65)(195,65)
\PText(250,72)(0)[]{k}
\PText(250,35)(0)[]{p}
%correcci\'on a un lazo
\ArrowLine(425,75)(375,50)
\ArrowLine(375,50)(425,25)
\Vertex(375,50){1}
\Photon(375,50)(325,50){-3}{4}
\Photon(410,68)(410,32){3}{4}
\PText(350,63)(0)[]{q}
\LongArrow(355,65)(345,65)
\PText(395,72)(0)[]{k-w}
\PText(395,35)(0)[]{p-w}
\PText(418,50)(0)[]{w}
\LongArrow(425,45)(425,55)
%Signos algebraicos
\PText(150,52)(0)[]{=}
\PText(300,52)(0)[]{-}
\end{picture}\\
\vspace{-10pt}
{\sl Diagram~8~: One-loop Correction to the Vertex.}
\end{center}
and can be expressed as~:
\be
\Gamma^{\mu}(k,p)=\,\gamma^{\mu}+\,\Lambda^{\mu} \;.  \label{01loopvertex}
\ee
Using the Feynman rules, $\Lambda^{\mu}$ to
$O(\alpha)$ is simply given by~:
\be
-ie\Lambda^{\mu}\,=\,\int_{M}\frac{d^3w}{(2\,\pi)^3}
(-ie\gamma^{\alpha})i S_F^0(p-w)(-ie\gamma^{\mu})
 i S_F^0(k-w)(-ie\gamma^{\beta})i
\Delta^0_{\alpha\beta}(w) \;,  \label{1loopvertex}
\ee
where the loop integration is to be performed in Minkowski space (in Euclidean space, these definitions are modified by the correspondig factors of $i$). $\Lambda^{\mu}$ can be expressed as~:
\bea
\Lambda^{\mu}&=&-\frac{{\it i}\,{\alpha}}{2\,{\pi}^2}
\Bigg\{ \left[ \gamma^{\alpha} \slsh{p}\,{\gamma^{\mu}}\,\slsh{k}
 \gamma_{\alpha}   + m ( 4 k^{\mu} + 4 p^{\mu} -  \slsh{p}
{\gamma^{\mu}} -  {\gamma^{\mu}}   \slsh{k} ) -m^2
{\gamma^{\mu}}       \right]
{\it J}^{(0)}  \nn \\
&& -  \left[ {\gamma^{\alpha}}
 \slsh{p}\,{\gamma^{\mu}}{\gamma^{\nu}} \gamma_{\alpha}
+ \gamma^{\alpha}  {\gamma^{\nu}} {\gamma^{\mu}} \slsh{k} \gamma_{\alpha}
+ 6 m g^{\mu \nu}     \right]
{{\it J}_{\nu}^{(1)}}
+{\gamma^{\alpha}}{\gamma^{\nu}}{\gamma^{\mu}}{\gamma^{\lambda}}
\gamma_{\alpha}{{\it J}_{\nu\lambda}^{(2)}}\nn\\
&&
+(\xi-1)\Bigg[ {\gamma^{\mu}}{\it K}^{(0)} -
\left[ {\gamma^{\nu}}\slsh{p}\,{\gamma^{\mu}}+\,
{\gamma^{\mu}}\slsh{k}\,{\gamma^{\nu}}+2m g^{\mu \nu}
\right]
{\it J}_{\nu}^{(1)}
\\
&&  +  \left[ {\gamma^{\nu}}\slsh{p}\,
{\gamma^{\mu}}\slsh{k}\,
{\gamma^{\lambda}}
+m(\,{\gamma^{\nu}}\slsh{p}\,{\gamma^{\mu}}\,{\gamma^{\lambda}}+
\,{\gamma^{\nu}}\,{\gamma^{\mu}}\,\slsh{k} \,{\gamma^{\lambda}})
+ m^2\,{\gamma^{\nu}}\,{\gamma^{\mu}}\,{\gamma^{\lambda}}  \right]
{{\it I}_{\nu\lambda}^{(2)}}
\Bigg]
\Bigg\}\nn\;, \label{1loopvertexevaluated}
\eea
where the integrals  $K^{(0)}$, $J^{(0)}$, $J^{(1)}_\mu$, $J^{(2)}_{\mu\nu}$, $I^{(0)}$, $I^{(1)}_\mu$ and $I^{(2)}_{\mu\nu}$ are~:
\bea
 K^{(0)}&=&\int_{M}\,d^3w\,\frac{1}{[(p-w)^2-m^2]\,[(k-w)^2-m^2]}
\nn \\
J^{(0)}&=&\int_{M}\,d^3w\,\frac{1}{w^2\,[(p-w)^2-m^2]\,[(k-w)^2-m^2]}
\nn \\
J^{(1)}_{\mu}&=&\int_{M}\,d^3w\,
\frac{w_{\mu}}{w^2\,[(p-w)^2-m^2]\,[(k-w)^2-m^2]}
\nn \\
J^{(2)}_{\mu\nu}&=&\int_{M}\,d^3w\,
\frac{w_{\mu}w_{\nu}}{w^2\,[(p-w)^2-m^2]\,[(k-w)^2-m^2]}
\nn \\
I^{(0)}&=&\int_{M}\,d^3w\,
\frac{1}{w^4\,[(p-w)^2-m^2]\,[(k-w)^2-m^2]}
\nn \\
I^{(1)}_{\mu}&=&\int_{M}\,d^3w\,
\frac{w_{\mu}}{w^4\,[(p-w)^2-m^2]\,[(k-w)^2-m^2]}
\nn \\
 I^{(2)}_{\mu\nu}&=&\int_{M}\,d^3w\,
\frac{w_{\mu}w_{\nu}}{w^4\,[(p-w)^2-m^2]\,[(k-w)^2-m^2]}
\label{integrals}  \;.
\eea

We evaluate these integrals  following the techniques developed in ~\cite{BC,BKP2,BKP3,Reenders}. The results are tabulated below, employing the notation $\Delta^2 = (k \cdot p)^2 - k^2 p^2$ and
$X_0=(2/i\pi^2)X^{(0)}$ for $X=I,J,K$.

\subsubsection{The Scalar Integral $J^{(0)}$}
\noindent
In arbitrary dimensions,
\be
J^{(0)}=\int\frac{d^dw}{[(k-w)^2-m^2][(p-w^2)-m^2]w^2}\;.
\ee
Employing Feynman parametrization,
\bea
J^{(0)}&=& \Gamma (3)\int d^dw\int_0^1d\alpha_1\int_0^1d\alpha_2\int_0^1d
\alpha_3\nn\\
&\times& \frac{\delta \left( \alpha_s-1\right)}
{[\alpha_1[(k-w)^2-m^2]+\alpha_2[(p-w)^2-m^2]+\alpha_3w^2]^3}\;,
\eea
with
\be
\alpha_s=\alpha_1+\alpha_2+\alpha_3\;.
\ee
We define $D$ as
\bea
D&=&\alpha_1[(k-w)^2-m^2]+\alpha_2[(p-w)^2-m^2]+\alpha_3w^2\nn\\
&=& \alpha_sw^2-2(\alpha_1k\cdot w+\alpha_2p\cdot w)
+\alpha_1(k^2-m^2)
+\alpha_2(p^2-m^2)\nn\\
&=&
\alpha_s\Bigg[w^2-\frac{2(\alpha_1k+\alpha_2p)\cdot
w}{\alpha_s}+\frac{(\alpha_1k+\alpha_2p)^2}{
\alpha_s^2}\nn\\
&&-\frac{(\alpha_1k+\alpha_2p)^2}{\alpha_s^2}+
\frac{\alpha_1(k^2-m^2)+\alpha_2(p^2-m^2)}{\alpha_s}\Bigg]
\nn\\
&=&
\alpha_s\Bigg[\left(w-\frac{\alpha_1k+\alpha_2p}
{\alpha_s}\right)^2+\nn\\
&&\frac{1}{\alpha_s^2}
%&&
\Bigg\{\alpha_s[\alpha_1(k^2-m^2)+\alpha_2(p^2-m^2)]
-(\alpha_1k+\alpha_2p)^2\Bigg\}\Bigg]\;.
\eea
Let
\be
w-\frac{\alpha_1k+\alpha_2p}{\alpha_s}\to w\;.
\ee
Then,
\bea
J^{(0)}&=&2\int_0^1d\alpha_1\int_0^1d\alpha_2\int_0^1d\alpha_3
\frac{\delta\left(\alpha_s-1\right)}
{\alpha_s^3}\\
&\times&\int \!\frac{d^dw}{\left[\! w^2\!+\!\alpha_s
^{-2}\!\left\{\!
\alpha_s [\alpha_1(k^2\!-\!m^2)\!+\!\alpha_2
(p^2\!-\!m^2)]\!-\!
(\alpha_1k\!+\!\alpha_2p)^2\right\}\right]^3}\;.\nn
\eea
Making use of the formulas
\bea
\int\frac{d^dw}{w^n}&=&0\nn\\
\int\frac{d^dw}{(w^2-s)^n}&=&(-1)^n i \pi^{\frac{d}{2}}
\frac{\Gamma\left(n-\frac{d}{2}\right)}{\Gamma(n)}s^{\frac{d}{2}-n}\;,
\label{magic1}
\eea
we can write
\bea
J^{(0)}&=&-i\pi^{\frac{d}{2}}\Gamma \left(3-\frac{d}{2}\right)\int_0^1d
\alpha_1\int_0^1d\alpha_2\int_0^1d\alpha_3 \\
&\times&\frac{\delta \left(\alpha_s-1\right)}
{\alpha_s^{d-3}\{-\alpha_s
[\alpha_1(k^2-m^2)+\alpha_2(p^2-m^2)]+(\alpha_1k+
\alpha_2p)^2\}^{3-\frac{d}{2}}}\;.\nn
\eea
At this point, we remind the Cheng-Wo Theorem \cite{Wu}~:
\newtheorem{wu}{Cheng-Wu Theorem}
\begin{wu}
If
\be
I=\int_0^1\prod_{i=1}^n d\alpha_i\delta\left(1-\sum_{i=1}^n
\alpha_i\right)F(\alpha)\;,
\ee
then we can write
\be
I=\int_0^\infty \prod ' d\alpha_i \int_0^1\prod '' d\alpha_i\delta\left(
1-\sum''\alpha_i\right)F(\alpha)\;,
\ee
where the set of $\alpha$ has been split into two nonempty sets
\be
\{\alpha '\}=\{ \alpha_1,\ldots ,\alpha_k\},\quad\{\alpha''\}=\{\alpha_{k+1},
\ldots,\alpha_n\}\;.
\ee
\end{wu}
Making use of this theorem,
\bea
J^{(0)}&=&-i \pi^{\frac{d}{2}}\Gamma\left(3-\frac{d}{2}\right)
\int_0^\infty d\alpha_3 \int_0^\infty d\alpha_2\int_0^1d\alpha_1\\
&\times&\frac{\delta (\alpha_1-1)}
{ \alpha_s^{d-3}\{-\alpha_s
[\alpha_1(k^2-m^2)+\alpha_2(p^2-m^2)]+(\alpha_1k+\alpha_2p)^2\}^{3-
\frac{d}{2}}}\;.\nn
\eea
After a trivial integration over $\alpha_1$, we obtain
\bea
D&=&(1+\alpha_2+\alpha_3)^{d-3}\left[m^2(1+\alpha_2+\alpha_3)
(1+\alpha_2)-k^2(1+\alpha_2+\alpha_3-1)\right.\nn\\
&&-p^2(1+\alpha_2+\alpha_3)\alpha_2
+\alpha_2^2p^2+2\alpha_2k\cdot p]^{3-\frac{d}{2}}\;.
\eea
Using $-2k\cdot p=q^2-k^2-p^2$,
\be
D=(1+\alpha_2+\alpha_3)^{d-3}
[m^2(1+\alpha_2+\alpha_3)
(1+\alpha_2)-\alpha_3k^2-\alpha_2\alpha_3p^2-\alpha_2q^2]^{3-\frac{d}{2}}\;.
\ee
Therefore,
\bea
J^{(0)}&=&-i \pi^{\frac{d}{2}}\Gamma\left(3-\frac{d}{2}\right)
\int_0^\infty d\alpha_3 \int_0^\infty \frac{d\alpha_2}
{(1+\alpha_2+\alpha_3)^{d-3}}\nn\\
&\times&\frac{1}
{ [m^2(1+\alpha_2+\alpha_3)(1+\alpha_2)-\alpha_3k^2-\alpha_2\alpha_3p^2-\alpha_2q^2]^{3-\frac{d}{2}}}.
\eea
In our three-dimensional case
\bea
J^{(0)}&=& -i \pi^{\frac{3}{2}}\frac{1}{2}\pi^{\frac{1}{2}}
\int_0^\infty d\alpha_3 \int_0^\infty d\alpha_2\nn\\
&\times&\frac{1}{[ m^2(1+\alpha_2+\alpha_3)(1+\alpha_2)-\alpha_3k^2-\alpha_2\alpha_3p^2-\alpha_2q^2]^{\frac{3}{2}}} \nn\\
&=&\frac{i\pi^2}{2}J_0\;,
\eea
where
\bea
J_0&=&-\int_0^\infty d\alpha_3\int_0^\infty
d\alpha_2\nn\\
&\times&\frac{1}{[ m^2(1+\alpha_2+\alpha_3)(1+\alpha_2)-\alpha_3k^2
-\alpha_2\alpha_3p^2-\alpha_2q^2]^{\frac{3}{2}}}\;.
\eea
Being simpler the integration over $\alpha_3$ first, we take
\bea
D&=&m^2(1+\alpha_2+\alpha_3)(1+\alpha_2)-\alpha_3k^2-\alpha_2\alpha_3p^2-\alpha_2q^2\nn\\
&=&\alpha_3[m^2(1+\alpha_2)-k^2-\alpha_2p^2]+m^2(1+\alpha_2)^2-\alpha_2q^2 \nn\\
&=&\alpha_3[\alpha_2(m^2-p^2)+(m^2-k^2)]+m^2(1+\alpha_2)^2-\alpha_2q^2 \;.
\eea
Therefore,
\bea
J_0&=&-\int_0^\infty d\alpha_2\int_0^\infty
\frac{d\alpha_3}{[\alpha_3[\alpha_2(m^2\!-\!p^2)\!+\!(m^2\!-\!k^2)]
\!+\!m^2(1\!+\!\alpha_2)^2\!-\!\alpha_2q^2]^\frac{3}{2}}\nn\\
&=&2\int_0^\infty d\alpha_2\frac{1}{[\alpha_2(m^2-p^2)+(m^2-k^2)]} \nn\\
&&\left.\times\frac{1}{\{ \alpha_3[\alpha_2(m^2-p^2)+(m^2-k^2)]+m^2(1+\alpha_2)^2
-\alpha_2q^2\}^\frac{1}{2}}\right|_{\alpha_3=0}^{\alpha_3=\infty}\nn \\
&=&-2\int_0^\infty d\alpha_2\frac{1}{[\alpha_2(m^2-p^2)+(m^2-k^2)]} \frac{1}{[m^2(1+\alpha_2)^2-\alpha_2q^2]^\frac{1}{2} }\nn\\
&=&-\frac{2}{(m^2-p^2)}\int_0^\infty d\alpha_2\frac{1}{\left[\alpha_2+\frac{m^2-k^2}{m^2-p^2}\right]}\frac{1}{[m^2(1+\alpha_2)^2-\alpha_2q^2]^\frac{1}{2} }.
\eea
Let
\be
\alpha_2+\frac{m^2-k^2}{m^2-p^2}=z\qquad \Rightarrow \qquad d\alpha_2=dz\;.
\ee
Integration limits are transformed in the following way~:
\be
\alpha_2=0\Rightarrow z=\frac{m^2-k^2}{m^2-p^2}\qquad\alpha_2\to\infty
\Rightarrow z\to\infty\;.
\ee
Now we rewrite the integrand as~:
\bea
m^2(1+\alpha_2)^2-\alpha_2q^2&=&m^2\left(1+z-\frac{m^2-k^2}{m^2-p^2}\right)^2
 -\left[z-\frac{m^2-k^2}{m^2-p^2}\right]q^2\nn\\
&=&m^2+\left[z^2+\frac{(k^2-p^2)^2}{(m^2-p^2)^2}+2z\frac{k^2-p^2}{m^2-p^2}
\right]\nn\\
&&-zq^2
+q^2\frac{m^2-k^2}{m^2-p^2}\nn\\
&=&m^2z^2+z\left[\frac{2m^2(k^2-p^2)-q^2(m^2-p^2)}{m^2-p^2}\right]\nn\\
&&+\frac{m^2(k^2-p^2)^2+q^2(m^2-k^2)(m^2-p^2)}{(m^2-p^2)^2} \nn\\
&=&cz^2+bz+a\equiv R\;,
\eea
where
\bea
c&=&m^2\;,\nn\\
b&=&\frac{2m^2(k^2-p^2)-q^2(m^2-p^2)}{m^2-p^2}\;,\nn\\
a&=&\frac{m^2(k^2-p^2)^2+q^2(m^2-k^2)(m^2-p^2)}{(m^2-p^2)^2}
=\frac{\chi}{(m^2-p^2)^2}\;.
\eea
Consequently,
\be
J_0=-\frac{2}{m^2\!-\!p^2}\int_{\frac{m^2\!-\!k^2}{m^2\!-\!p^2}}^\infty
 \frac{dz}{z}
\frac{1}{\sqrt{R}}=\left.\frac{2}{(m^2\!-\!p^2)}\frac{1}{\sqrt{-a}}
\arctan{ \frac{2a\!+\!bz}{2\sqrt{-a}\sqrt{R}} }\right|_{
 \frac{m^2\!-\!k^2}{m^2\!-\!p^2}}^\infty.
\ee
We consider
\begin{enumerate}
\item Evaluation at infinity
\bea
\lim_{z\to\infty}\frac{2a+bz}{2\sqrt{-a}\sqrt{R}}&=&\frac{b}{2\sqrt{-a}\sqrt{c}} \nn\\
&=&\frac{m^2(k^2-p^2)[2m^2-k^2-p^2]+\chi}{2m\sqrt{-\chi}(m^2-k^2)}\;.
\eea

\item Evaluation at the lower limit
\be
\left.\frac{2a+bz}{2\sqrt{-a}\sqrt{R}}\right|_{z=\frac{m^2-k^2}{m^2-p^2}}
=\frac{-\chi+m^2(k^2-p^2)(2m^2-k^2-p^2)}{2m\sqrt{-\chi}(m^2-p^2)}\;,
\ee
\end{enumerate}
and then,
\bea
J_0&=&\frac{-2}{\sqrt{-\chi}}\Bigg\{\arctan{\left[
\underbrace{\frac{m^2(k^2-p^2)(2m^2-k^2-p^2)+\chi}{2m\sqrt{-\chi}(m^2-k^2)}}_a
\right]}\nn\\
&-&\arctan{\left[ \underbrace{\frac{-\chi+m^2(k^2-p^2)(2m^2-k^2-p^2)}
{2m\sqrt{-\chi}(m^2-p^2)}}_b \right]}\Bigg\}\;.\label{j_0}
\eea
Since we difined
\be
I(y)= \frac{1}{\sqrt{-y^2}}\arctan{\sqrt{\frac{-y^2}{m^2}}}\;,
\ee
we can write
\bea
I(\eta_i^2\chi)&=&\frac{1}{\sqrt{-\eta_i^2\chi}}
\arctan{\sqrt{\frac{-\eta_i^2\chi}{m^2}}}=
\frac{1}{\eta_i\sqrt{-\chi}} \arctan{\frac{\eta_i}{m}\sqrt{-\chi}}\;.
\eea
Rewriting then eq.~(\ref{j_0}) with the identifications
\bea
a&=&\frac{m^2(k^2-p^2)(2m^2-k^2-p^2)+\chi}{2m\sqrt{-\chi}(m^2-k^2)} \nn\\
&=&\frac{\sqrt{-\chi}}{2m}\left[- \left\{\frac{m^2(k^2-p^2)(2m^2-k^2-p^2)+\chi}
{2\chi(m^2-k^2)} \right\}\right]\nn\\
&\equiv&\frac{\sqrt{-\chi}}{2m}\eta_1\;,
\eea
where
\be
\eta_1=- \left\{\frac{m^2(k^2-p^2)(2m^2-k^2-p^2)
+\chi}{2\chi(m^2-k^2)} \right\}\;.
\ee
In a similar way
\bea
b&=&\frac{-\chi+m^2(k^2-p^2)(2m^2-k^2-p^2)}{2m\sqrt{-\chi}(m^2-p^2)} \nn\\
&=&\frac{\sqrt{-\chi}}{2m}\left[
\frac{\chi-m^2(k^2-p^2)(2m^2-k^2-p^2)}{\chi
(m^2-p^2)}\right]\nn\\
&\equiv&
\frac{\sqrt{-\chi}}{2m}\eta_2\;,
\eea
where
\be
\eta_2=\left[ \frac
{\chi-m^2(k^2-p^2)(2m^2-k^2-p^2)}{\chi (m^2-p^2)}\right]\;.
\ee
Finally
\be
J_0=\left[-\eta_1I\left(\frac{\eta_1\sqrt{\chi}}{2}\right)
+\eta_2I\left(\frac{\eta_2\sqrt{\chi}}{2}\right)\right]\;.
\ee

\subsubsection{The Scalar Integral $K^{(0)}$}
\noindent
We have,
\be
K^{(0)}=\int \frac{d^3w}{[(p-w)^2-m^2][(k-w)^2-m^2]}\;.
\ee
Employing Feynamn parametrization~:
\be
\frac{1}{ab}=\int_0^1\frac{dz}{[az+b(1-z)]^2}\;.
\ee
We take
\bea
a&=&[(p-w)^2-m^2]\;,\nn\\
b&=&[(k-w)^2-m^2]\;.
\eea
We define now
\bea
D&=&z[(p-w)^2-m^2]+(1-z)[(k-w)^2-m^2]\nn\\
&=&z(p^2-2p\cdot w)+(1-z)(k^2-2k\cdot w)+w^2-m^2\;.\nn\\
\eea
We perform now the following change of variable
\be
w\to w'=w-k(1-z)\;.
\ee
Then, we obtain
\be
D=zp^2-2zp\cdot w'-2p\cdot k z(1-z)+k^2z(1-z)+w'^2-m^2\;.
\ee
A second change of variable
\be
w'\to w=w'-zp\;,
\ee
implies
\bea
D&=&w^2+(k-p)^2z(1-z)-m^2\nn\\
&=&w^2+q^2z(1-z)-m^2\;.
%\qquad q=k-p\;.
\eea
Therefore,
\be
K^{(0)}=\int_0^1dz\int d^3w\frac{1}{[w^2+q^2z(1-z)-m^2]^2}\;.
\ee
Using eqs. ~(\ref{magic1}), we arrive at~:
\bea
K^{(0)}&=& \int_0^1dz(-1)^2i\pi^\frac{3}{2}
\frac{\Gamma \left(\frac{1}{2}\right)}{\Gamma(2)}[m^2-q^2z(1-z)]^
{-\frac{1}{2}}
\nn\\
&=&i\pi^2\int_0^1 dz[m^2-q^2z(1-z)]^{-\frac{1}{2}} \nn\\
&=&\frac{2i\pi^2}{\sqrt{-q^2}}\arctan{\sqrt{\frac{-q^2}{4m^2}}}\nn \\
&=&i\pi^2I\left(\frac{q}{2}\right)\;,
\eea
which is our final expression.

\subsubsection{The Tensor Integral $J_\mu^{(1)}$}
\noindent
We have
\be
J^{(1)}_{\mu}=\int\,d^3w\,
\frac{w_{\mu}}{w^2\,[(p-w)^2-m^2]\,[(k-w)^2-m^2]}\;.
\ee
We write this integral in its most general form as~:
\be
J^{(1)}_{\mu}=\frac{{\it i}\pi^2}{2}\left\{
k_{\mu}J_{A}(k,p)+p_{\mu}J_{B}(k,p)\right\} \;.
\label{j_mu_ans}
\ee
Contracting with $k^\mu$ and $p^\mu$, we obtain the following
system of equations~:
\bea
k^\mu J_\mu^{(1)}&=&\frac{i\pi^2}{2}[k^2J_A+(k\cdot p)J_B]\nn\\
p^\mu J_\mu^{(1)}&=&\frac{i\pi^2}{2}[(k\cdot  p)J_A+p^2J_B]\;.
\label{j_mu_sys}
\eea
On the other hand,
\bea
p^\mu J_\mu^{(1)}&=&\frac{(p^2-m^2)}{2}\int d^3w\frac{1}
{w^2[(p-w)^2-m^2][(k-w)^2-m^2]}\nn\\
&-&\frac{1}{2}\int d^3w\frac{1}
{[(k-w)^2-m^2]w^2}\nn\\
&+&\frac{1}{2}\int d^3w\frac{1}{[(p-w)^2-m^2][(k-w)^2-m^2]}
\nn\\
&=&\frac{(p^2-m^2)}{2}J^{(0)}+\frac{1}{2}K^{(0)}-\frac{1}{2}I(0,1,1)\;.
\eea
$I(0,1,1)$ will be difined as a Master Integral when we calculate
$I^{(0)}$, and we use the relation~:
\be
p\cdot w=\frac{1}{2}(p^2+w^2-(p-w)^2-m^2+m^2)\;.
\ee
In a similar way, we also have that~:
\be
k^\mu J_\mu^{(1)}=\frac{(k^2-m^2)}{2}J^{(0)}+\frac{1}{2}K^{(0)}-
\frac{1}{2}I(1,0,1)\;.
\ee
We can now solve the system of equations~(\ref{j_mu_sys}). On solving, we find~:
\bea
J_{A}(k,p)&=&-\frac{2}{\Delta^2}   \Bigg\{  \left[
p^2 (k^2-k \cdot p) - m^2 (p^2-k \cdot p) \right] \frac{ J_{0}}{4}
  + k \cdot p  I(k)  \nn\\
  &&- p^2  I(p)
  +  \frac{1}{2} \, (p^2-k \cdot p)  I(q/2)
   \Bigg\} \; ,    \nn  \\
J_{B}(k,p)&=&J_{A}(p,k) \;.\label{j_mu_ans_2}
\eea
Equations~(\ref{j_mu_ans}) and~(\ref{j_mu_ans_2}) form the final answer.

\subsubsection{The Tensor Integral $J_{\mu\nu}^{(2)}$}
\noindent
We have
\be
J^{(2)}_{\mu\nu}=\int\,d^3w\,
\frac{w_{\mu}w_{\nu}}{w^2\,[(p-w)^2-m^2]\,[(k-w)^2-m^2]}\;.
\ee
We write this integral in its most general form as~:
\bea
J_{\mu\nu}^{(2)}&=&\frac{i\pi^3}{2}\left\{\frac{g_{\mu\nu}}{3}K_0\!+\!
\left(k_\mu k_\nu\!-\!g_{\mu\nu}\frac{k^2}{3}\right)J_C\!+\!
\left(p_\mu k_\nu\!+\!k_\mu p_\nu\!-\!g_{\mu\nu}\frac{2(k\cdot p)}{3}\right)J_D\right.
\nn\\
&+&\left.\left(p_\mu p_\nu-g_{\mu\nu}\frac{p^2}{3}\right)J_E\right\}\;.
\eea
Contracting with $p^\mu$,
\bea
p^\mu J_{\mu\nu}^{(2)}&=&\frac{i\pi^3}{2}\left\{ \frac{p_\nu}{3}K_0\!+\!
\left((k\cdot p)k_\nu\!-\!p_\nu\frac{k^2}{3}\right)J_C\right.\\
&+&\left.\left(
p^2k_\nu\!+\!(k\cdot p)
p_\nu\!-\!p_\nu\frac{2(k\cdot p)}{3}\right)J_D
\!+\!\left(p^2p_\nu\!-\!p_\nu\frac{p^2}{3}
\right)J_E\right\}\;.\nn
\eea
Performing the remaining contractions we obtain~:
\be
p^\nu p^\mu J_{\mu\nu}^{(2)}=\frac{i\pi^3}{2}\left\{ \frac{p^2}{3}K_0\!+\!
\left(
(k\cdot p)^2\!-\!\frac{k^2p^2}{3}J_C\right)\!+\!\frac{4}{3}p^2(k\cdot p)J_D\!+\!
\frac{2}{3}p^4J_E\right\} \;.
\ee
Other possible contractions are~:
\bea
k^\nu p^\mu J_{\mu\nu}^{(2)}&=&\frac{i\pi^3}{2}\left\{\frac{(k\cdot p)}{3}
K_0+\frac{2}{3}k^2(k\cdot p)J_C
%\right.\nn\\
%&+&\left.
+\left(k^2p^2+\frac{(k\cdot p)^2}{3}
\right)J_D\right.\nn\\
&&\left. +\frac{2}{3}p^2(k\cdot p)J_E\right\}
\eea
and
\bea
k^\mu J_{\mu\nu}^{(2)}&=&\frac{i\pi^3}{2}\left\{\frac{k_\nu}{3}K_0+
\frac{2}{3}k^2k_\nu J_C+\left( (k\cdot p)k_\nu+k^2p_\nu-\frac{2}{3}k_\nu
(k\cdot p)\right) J_D\right.\nn\\
&+&\left.\left( (k\cdot p)p_\nu-k_\nu\frac{p^2}{3}\right)J_E\right\}\;,
\eea
wich, after a second contraction, yield~:
\be
k^\nu k^\mu J_{\mu\nu}^{(2)}=\frac{i\pi^3}{2}\left\{\frac{k^2}{3}K_0+
\frac{2}{3} k^4J_C+\frac{4}{3}k^2 (k\cdot p) J_D + \left( (k\cdot p)^2-
\frac{k^2p^2}{3}\right)J_E\right\}
\ee
and
\bea
p^\nu k^\mu J_{\mu\nu}^{(2)}&=&\frac{i\pi^3}{2}\left\{
\frac{(k\cdot p)}{3}K_0+ \frac{2}{3}k^2 (k\cdot p)J_C+
\left( k^2p^2+\frac{1}{3}(k\cdot p)^2\right)J_D\right.\nn\\
&+&\left.\frac{2}{3}p^2 (k\cdot p)J_E\right\}\;.
\eea
On the other hand,
\bea
p^\mu J_{\mu\nu}^{(2)}&=&\int d^3w \frac{p\cdot w\; w_\nu}
{w^2[(p-w)^2-m^2][(k-w)^2-m^2]}\nn\\
&=&\frac{(p^2-m^2)}{2}\int d^3w\frac{w_\nu}{ w^2[(p-w)^2-m^2]
[(k-w)^2-m^2]}\nn\\
&+& \int d^3w\frac{w_\nu}{[(p-w)^2-m^2][(k-w)^2-m^2]}\nn\\
&-&\frac{1}{2}\int d^3w\frac{w_\nu}{w^2[(k-w)^2-m^2]}\nn\\
&=&\frac{(p^2-m^2)}{2}J_\nu^{(1)}+\frac{1}{2}A_\nu(k,p)-
\frac{1}{2}E_\nu(k)\;.
\eea
Similarly
\be
k^\mu J_{\mu\nu}^{(2)}=\frac{(k^2-m^2)}{2}J_\nu^{(1)}+
\frac{1}{2}A_\nu(k,p)-\frac{1}{2}E_\nu(p)\;.
\ee
The new integrals needed are  $E_\nu(p)$ and $A_\nu(k,p)$.  Before proceding with the calculation of $J_{\mu\nu}^{(2)}$, let us pay attention to them. First,
\be
E_\nu(p) = \int d^3w \frac{w_\nu}{w^2[(p-w)^2-m^2]}\;.
\ee
Using Feynam Parametrization,
\bea
E_\nu(p) &=& \int_0^1 dx \int d^3w \frac{w_\nu}{[x[(p-w)^2-m^2]
+(1-x)w^2]^2}\;.
\eea
Let
\bea
D&=&x[(p-w)^2-m^2]+(1-x)w^2\nn\\
&=&w^2+xp^2-2x\pw -xm^2\;.
\eea
We take now
\be
w=w'+xp\;,
\ee
such that
\be
D=w'^2+p^2x(1-x)-m^2x\;,
\ee
and, consequently,
\bea
E_\nu(p)&=&\int_0^1 dx \int d^3w' \frac{w'_\nu+p_\nu x}
{[w'^2+p^2x(1-x)-m^2x]^2}\nn\\
&=&p_\nu\int_0^1dx\, x(-1)^2i\pi^{\frac{3}{2}} \frac{\Gamma\left(
2-\frac{3}{2}\right)}{\Gamma(2)}[-p^2x(1-x)+m^2x]^{\frac{3}{2}-2}\nn\\
&=&i\pi^2p_\nu\int_0^1
dx\,x^{\frac{1}{2}}[-p^2(1-x)+m^2]^{-\frac{1}{2}}\nn\\
&=&\frac{i\pi^2p_\nu}{p^2}\left[m-\frac{(m^2-p^2)}{\sqrt{-p^2}}
\arctan{\sqrt{\frac{-p^2}{m^2}}}\right]\nn\\
&\ktop&E_\nu(k)\;.
\eea
The other integral we need to calculate is
\be
A_\nu(k,p)=\int d^3w \frac{w_\nu}{[(p-w)^2-m^2][(k-w)^2-m^2]}\;.
\ee
After Feynman parametrization, we write~:
\be
A_\nu(k,p)=\int_0^1dz\int d^3w\frac{w_\nu}
{\{ z[(p-w)^2-m^2]+ (1-z)[(k-w)^2-m^2]\}^2}\;.
\ee
We define
\be
D = z[(p-w)^2-m^2]+ (1-z)[(k-w)^2-m^2]\;.
\ee
After the change of variable
\be
w'=w-zp-(1-z)k\;,
\ee
we have
\be
D=w'^2+q^2z(1-z)-m^2\;,
\ee
and, therefore,
\be
A_\nu(k,p)=\int_0^1dz\int d^3w \frac{w_\nu'+p_\nu z+k_\nu(1-z)}
{[ w'^2+q^2z(1-z)-m^2]^2}\;.
\ee
On simplifying,
\bea
A_\nu(k,p)&=&\int_0^1dz i\pi^2\frac{p_\nu z+k_\nu (1-z)}{\sqrt{m^2-q^2z(1-z)}}\nn\\
&=&i\pi^2(k+p)_\nu\int_0^1dz\frac{z}{\sqrt{m^2-q^2z(1-z)}}\nn\\
&=&\frac{i\pi^2(k+p)_\nu}{\sqrt{-q^2}}\arctan{\sqrt{\frac{-q^2}{4m^2}}}
\equiv \frac{1}{2}K^{(0)}(k+p)_\nu\;.
\eea
Let us turn back to the claculation of $J_{\mu\nu}^{(2)}$. Second contraction yields~:
\bea
p^\nu p^\mu J_{\mu\nu}^{(2)}&=&\frac{(p^2-m^2)}{2}\frac{i\pi^3}{2}
[(k\cdot p)J_A\!+\!p^2J_B]\!+\!\frac{1}{2}p^\nu A_\nu\!-\!\frac{p^\nu}{2}
E_\nu(k)\nn\\
k^\nu p^\mu J_{\mu\nu}^{(2)}&=&\frac{(p^2-m^2)}{2}\frac{i\pi^3}{2}
[k^2J_A\!+\!(k\cdot p)J_B]\!+\!\frac{1}{2}k^\nu A_\nu\!-\!\frac{k^\nu}{2}
E_\nu(k)\nn\\
k^\nu k^\mu J_{\mu\nu}^{(2)}&=&\frac{(k^2-m^2)}{2}\frac{i\pi^3}{2}
[k^2J_A\!+\!(k\cdot p)J_B]\!+\!\frac{1}{2}k^\nu A_\nu\!-\!\frac{k^\nu}{2}
E_\nu(p)\nn\\
p^\nu k^\mu J_{\mu\nu}^{(2)}&=&\frac{(k^2-m^2)}{2}\frac{i\pi^3}{2}
[(k\cdot p)J_A\!+\!p^2J_B]\!+\!\frac{1}{2}p^\nu A_\nu\!-\!\frac{p^\nu}{2}
E_\nu(p)\;.
\eea
We arrive then to the following system of equations~:
\bea
\left( (k\cdot p)^2-\frac{k^2p^2}{3}\right)J_C+ \frac{4}{3}p^2(k\cdot p)J_D+\frac{2}{3}p^4J_E&=&a\nn\\
\frac{2}{3}k^4J_C+\frac{4}{3}k^2(k\cdot p)J_D+ \left( (k\cdot p)^2-\frac{k^2p^2}{3}\right)J_E&=&b\nn\\
\frac{2}{3}k^2 (k\cdot p)J_C+ \left(k^2p^2+\frac{(k\cdot p)^2}{3}\right)J_D
+\frac{2}{3}p^2(k\cdot p)J_E&=&c\;,
\eea
where
\bea
a&=&\frac{2}{i\pi^3}p^\mu p^\nu J_{\mu\nu}^{(2)}-\frac{p^2}{3}K_0\nn\\
b&=&\frac{2}{i\pi^3}k^\mu k^\nu J_{\mu\nu}^{(2)}-\frac{k^2}{3}K_0\nn\\
c&=&\frac{1}{i\pi^3}[k^\nu p^\mu J_{\mu\nu}^{(2)}+p^\nu k^\mu J_{\mu\nu}^{(2)}
-\frac{(k\cdot p)}{3}K_0]\;.
\eea
Solutions to this system are~:
\bea
J_{C}(k,p)&=&\frac{1}{\Delta^2}
\Bigg\{ \left[ p^2 (k \cdot p \!-\! 2 k^2) \!-\! m^2 (k \cdot p \!-\! 2 p^2)
\right]
\frac{J_A}{2}
 \!-\!p^2(p^2-m^2)  \frac{J_B}{2} \;  \nn\\
&&  + \frac{k \cdot p}{k^2} \, (m^2-k^2)  \; I(k)
+  \frac{1}{2} \, (k \cdot p + p^2) \; I(q/2)  \;
-m \frac{ k\cdot p}{k^2}   \Bigg\} \;,
   \nn \\
J_{D}(k,p)&=&\frac{1}{2\Delta^2}
\Bigg\{ \left[k^2 (3 k \cdot p - p^2) - m^2 (3 k \cdot p - k^2)
\right]
\; \frac{J_A}{2} \nn\\
 &&+  \left[p^2 (3 k \cdot p - k^2) - m^2 (3 k \cdot p - p^2)
\right]
\; \frac{J_B}{2}   \nn   \\
&& - (m^2\!-\!k^2)  I(k) \!-\! (m^2\!-\!p^2) I(p) \!-\! \frac{1}{2}
(k+p)^2
 I(q/2) \!+\! 2m
\Bigg\} \;,
\nn \\
J_{E}(k,p)&=&J_{C}(p,k) \;,
\eea
which completes or calculation of $J_{\mu\nu}^{(2)}$.

\subsubsection{The Scalar Integral $I^{(0)}$}
\noindent
We have
\be
I^{(0)}=\int d^3w\frac{1}{w^4[(p-w)^2-m^2][(k-w)^2-m^2]}\;.
\ee
In general, let us define the family of integrals~:
\be
I(\nu_1,\nu_2,\nu_3)=\int d^dw\frac{1}{[(p-w)^2-m^2]^{\nu_1}
[(k-w)^2-m^2]^{\nu_2}[w^2]^{\nu_3}}\;.
\ee
We use the following Integration by Parts (IBP) identity ~:
\be
\int d^dw\frac{\partial}{\partial w_\mu}\left[
\frac{(q_i-w)_\mu}{[(p-w)^2-m^2]^{\nu_1}[(k-w)^2-m^2]^
{\nu_2}[w^2]^{\nu_3}}\right]=0\;.\label{IBP3d}
\ee
After the differentiation we obtain~:
\bea
&&\int d^dw (q_i-m)\Bigg[\frac{2\nu_1(p-w)^\mu}{A^{\nu_1+1}B^{\nu_2}
C^{\nu_3}} +\frac{2\nu_2(k-w)^\mu}{A^{\nu_1}B^{\nu_2+1}C^{\nu_3}}-
\frac{2\nu_2w^\mu}{A^{\nu_1}B^{\nu_2}}C^{\nu_3+1}\Bigg]\nn\\
&+&\int d^dw\frac{(-d)}{A^{\nu_1}B^{\nu_2}C^{\nu_3}}=0\;,
\eea
since
\bea
\frac{\partial}{\partial w_\mu}w_\mu&=&g_\mu^\mu=d\;.
\eea
We have defined
\be
A=(p-w)^2-m^2\;,\quad B=(k-w)^2-m^2\;,\quad C=w^2\;.
\ee
Therefore,
\bea
dI(\nu_1,\nu_2,\nu_3)&=& \int d^dw (q_i-w)\nn\\
&&\hspace{-8mm}
\Bigg[\frac{2\nu_1(p-w)^\mu}
{A^{\nu_1+1}B^{\nu_2}C^{\nu_3}} +\frac{2\nu_2(k-w)^\mu}{A^{\nu_1}
B^{\nu_2+1}C^{\nu_3}}-\frac{2\nu_2w^\mu}{A^{\nu_1}B^{\nu_2}C^{\nu_3+1}}\Bigg]\;.
\eea
Assuming that $q_i$ is defined such that $q_1=p$, $q_2=k$ and $q_3=0$. Then, for $i=0$~:
\bea
dI(\nu_1,\nu_2,\nu_3)&=& \int d^dw (p-w)\nn\\
&&\hspace{-8mm}
\Bigg[\frac{2\nu_1(p-w)^\mu}
{A^{\nu_1+1}B^{\nu_2}C^{\nu_3}} +\frac{2\nu_2(k-w)^\mu}{A^{\nu_1}
B^{\nu_2+1}C^{\nu_3}}-\frac{2\nu_2w^\mu}{A^{\nu_1}B^{\nu_2}C^{\nu_3+1}}\Bigg]
\;.\label{di0}
\eea
Let us observe the following products~:
\begin{enumerate}
\item
\bea
2(p-w)\cdot (k-w)&=&(p-w)^2+(k-w)^2-(p-k)^2-m^2-m^2+2m^2\nonumber\\
&=&A+B-[(p-k)^2-2m^2]\;.
\eea
\item
\be
2(p-w)\cdot (-w)=(p-w)^2-m^2+w^2-p^2+m^2=A+C-p^2+m^2\;.
\ee
\item
\be
(p-w)^2=(p-w)^2-m^2+m^2=A+m^2\;.
\ee
\end{enumerate}
Substituting these products into eq.~(\ref{di0}), we obtain~:
\bea
dI(\nu_1,\nu_2,\nu_3)&=&\int d^dw\Bigg[\frac{2\nu_1[A+m^2]}{A^{\nu_1+1}
B^{\nu_2}C^{\nu_3}}+\frac{\nu_2[A+B-[(p-k)^2-2m^2]]}{A^{\nu_1}B^{\nu_2+1}
C^{\nu_3}}\nonumber\\
&&+\frac{\nu_3[A+C-(p^2-m^2)]}{A^{\nu_1}B^{\nu_2}C^{\nu_3+1}}\Bigg]\nonumber\\
&=&\int d^dw\Bigg[ \frac{2\nu_1+\nu_2+\nu_3}{A^{\nu_1}B^{\nu_2}C^{\nu_3}} +
\frac{\nu_2}{A^{\nu_1-1}B^{\nu_2+1}C^{\nu_3}} +
\frac{2\nu_1m^2}{A^{\nu_1+1}B^{\nu_2}C^{\nu_3}}\nn\\
&&-\frac{\nu_2[(p-k)^2-2m^2]}{A^{\nu_1}B^{\nu_2+1}C^{\nu_3}}+
\frac{\nu_3}{A^{\nu_1-1}B^{\nu_2}C^{\nu_3+1}}-
\frac{\nu_3(p^2-m^2)}{A^{\nu_1}B^{\nu_2}C^{\nu_3+1}}\Bigg]\nonumber\\
&=&(2\nu_1\!+\!\nu_2\!+\!\nu_3)I(\nu_1,\nu_2,\nu_3)+
\nu_2I(\nu_1\!-\!1,\nu_2\!+\!2,\nu_3)\nn\\
&&+ 2m^2\nu_1I(\nu_1\!+\!1,\nu_2,\nu_3)
-\nu_2[(p-k)^2-2m^2]I(\nu_1,\nu_2\!+\!1,\nu_3)\nn\\
&&+ \nu_3I(\nu_1\!-\!1,\nu_2,\nu_3\!+\!1)-
\nu_3(p^2-m^2)I(\nu_1,\nu_2,\nu_3+1)\;.
\eea
Rearranging the terms in such a way that those integrals with $\nu_1+\nu_2+\nu_3$ as the sum of their arguments are on the right hand side, while those with $\nu_1+\nu_2+\nu_3+1$ on the left hand side, we get~:
\bea
&&\hspace{-5mm}
-2\nu_1m^2I(\nu_1\!+\!1,\nu_2,\nu_3)+\nu_2[(p-k)^2-2m^2]I(\nu_1,\nu_2\!
+\!1,\nu_3)\nn\\
&&+ \nu_3(p^2-m^2)I(\nu_1,\nu_2,\nu_3+1)=
(2\nu_1\!+\!\nu_2\!+\!\nu_3\!-\!d)I(\nu_1,\nu_2,\nu_3)\nn\\
&&\hspace{15mm}
+ \nu_2I(\nu_1-1,\nu_2+1,\nu_3)+\nu_3I(\nu_1-1,\nu_2,\nu_3+1)\;.
\eea
For $i=2$ we have~:
\bea
dI(\nu_1,\nu_2,\nu_3)&=&\int d^dw(k-w)_\mu\nn\\
&&\Bigg[
\frac{2\nu_1(p-w)^\mu}{A^{\nu_1+1}B^{\nu_2}C^{\nu_3}}+
\frac{2\nu_2(k-w)^\mu}{A^{\nu_1}B^{\nu_2+1}C^{\nu_3}}-
\frac{2\nu_3w^\mu}{A^{\nu_1}B^{\nu_2}C^{\nu_3+1}}\Bigg]\nonumber\\
&=&(\nu_1+2\nu_2+\nu_3)I(\nu_1,\nu_2,\nu_3)+ \nu_1I(\nu_1+1,\nu_2-1,\nu_3)
\nn\\
&&+\nu_3I(\nu_1,\nu_2-1,\nu_3+1)\nn\\
&&-[(p-k)^2-2m^2]\nu_1I(\nu_1+1,\nu_2,\nu_3)
\nn\\
&&+ 2\nu_2m^2I(\nu_1,\nu_2+1,\nu_3)\nn\\
&&-\nu_3(k^2-m^2)I(\nu_1,\nu_2,\nu_3+1)
\eea
or
\bea
&&\hspace{-5mm}
\nu_1[(p-k)^2-2m^2]I(\nu_1+1,\nu_2,\nu_3)-2\nu_2m^2I(\nu_1,\nu_2+1,\nu_3)\nn\\
&&+ \nu_3(k^2-m^2)I(\nu_1,\nu_2,\nu_3+1)=
(\nu_2+2\nu_2+\nu_3-d)I(\nu_1,\nu_2,\nu_3)\nn\\
&&\hspace{15mm}
+\nu_1I(\nu_1+1,\nu_2-1,\nu_3)
+\nu_3I(\nu_1,\nu_2,\nu_3+1)\;,
\eea
and for $i=3$~:
\bea
dI(\nu_1,\nu_2,\nu_3)&=&\int d^dw (-w_\mu)\nn\\
&&\Bigg[\frac{2\nu_1(p-w)^\mu}{A^{\nu_1+1}B^{\nu_2}C^{\nu_3}}
+ \frac{2\nu_2(k-w)^\mu}{A^{\nu_1}B^{\nu_2+1}C^{\nu_3}}-
\frac{2\nu_3w^\mu}{A^{\nu_1}B^{\nu_2}C^{\nu_3+1}}\Bigg]\nn\\
&=&(\nu_1+\nu_2+2\nu_3)I(\nu_1,\nu_2,\nu_3)+\nu_1I(\nu_1+1,\nu_2,\nu_3-1)\nn\\
&&+\nu_2I(\nu_1+\nu_2+1,\nu_3-1)\nn\\
&&-\nu_1(p^2-m^2)I(\nu_1+1,\nu_2,\nu_3)\nn\\
&&-\nu_2(k^2-m^2)I(\nu_1,\nu_2+1,\nu_3)
\eea
or
\bea
&&\hspace{-5mm}
\nu_1(p^2-m^2)I(\nu_1+1,\nu_2,\nu_3)+\nu_2(k^2-m^2)I(\nu_1,\nu_2+1,\nu_3)=\nn\\
&&(\nu_1+\nu_2+2\nu_3-d)I(\nu_1,\nu_2,\nu_3)+\nu_1I(\nu_1+1,\nu_2,\nu_3-1)\nn\\
&&\hspace{15mm}
+\nu_2I(\nu_1,\nu_2+1,\nu_3-1)\;.
\eea
Therefore, we need to solve the system of equations~:
\bea
\hspace{-1cm}-2\nu_1m^2I(\nu_1+1,\nu_2,\nu_3)+\nu_2(q^2-2m^2)I(\nu_1+\nu_2+1,\nu_3)&&\nn\\
\hspace{-1cm}+\nu_3(p^2-m^2)I(\nu_1,\nu_2,\nu_3+1)&=&a\nn\\
\hspace{-1cm}\nu_1(q^2-2m^2)I(\nu_1+1,\nu_2,\nu_3)-2\nu_2m^2I(\nu_1,\nu_2+1,\nu_3)&&\nn\\
\hspace{-1cm}+\nu_3(k^2-m^2)I(\nu_1,\nu_2,\nu_3+1)&=&b\nn\\
\hspace{-1cm}\nu_1(p^2-m^2)I(\nu_1+1,\nu_2,\nu_3)+\nu_2(k^2-m^2)I(\nu_1,\nu_2+1,\nu_3)&&\nn\\
\hspace{-1cm}+0 I(\nu_1,\nu_2,\nu_3\!+\!1)&=&c\;,
\eea
where
\bea
a&=&(2\nu_1+\nu_2+\nu_3-d)I(\nu_1+\nu_2+\nu_3)+
\nu_2I(\nu_1-1,\nu_2+1,\nu_3)\nn\\
&&+\nu_3I(\nu_1-1,\nu_2,\nu_3+1)\nn\\
b&=&(\nu_1+2\nu_2+\nu_3-d)I(\nu_1+\nu_2+\nu_3)+
\nu_1I(\nu_1+1,\nu_2-1,\nu_3)\nn\\
&&+\nu_3I(\nu_1,\nu_2-1,\nu_3+1)\nn\\
c&=&(\nu_1+\nu_2+2\nu_3-d)I(\nu_1+\nu_2+\nu_3)+
\nu_1I(\nu_1+1,\nu_2,\nu_3-1)\nn\\
&&+\nu_2I(\nu_1,\nu_2+1,\nu_3-1)\;.
\eea
In obvious notation~:
\be
\left( \begin{array}{ccc}
-2m^2\nu_1 & \nu_2(q^2-2m^2) & \nu_3(p^2-m^2)\\
\nu_1(q^2-2m^2) & -2m^2\nu_2 & \nu_3(k^2-m^2)\\
\nu_1(p^2-m^2) & \nu_2(k^2-m^2) & 0 \end{array} \right)
\left( \begin{array}{c}
I_1\\I_2\\I_3\end{array}\right) =
\left( \begin{array}{c}
a\\b\\c\end{array}\right)\;,
\ee
which formally we can write as~:
\be
SI=A\;.
\ee
Then,
\be
\abs{S} = 2\nu_1\nu_2\nu_3
[m^2(k^2-m^2)^2+m^2(p^2-m^2)^2+(k^2-m^2)(p^2-m^2)(q^2-2m^2)]\;,
\ee
and besides
\bea
\abs{S} I_3&=&
\nu_1\nu_2[(q^2-2m^2)(k^2-m^2)+2m^2(p^2-m^2)]a\nn\\
&&+\nu_1\nu_2[(q^2-2m^2)(p^2-m^2)+2m^2(k^2-m^2)]b\nn\\
&&+\nu_1\nu_2[4m^4-(q^2-2m^2)^2]c\;.\label{I0rsolved}
\eea
For $\nu_1=\nu_2=\nu_3=1$,
\bea
a&=&I(1,1,1)+I(0,2,1)+I(0,1,2)\nn\\
b&=&I(1,1,1)+I(2,0,1)+I(1,0,2)\nn\\
c&=&I(1,1,1)+I(2,1,0)+I(1,2,0)\,
\eea
with  $I^{(0)}=I(1,1,2)$ and $J^{(0)}=I(1,1,1)$.  So, we need to calculate the integrals $I(0,2,1)\ktop I(2,0,1)$, $I(0,1,2)\ktop I(1,0,2)$ and $I(2,1,0)\ktop I(1,2,0)$.  For that purpose, we define the integral
\be
I_{pn}(k,p,m_1,m_2)=\int d^dw
\frac{1}{[(k-w)^2-m_1^2]^p[(p-w)^2-m_2^2]^n} \;.\label{MII0}
\ee
Using Feynam parametrization, we know that
\be
\frac{1}{A^nB^p}=\frac{\Gamma(n+p)}{\Gamma(n)\Gamma(p)} \int_0^1
dx \,x^{n-1}(1-x)^{p-1}\frac{1}{[xA+(1-x)B]^{n+p}}.\label{magic2}
\ee
Let
\bea
A&=&[(p-w)^2-m^2]\;,\nn\\
B&=&[(k-w)^2-m^2]\;,
\eea
and
\bea
\hspace{-5mm}
D&=&xA+(1-x)B\nn\\
\hspace{-5mm}
&=&x[p^2+w^2-2\pw-m_2^2]+(1-x)[k^2+w^2-2\kw-m_1^2]\nn\\
\hspace{-5mm}
&=&w^2\!-\!2w\cdot [px\!+\!k(1\!-\!x)]\!+\!p^2x
\!+\!k^2(1\!-\!x)\!-\!m_2^2x\!-\!m_1^2(1\!-\!x)\;.
\eea
We make the change of variable
\be
w'= w-[px+k(1-x)]\;,
\ee
such that
\bea
w^2-2w\cdot [px+k(1-x)] &=&w'^2-[px+k(1-x)]^2 \;.
\eea
In this way, $D$ is rewritten as~:
\bea
D&=&w'^2-[px+k(1-x)]^2+p^2x+k^2(1-x)-m_2^2x-m_1^2(1-x)\nn\\
&=&w'^2+p^2x(1-x)+k^2x(1-x)-2\kp x(1-x)-m_2^2x -m_1^2(1-x)\nn\\
&=&w'^2+q^2x(1-x)-m_2^2x-m_1^2(1-x)\;.
\eea
Therefore,
\bea
I_{pn}(k,p,m_1,m_2)&=&\frac{\Gamma(n+p)}{\Gamma(n)\Gamma(p)}
\int_0^1 dx\, x^{n-1}(1-x)^{p-1} \int d^dw' \frac{1}{D^{n+p}}\nn\\
&&\hspace{-3cm}
=\frac{\Gamma(n+p)}{\Gamma(n)\Gamma(p)}
\int_0^1 dx \,x^{n-1}(1-x)^{p-1}
(-1)^{n+p}i\pi^{d/2}
\frac{\Gamma(n+p-d/2)}{\Gamma(n+p)}s^{\frac{d}{2}-n-p} \nn\\
&&\hspace{-3cm}
=i(-1)^{n+p}\pi^{d/2}\frac{\Gamma(n+p-d/2)}{\Gamma(n)\Gamma(p)}
\int_0^1 dx x^{n-1}(1-x)^{p-1}s^{\frac{d}{2}-n-p}\;,
\eea
with
\be
s=-q^2x(1-x) +m_2^2x+m_1^2(1-x)\;.
\ee
For the integrals of our interest, we help ourselves with
\bea
I_{11}(k,p,m_1,m_2)&=&\int \frac{d^dw}{[(p-w)^2-m_2^2][(k-w)^2-m_1^2]}\nn\\
&=&i\pi^{\frac{d}{2}}\Gamma\left(2-\frac{d}{2}\right) \int_0^1 dx
s^{\frac{d}{2}-2}\nn\\
&=&i\pi^2\int_0^1 dx s^{-\frac{1}{2}}\;,
\eea
where, in the last line, we set $d=3$. So
\be
I(2,1,0)=\left.\frac{1}{2m_1}\frac{\partial}{\partial m_1}
I_{11}(k,p,m_1,m_2)\right|_{m_1=m_2=m}\;.
\ee
Now,
\be
\frac{\partial}{\partial m_1}s^{-\frac{1}{2}}=-m_1(1-x)s^{-\frac{3}{2}}\;,
\ee
and therefore,
\bea
I(2,1,0)&=&-\frac{i\pi^2}{2} \int_0^1 dx
\frac{(1-x)}{[-q^2x(1-x)+m^2]^{\frac{3}{2}}}\nn\\
&=&\frac{-i\pi^2}{m[4m^2-q^2]}\nn\\
&\ktop&I(1,2,0)\;.
\eea
On the other hand
\bea
I(2,0,1)&=&\left.\frac{1}{2m_1}\frac{\partial}{\partial m_1}
I_{11}(k,p,m_1,m_2)\right|_{k=m_1=0,m_2=m}\nn\\
&=&-\frac{i\pi^2}{2}\int_0^1 dx
\frac{x}{[-p^2x(1-x)+m^2x]^{\frac{3}{2}}}\nn\\
&=&-\frac{i\pi^2}{m(m^2-p^2)}\nn\\
&\ktop&I(0,2,1)\;.
\eea
To solve $I(1,0,2)$, we take the integral
\be
I^\nu=\int d^dw \frac{w^\nu}{w^4[(p-w)^2-m^2]}\;.
\ee
Then,
\bea
p_\nu I^\nu &=& \frac{1}{2}(p^2-m^2)\int d^dw
\frac{1}{w^4[(p-w)^2-m^2]}\nn\\
&&+\frac{1}{2} \int d^dw
\frac{1}{w^2[(p-w)^2-m^2]}-\frac{1}{2}\int \frac{d^dw}{w^4}\nn\\
&=&\frac{1}{2}(p^2-m^2)I(1,0,2)+\frac{1}{2}I(1,0,1)\;,
\eea
since tha last term vanishes in the dimensional regularization scheme. It is easy to see that
\bea
I(1,0,1)&=&I_{11}(k,p,m_1,m_2)\Big|_{k=m_1=0,m_2=m}\nn\\
&=&i\pi^2\int_0^1[-p^2x(1-x)+m^2x]^{-\frac{1}{2}}\nn\\
&=&\frac{2i\pi^2}{\sqrt{-p^2}}\arctan{\sqrt{\frac{-p^2}{m^2}}}\nn\\
&\ktop&I(0,1,1)\;.
\eea
To calculate explicitly $I^\nu$, we take in eq.~(\ref{magic2})
\be
A=[(p-w)^2-m^2]\, ,\quad B=w^2\, ,\quad n=1\, ,\quad p=2\;.
\ee
So, in three dimensions,
\bea
I^\nu&=&\frac{\Gamma(3)}{\Gamma(1)\Gamma(2)} \int_0^1 dx\, (1-x)
\int d^3w w^\nu
\frac{1}{[Ax+B(1-x)]^3}\;.
\eea
Let
\be
w'=w-p x\;,
\ee
Then,
\bea
D&=&Ax+B(1-x)\nn\\
&=&w^2+p^2x-2\pw x-xm^2\nn\\
&=&w'^2+p^2x-p^2x^2-m^2x \nn\\
&=&w'^2+p^2x(1-x)-m^2x\;,
\eea
and therefore
\be
I^\nu=2\int_0^1 dx (1-x)\int(d^3w) \frac{p^\nu x}{[w^2+p^2x(1-x)-m^2x]^3}
\;.
\ee
Using now~(\ref{magic1}),
\bea
\hspace{-5mm}
I^\nu&=&2p^\nu\int_0^1 dx\, x (1-x) (-1)^3 i\pi^{\frac{3}{2}}
\frac{\Gamma\left(\frac{3}{2}\right)}{\Gamma(3)}
[-p^2x(1-x)+m^2x]^{\frac{3}{2}}\nn\\
\hspace{-5mm}
&=&-\frac{i\pi^2}{2}p^\nu\int_0^1 dx\,x^{-\frac{1}{2}}(1-x)
[-p^2(1-x)+m^2]^{-\frac{3}{2}}\nn\\
\hspace{-5mm}
&=&-\frac{i\pi^2}{2}p^\nu\left[\int_0^1 dx
\left\{x^{-\frac{1}{2}}-x^{\frac{1}{2}} \right\}
[-p^2(1-x)+m^2]^{-\frac{3}{2}} \right]\nn\\
\hspace{-5mm}
&=&-\frac{i\pi^2}{2}[-p^2]^{-\frac{3}{2}}p^\nu\left[
-\frac{2m^2}{m^2-p^2} \sqrt{-\frac{p^2}{m^2}}+
2\arctan{\sqrt{\frac{-p^2}{m^2}}}\right]\;.
\eea
So that
\bea
I(1,0,2)&=&\frac{2}{p^2-m^2}\left[p_\nu I^\nu-\frac{1}{2}I(1,0,1)
\right]\nn\\
&=&\frac{2m}{(m^2-p^2)^2} i\pi^2\nn\\
&\ktop&I(0,1,2)\;.
\eea
Finally, substiututing into~(\ref{I0rsolved})
\bea
   I^{(0)} &=& \frac{1}{\chi} \; \left\{ q^2 (m^2+k\cdot p)
 J^{(0)}  \; + \; i \pi^2 m L  \right\}  \;,
\eea
where
\bea
  L&=& \frac{q^2(k^2-m^2)-(k^2-p^2)(k^2+m^2)}{(k^2-m^2)^2}  \nn\\
  &&+
       \frac{q^2(p^2-m^2)+(k^2-p^2)(p^2+m^2)}{(p^2-m^2)^2}  \;.
\eea
This is our final expression.

\subsubsection{The Tensor Integral $I_\mu^{(1)}$}
\noindent
We have
\bea
I^{(1)}_{\mu}&=&\int_{M}\,d^3w\,
\frac{w_{\mu}}{w^4\,[(p-w)^2-m^2]\,[(k-w)^2-m^2]}\;.
\eea
We write this integral in its most general form as~:
\bea
I^{(1)}_{\mu}&=&\frac{{\it i}\pi^2}{2}\left[
 k_{\mu}I_{A}(k,p)+p_{\mu}I_{B}(k,p)\right]  \;.\label{i_mu_ans_1}
\eea
Contracting with $k^\mu$ and $p^\mu$ we obtain the following system of equations~:
\bea
k^\mu I_\mu^{(1)}&=&\frac{i\pi^2}{2}[k^2I_A+(k\cdot p)I_B]\nn\\
p^\mu I_\mu^{(1)}&=&\frac{i\pi^2}{2}[(k\cdot  p)I_A+p^2I_B]\;. \label{I_mu_sys}
\eea
On the other hand
\bea
p^\mu I_\mu^{(1)}&=&\frac{(p^2-m^2)}{2}\int
d^3w\frac{1}{w^4[(p-w)^2-m^2][(k-w)^2-m^2]}\nn\\
&-&\frac{1}{2}\int d^3w
\frac{1}{[(k-w)^2-m^2]w^4}\nn\\
&+&\frac{1}{2}\int d^3w\frac{1}
{w^2[(p-w)^2-m^2][(k-w)^2-m^2]}\nn\\
&=&\frac{(p^2-m^2)}{2}I^{(0)}+\frac{1}{2}J^{(0)}-\frac{1}{2}I(0,1,2)\;,
\eea
where we have used~:
\be
p\cdot w=\frac{1}{2}(p^2+w^2-(p-w)^2-m^2+m^2)\;.
\ee
We also have that~:
\be
k^\mu I_\mu^{(1)}=\frac{(k^2-m^2)}{2}I^{(0)}+\frac{1}{2}J^{(0)}
-\frac{1}{2}I(1,0,2)\;.
\ee
We solve the system of equations~(\ref{I_mu_sys}), to find~:
\bea
I_{A}(k,p)&=&\frac{2}{\Delta^2}
\Bigg\{  \left[k\cdot p (p^2-m^2) - p^2(k^2-m^2) \right]
\; \frac{I_{0}}{4} + p\cdot q \; \frac{J_{0}}{4}\nn\\
&+& \frac{mp^2}{(m^2-p^2)^2} -
\frac{mk\cdot p}{(m^2-k^2)^2} \Bigg\} \;,
\nn \\ \nn \\
 I_{B}(k,p)&=&I_{A}(p,k)  \;.\label{i_mu_ans_2}
\eea
Equations~(\ref{i_mu_ans_1}) and~(\ref{i_mu_ans_2}) form the complete solution.

\subsubsection{The Tensor Integral $I_{\mu\nu}^{(2)}$}
\noindent
We have,
\bea
 I^{(2)}_{\mu\nu}&=&\int_{M}\,d^3w\,
\frac{w_{\mu}w_{\nu}}{w^4\,[(p-w)^2-m^2]\,[(k-w)^2-m^2]}\;.
\eea
We express this integral in its most general form as~:
\bea
I_{\mu\nu}^{(2)}&=&\frac{i\pi^3}{2}\left\{\frac{g_{\mu\nu}}{3}J_0
\!+\!\left(k_\mu k_\nu\!-\!g_{\mu\nu}\frac{k^2}{3}\right)I_C
\!+\!\left(p_\mu k_\nu\!+\!k_\mu p_\nu\!-\!g_{\mu\nu}
\frac{2(k\cdot p)}{3}\right)I_D\right.
\nn\\
&+&\left.\left(p_\mu p_\nu-g_{\mu\nu}\frac{p^2}{3}\right)I_E\right\}\;.
\eea
Contracting with $p^\mu$ we obtain~:
\bea
\hspace{-7mm}p^\mu I_{\mu\nu}^{(2)}&=&\frac{i\pi^3}{2}\left\{
\frac{p_\nu J_0}{3}+\left((k\cdot
p)k_\nu-p\nu\frac{k^2}{3}\right)I_C\right.\nn\\
\hspace{-7mm}&+&\left.\left(p^2k_\nu\!+\!(k\cdot p)p_\nu\!-\!p_\nu\frac{2(k\cdot p)}{3}
\right)I_D
\!+\!\left(p^2p_\nu\!-\!p_\nu\frac{p^2}{3}\right)I_E\right\}\;.
\eea
Performing the remaining contractions,
\be
p^\nu p^\mu I_{\mu\nu}^{(2)}=\frac{i\pi^3}{2}\left\{ \frac{p^2}{3}J_0+
\left((k\cdot p)^2-\frac{k^2p^2}{3}\right)I_C+
\frac{4}{3}p^2(k\cdot p)I_D+\frac{2}{3}p^4I_E\right\}
\ee
and
\bea
k^\nu p^\mu I_{\mu\nu}^{(2)}&=&\frac{i\pi^3}{2}\left\{\frac{(k\cdot p)}{3}J_0+
\left(\frac{2}{3}k^2(k\cdot p)\right)I_C+
\left(k^2p^2+\frac{(k\cdot p)^2}{3}\right)I_D\right.\nn\\
&+&\left.\frac{2}{3}p^2(k\cdot p)I_E\right\}\;.
\eea
In the same way,
\bea
k^\mu I_{\mu\nu}^{(2)}&=&\frac{i\pi^3}{2}\left\{\frac{k_\nu}{3}J_0+
 \frac{2}{3}k^2k_\nu I_C+\left( (k\cdot p)k_\nu+k^2p_\nu-\frac{2}{3}k_\nu
 (k\cdot p)\right) I_D\right.\nn\\
 &+&\left.\left( (k\cdot p)p_\nu-k_\nu\frac{p^2}{3}\right)I_E\right\}\;,
\eea
from where, after a second contraction, we obtain~:
\be
k^\nu k^\mu I_{\mu\nu}^{(2)}=\frac{i\pi^3}{2}\left\{\frac{k^2}{3}J_0
+\frac{2}{3} k^4I_C+\frac{4}{3}k^2 (k\cdot p) I_D
+ \left( (k\cdot p)^2-\frac{k^2p^2}{3}\right)I_E\right\}
\ee
and
\bea
p^\nu k^\mu I_{\mu\nu}^{(2)}&=&\frac{p\pi^3}{2}\left\{ \frac{(k\cdot p)}{3}J_0
+ \frac{2}{3}k^2 (k\cdot p)I_C
+\left( k^2p^2+\frac{1}{3}(k\cdot p)^2\right)I_D\right.\nn\\
&&+\left.\frac{2}{3}p^2 (k\cdot p)I_E\right\}\;.
\eea
On the other hand,
\bea
p^\mu I_{\mu\nu}^{(2)}&=&\int d^3w \frac{p\cdot w w_\nu}{w^4[(p-w)^2-m^2][(k-w)^2-m^2]}\nn\\
&=&\frac{(p^2-m^2)}{2}\int d^3w\frac{w_\nu}{ w^4[(p-w)^2-m^2]
[(k-w)^2-m^2]}\nn\\
&&+ \int d^3w\frac{w_\nu}{ w^2[(p-w)^2-m^2][(k-w)^2-m^2]}\nn\\
&&-
\frac{1}{2}\int d^3w\frac{w_\nu}{w^4[(k-w)^w-m^2]}\nn\\
&=&\frac{(p^2-m^2)}{2}I_\nu^{(1)}+\frac{1}{2}J_\nu^{(1)}-
\frac{1}{2}E_\nu(k)\;.
\eea
Similarly,
\be
k^\mu I_{\mu\nu}^{(2)}=\frac{(k^2-m^2)}{2}I_\nu^{(1)}
+\frac{1}{2}J_\nu^{(1)}-\frac{1}{2}E_\nu(p)\;.
\ee
From the second contraction we obtain~:
\bea
\hspace{-5mm}p^\nu p^\mu I_{\mu\nu}^{(2)}&=&\frac{(p^2\!-\!m^2)}{2}\frac{i\pi^3}{2}
[(k\cdot p)I_A\!+\!p^2I_B]\!+\!\frac{1}{2}p^\nu J_\nu^{(1)}\!-\!
\frac{p^\nu}{2}E_\nu(k)\nn\\
\hspace{-5mm}k^\nu p^\mu
I_{\mu\nu}^{(2)}&=&\frac{(p^2\!-\!m^2)}{2}\frac{i\pi^3}{2}
[k^2I_A\!+\!(k\cdot p)I_B]\!+\!\frac{1}{2}k^\nu J_\nu^{(1)}\!-\!
\frac{k^\nu}{2}E_\nu(k)\nn\\
\hspace{-5mm}k^\nu k^\mu I_{\mu\nu}^{(2)}&=&\frac{(k^2\!-\!m^2)}{2}\frac{i\pi^3}{2}
[k^2I_A\!+\!(k\cdot p)I_B]\!+\!\frac{1}{2}k^\nu J_\nu^{(1)}\!-\!
\frac{k^\nu}{2}E_\nu(p)\nn\\
\hspace{-5mm}p^\nu k^\mu I_{\mu\nu}^{(2)}&=&\frac{(k^2\!-\!m^2)}{2}\frac{i\pi^3}{2}
[(k\cdot p)I_A\!+\!p^2I_B]\!+\!\frac{1}{2}p^\nu J_\nu^{(1)}\!-\!
\frac{p^\nu}{2}E_\nu(p)\;.
\eea
In this way, we arrive at the following system of equations~:
\bea
\left( (k\cdot p)^2-\frac{k^2p^2}{3}\right)I_C+ \frac{4}{3}p^2(k\cdot p)
I_D+\frac{2}{3}p^4I_E&=&a\nn\\
\frac{2}{3}k^4I_C+\frac{4}{3}k^2(k\cdot p)I_D+ \left( (k\cdot p)^2-
\frac{k^2p^2}{3}\right)I_E&=&b\nn\\
\frac{2}{3}k^2 (k\cdot p)I_C+ \left(k^2p^2+\frac{(k\cdot p)^2}{3}\right)
I_D+\frac{2}{3}p^2(k\cdot p)I_E&=&c\;,
\eea
where
\bea
a&=&\frac{2}{i\pi^3}p^\mu p^\nu I_{\mu\nu}^{(2)}-\frac{p^2}{3}J_0\nn\\
b&=&\frac{2}{i\pi^3}k^\mu k^\nu I_{\mu\nu}^{(2)}-\frac{k^2}{3}J_0\nn\\
c&=&\frac{1}{i\pi^3}\left[k^\nu p^\mu I_{\mu\nu}^{(2)}+
p^\nu k^\mu I_{\mu\nu}^{(2)}-\frac{(k\cdot p)}{3}J_0\right]\;.
\eea
Solution to the system are~:
\bea
I_{C}(k,p)&=&\frac{1}{\Delta^2}   \Bigg\{ p^2\;J_{0}  +
\left[ p^2 (k \cdot p - 2 k^2) -m^2(k\cdot p-2p^2)\right] \frac{I_A}{2}\nn\\
&&- p^2(p^2-m^2)  \frac{I_B}{2}
  + (k \cdot p - 2 p^2) \; \frac{J_A}{2}  - p^2 \;
\frac{J_B}{2} \nn - \frac{k \cdot p}{k^2 } \; I(k)  \nn\\
&&+ \frac{mk\cdot p}{k^2(m^2-k^2)}
   \Bigg\}  \;,  \nn
\\
I_{D}(k,p)&=&\frac{1}{2\Delta^2}   \Bigg\{ - 2 k \cdot p \; J_{0}  +
\left[ k^2 (3 k \cdot p - p^2)-m^2(3k\cdot p-k^2) \right] \; \frac{I_A}{2} \nn\\
&&+  \left[p^2 (3 k \cdot p - k^2)-m^2(3k\cdot p-p^2) \right]
\; \frac{I_B}{2}   + (3 k \cdot p - k^2) \; \frac{J_A}{2} \nn\\
&&  +
  (3 k \cdot p - p^2) \; \frac{J_B}{2}
\; +\; I(k) + I(p) \;
-   \;\frac{m}{m^2-k^2} \;
- \; \frac{m}{m^2-p^2}  \Bigg\} \,, \nn
\\
I_{E}(k,p)&=&I_{C}(p,k) \;.
\eea
This completes the calculation of the vertex to ${\cal O}(\alpha)$.

\subsection{Transverse Vertex to One loop}
\noindent
We can subtract from the full vertex, Eq.~(\ref{1loopvertexevaluated}), the longitudinal vertex,
Eqs.~(\ref{1loopLvertex}) and~(\ref{sigmas}), and obtain the transverse vertex to ${\cal O}(\alpha)$. Let us recall that  the transverse vertex ${\Gamma^{\mu}_{T}(k,p)}$ can be written in terms of 8
basis vectors as follows~:
\begin{eqnarray*}
\Gamma^{\mu}_{T}(k,p)=\sum_{i=1}^{8} \tau_{i}(k^2,p^2,q^2)T^{\mu}_{i}(k,p)
\;,
\end{eqnarray*}
with the basis~:
\bea
&T^{\mu}_{1}&=\left[p^{\mu}(k\cdot q)-k^{\mu}(p\cdot q)\right]\nn\\
&T^{\mu}_{2}&=\left[p^{\mu}(k\cdot q)-k^{\mu}(p\cdot q)\right](\slsh{k}
+\slsh{p})\nn\\
&T^{\mu}_{3}&=q^2\gamma^{\mu}-q^{\mu}\slsh{q}\nn\\
&T^{\mu}_{4}&=q^2\left[\gamma^{\mu}(\slsh{k}+\slsh{p})-k^{\mu}
-p^{\mu}\right]-2(k-p)^{\mu}k^{\lambda}p^{\nu}\sigma_{\lambda\nu}\nn\\
&T^{\mu}_{5}&=q_{\nu}\sigma^{\nu\mu}\nn\\
&T^{\mu}_{6}&=-\gamma^{\mu}(k^2-p^2)+(k+p)^{\mu}\slsh{q}\nn\\
&T^{\mu}_{7}&=-\frac{1}{2}(k^2-p^2)\left[\gamma^{\mu} (\slsh{k}+\slsh{p})
-k^{\mu}-p^{\mu} \right]
+(k+p)^{\mu}k^{\lambda}p^{\nu}\sigma_{\lambda\nu}\nn\\
&T^{\mu}_{8}&=-\gamma^{\mu}k^{\nu}p^{\lambda}{\sigma_{\nu\lambda}}
+k^{\mu}\slsh{p}-p^{\mu}\slsh{k} \;,\nn\\
\mbox{with}\;\;\;\;\;\;\;\;\;
&\sigma_{\mu\nu}&=\frac{1}{2}[\gamma_{\mu},\gamma_{\nu}]\;.
\eea
After a lengthy but straightforward algebra, the coefficients $\tau_{\it i}$ can be identified. We prefer to write these out in the following form~:
\bea
  \hspace{-3mm} \tau_i(k,p) &=& \alpha g_i \left[ \sum_j^5
  a_{ij}(k,p) I(l_j)
+ \frac{a_{i6}(k,p)}{k^2 p^2}  \right]  \hspace{5 mm} i=1, \cdots 8  \;,
\label{taui}
\eea
where $l_1^2 = \eta_1^2\chi/4$, $l_2^2 = \eta_2^2\chi/4$, $l_3^2 = k^2$, $l_4^2 = p^2$ and
$l_5^2 = q^2/4$. Functions $\eta_1$, $\eta_2$ and $\chi$ have been defined above. Similarly, the factors $g_i$ are $-g_1 = m \Delta^2 g_2 = 2m \Delta^2 g_3 = 2 \Delta^2 g_4 = g_5 = 2 m \Delta^2 g_6 = \Delta^2 g_7 = m g_8 = m/ 4 \Delta^2$. The coefficients $a_{ij}$ in the one loop perturbative expansion of the $\tau_i$, Eq.~(\ref{taui}),  are tabulated below~:
\bea
a_{11}(k,p) &=& -(\xi+2)\eta_1(m^2 + k\cdot p)        \nn \\
a_{12}(k,p) &=& a_{11}(p,k)                           \nn \\
a_{13}(k,p) &=& 4(\xi+2)\frac{(k^2 + k\cdot p)}{(k^2 - p^2)}
                                                      \nn \\
a_{14}(k,p) &=&a_{13}(p,k)                           \nn \\
a_{15}(k,p) &=& -2(\xi + 2)                           \nn \\
a_{16}(k,p) &=& 0                                     \nn \\
a_{21}(k,p) &=& -\eta_1\Bigg\{  \left[-\frac{q^{2}}{2}
m^{4}+
\left\{(k \cdot p)^2-(k^{2}+p^{2})(k \cdot p)+k^{2}p^{2}
\right\} m^{2}\right.\nn\\
&&-\left.\frac{q^{2}}{4} \left\{ (k \cdot p)^{2}+ k^{2}p^{2}\right\}
\right]                                                \nn \\
&&+ \frac{(\xi-1)}{2\chi}\Big[-q^{4}  m^{8}- q^2 \left\{
(k\cdot p)^{2}+2(k^{2}+p^{2})k\cdot p-5k^{2}p^{2}\right\}m^{6}\nn\\
&& +\frac{3}{2} q^2 (k^2+p^2) \Delta^2  m^{4}            \nn \\
&&+ \Big\{2(k^{4}+p^{4}+k^2 p^2)(k\cdot p)^{3}
-7k^{2}p^{2}(k^2+p^2)(k\cdot p)^{2}\nn\\
&& +10 k^{4}p^{4}k\cdot p
-k^{4}p^{4}(k^2+p^2)\Big\}  m^{2}                      \nn \\
&&+ \frac{1}{2} k^2 p^2 q^2 \left\{ (k^2+p^2) (k \cdot p)^2 -
4k^2 p^2 k \cdot p + k^2 p^2 (k^2 + p^2)    \right\}
\Big] \Bigg\}                                           \nn \\
a_{22}(k,p) &=& a_{21}(p,k)                           \nn \\
a_{23}(k,p) &=& \frac{1}{(k^2 - p^2)}\Bigg[ \xi\left\{(k\cdot p)^{3}+
k^{2}(k\cdot p)^{2}-3k^{2}p^{2}k\cdot p \right.\nn\\
&&\left.+2k^{4}k\cdot p
+k^{4}p^{2}-2k^{6}\right\}  m^2/k^2                     \nn \\
&&+(k\cdot p)^3 +(2k^2-p^2)(k\cdot p)^2+k^2p^2k\cdot p -2k^4 k\cdot p
-k^2p^4
                                                        \nn \\
&&+\!(\xi\!-\!1)\left\{(k\cdot p)^3\!+\!p^2(k\cdot p)^2\!-\!3k^2 p^2k\cdot p
\right.\nn\\
&&\left.
\!+\!  2k^4k\cdot p \!+\!k^2p^4\!-\!2k^4 p^2\right\} \Bigg] \nn\\
a_{24}(k,p) &=& a_{23}(p,k)                             \nn \\
a_{25}(k,p) &=& q^2(m^2 + k\cdot p) \nn\\
&&+ (\xi-1)(q^2 m^2 + (k\cdot p)^2 -
(k^2 + p^2)(k\cdot p) + k^2 p^2)                         \nn \\
a_{26}(k,p) &=& m \Delta^2\Bigg\{ k\cdot p + \frac{(\xi-1)}{\chi}
\Big[ q^2 k\cdot p m^4 +2(k^2 + p^2)\Delta^2 m^2\nn\\
&& - k^2p^2 \left\{ 2(k\cdot p)^2+(k^2 + p^2)k\cdot p -4 k^2p^2
\right\}\Big] \Bigg\}                                    \nn \\
a_{31}(k,p) &=& -\frac{\eta_1}{2}\Bigg\{\Bigg[\left\{-2(k\cdot p)^{2}
+k^{4}+p^{4}\right\}
 m^{4}      \nn\\
&&+2\left\{ (k^{2}+p^{2})(k\cdot p)^{2}
+(k^2-p^2)^2\,k\cdot p- k^{2}p^{2}(k^{2}+p^{2}) \right\}m^{2}                   \nn \\
&&+\frac{1}{2} \left\{ -4 (k \cdot p)^4 +
(k^{2}+p^{2})^2 (k\cdot p)^{2} + k^{2}p^{2}(k^{2}-p^{2})^2   \right\}\Bigg]
                                                         \nn \\
&&+\frac{(\xi-1)}{\chi}\Bigg[q^2
\left\{-2(k\cdot p)^{2}+k^{4}+p^{4}\right\} m^{8}         \nn \\
&&+ 2 \Big\{ (k^2+p^2) [-2 (k \cdot p)^3
+ (k^2+p^2) (k \cdot p)^2 + (k^4+p^4) k \cdot p ] \nn\\
&& - k^2 p^2 (3 k^4 + 3 p^4 -2 k^2 p^2) \Big\} m^{6}        \nn \\
&&-\frac{3}{2} q^2 \Delta^2 (k^2-p^2)^2
 m^{4} -2\Big\{ (k^2 + p^2) (k^4+p^4-4k^2p^2) (k \cdot p)^3\nn\\
&&- k^2 p^2 (k^4 + p^4 - 6 k^2 p^2 ) (k \cdot p)^2          \nn \\
&&+ 2 k^4 p^4 (k^2+p^2) k \cdot p - k^4 p^4 (k^2+p^2)^2  \Big\} m^{2}\nn\\
&&-\frac{1}{2} k^2 p^2 q^2
\Big\{ (k^4+p^4-6k^2 p^2) (k \cdot p)^2
+ k^2 p^2 (k^2 + p^2)^2 \Big\} \Bigg]\Bigg\}              \nn \\
a_{32}(k,p) &=& a_{31}(p,k)                              \nn \\
a_{33}(k,p) &=&\xi\left\{ (k\cdot p)^{3}\! -\!k^{2}(k\cdot p)^{2}\!-\!
3k^{2}p^{2}k\cdot p\!+\!2k^{4}k\cdot p\! -\!k^{4}p^{2}+2k^{6}\right\}
m^2/k^2                                                   \nn \\
&&+(\xi-2)\left\{ (k\cdot p)^{3}-(2k^{2}-p^{2})(k\cdot
p)^{2}\right.\nn\\
&&\left.+k^{2}p^{2}k\cdot p-2k^{4}k\cdot p+k^{2}p^{4}\right\}     \nn \\
a_{34}(k,p) &=& a_{33}(k,p)                               \nn \\
a_{35}(k,p) &=& -(k^4 + p^4 -2(k\cdot p)^2)\left[\xi m^2
+(\xi-2)k\cdot p\right]                                   \nn \\
a_{36}(k,p) &=& -m\Delta^2\Bigg\{ k\cdot p (k^2 \!+\! p^2)\!+\!2k^2p^2
\nn\\
&&+\frac{(\xi\!-\!1)}{\chi}\Big[q^2\left\{ (k^2\! +\!p^2)
k\cdot p\!+\!2k^2 p^2\right\}m^4
\nn\\
&&+2(k^2 + p^2)^2\Delta^2m^2                                \
-k^2p^2(k+p)^2 \left\{ (k^2 + p^2)k\cdot p-2k^2 p^2\right\}\Big]\Bigg\}
                                                          \nn \\
a_{41}(k,p) &=& -\eta_1(\xi\!-\!1)\frac{(k^2\! - \!p^2)}{2\chi}
\Bigg[ -q^{4}m^{6}\!+\!3q^{2} \left\{ -(k^{2}\!+\!p^{2}) k\cdot p \!+\! 2
k^{2}p^{2} \right\} m^{4}                                  \nn \\
 &&+\Big\{  (k\cdot p)^{2} [ 4 (k\cdot p)^{2} -
3k^4-3p^4-26k^2p^2]  \nn\\
&&+  k^2 p^2 [
24 (k^{2}+p^{2})  k \cdot p - 3k^4-3p^4-14k^2 p^2)]  \Big\}
\frac{m^{2}}{2}                                            \nn \\
&&+\frac{q^2}{2}
\Big\{ (k^2+p^2) (k \cdot p)^3 +  2 k^2 p^2 (k \cdot
p)^2\nn\\
&&-
3 k^2 p^2 (k^2 + p^2) k \cdot p + 2 k^4 p^4 \Big \}\Bigg]  \nn \\
a_{42}(k,p) &=& -a_{41}(p,k)                                \nn \\
a_{43}(k,p) &=& \frac{(\xi\!-\!1)}{k^2}\left[ (k^2 \!+\! k\cdot
p)(k\cdot p)^2
\!+\!k^2(2k^2\!-\!3p^2)k\cdot p\!+\!k^4(p^2\! -\!2k^2)\right]              \nn \\
a_{44}(k,p) &=& - a_{43}(p,k)                              \nn \\
a_{45}(k,p) &=& (\xi-1)(k^2-p^2)q^2                        \nn \\
a_{46}(k,p) &=& m (\xi-1)(k^2 - p^2)\frac{\Delta^2}{\chi}
\left[ q^2 k\cdot p m^2+2(k^2 + p^2)(k\cdot p)^2\right.\nn\\
&&\left.-2k^2 p^2k\cdot p-k^2p^2(k^2+p^2)\right]                   \nn \\
a_{51}(k,p)&=&-\eta_1\Bigg\{ \Delta^2
+\frac{(\xi-1)}{4\chi}\Bigg[- 2 q^4 m^6 \nn\\
&&+ 6 q^2\left\{ 2 k^2 p^2 - (k^2+p^2) k \cdot p
\right\} m^4 - 6 k^2 p^2 q^4 m^2                           \nn \\
&&- q^2 \left\{ (k^2\!-\!p^2)^2 (k \cdot p)^2
\!+\!2 k^2 p^2 (k^2 \!+\! p^2) k \cdot p \!-\! k^2 p^2 (k^2\!+\!p^2)^2
\right\} \Bigg]\Bigg\}
                                                            \nn \\
a_{52}(k,p) &=& a_{51}(p,k)                                \nn \\
a_{53}(k,p) &=& \frac{(\xi-1)}{k^2}\left[(k\cdot p)^2+2k^2k\cdot p
-k^2(2k^2+p^2)\right]                                       \nn \\
a_{54}(k,p) &=& a_{53}(p,k)                                \nn \\
a_{55}(k,p) &=& (\xi-1)q^2                                  \nn \\
a_{56}(k,p) &=& -m(\xi\!-\!1)\frac{\Delta^2}{\chi}
\left[ q^2(k^2\!+\!p^2)m^2\!+\!2(k^4\!+\! p^4)k\cdot p\!-\!
2k^2 p^2(k^2\!+\!p^2) \right]
                                                            \nn \\
a_{61}(k,p)&=& -\eta_1\frac{(k^2-p^2)}{2}\Bigg[ {q^{2}m^{4}}
-2\left\{(k\cdot p)^{2}-(k^{2}+p^{2})k\cdot p+k^{2}p^{2}\right\}m^{2}
\nn\\
&&+\frac{q^{2}}{2}\left\{(k\cdot p)^{2}+k^{2}p^{2}\right\}  \nn \\
&&+\frac{q^2(\xi-1)}{\chi}\Bigg[q^2 m^8 +
2 \left\{ (k \cdot p)^2 + (k^2+p^2) k \cdot p - 3 k^2 p^2 \right\}
m^{6}\nn\\
&& - \frac{3}{2} q^{2} \Delta^2 m^{4}
- 2 \left\{ (k^2+p^2) (k \cdot p)^3 - k^2 p^2 (k \cdot p)^2
- k^4 p^4\right\}  m^{2}\nn\\
&& - \frac{1}{2} k^2 p^2 q^2 \left\{ (k \cdot p)^2
+ k^2 p^2\right\}\Bigg]\Bigg]                               \nn \\
a_{62}(k,p) &=&- a_{61}(p,k)                               \nn \\
a_{63}(k,p) &=& -\Bigg[ \xi\left\{ (k^{2}+k\cdot p)(k\cdot p)^{2}
+k^{2}(2k^{2}-3p^{2})k\cdot p \right.\nn\\
&&\left.-k^{4}(2k^{2}-p^{2})\right\}m^2/k^2                           \nn \\
&&+(\xi\!-\!2)\left\{(2k^{2}\!-\!p^{2}\!+\!k\cdot p) (k\cdot
p)^{2} \!-\!k^{2}(2k^{2}\!-\!p^{2})
k\cdot p \!-\!k^{2}p^{4}\right\}\Bigg]                           \nn \\
a_{64}(k,p) &=& -a_{63}(p,k)                                 \nn \\
a_{65}(k,p) &=& -q^2(k^2 -p^2)\left[\xi m^2 -(\xi-2)k\cdot p \right]
                                                             \nn \\
a_{66}(k,p) &=& -m(k^2-p^2)\Delta^2\Bigg[ k\cdot p +\frac{(\xi-1)}{\chi}
\left( q^2k\cdot p m^4 +2(k^2 + p^2)\Delta^2 m^2\right.\nn\\
&&\left.-k^2 p^2 q^2 k\cdot p\right) \Bigg]                          \nn \\
a_{71}(k,p) &=&-\frac{\eta_1(\xi-1)}{4\chi}\Bigg[- 2 q^{6}m^{6}
- 6 q^{4}  \left\{ (k^2+p^2) (k\cdot p) - k^{2} p^{2} \right\} m^4
                                                             \nn \\
&&- 3 q^2 \left\{ ((k\cdot p)^2 \!+\! k^2 p^2) (k^4\!+\! p^4 \!+\! 6 k^2 p^2)
\!-\! 8 k^2 p^2 (k^2\!+\!p^2) k\cdot p \right\} m^2                  \nn \\
&&+q^2 \Big\{ (k^2-p^2)^2 (k \cdot p)^3 +
4k^2 p^2 (k^2+p^2) (k \cdot p)^2 \nn\\
&&-k^2 p^2 (3k^4+3 p^4+10k^2 p^2) k \cdot p +
4 k^4 p^4 (k^2+p^2) \Big\}\Bigg]                             \nn \\
a_{72}(k,p) &=& a_{71}(p,k)                                 \nn \\
a_{73}(k,p) &=& (\xi-1)\frac{(k^2-k\cdot p)}{k^2}\left((k\cdot p)^2
+4k^2 k\cdot p -2k^4-3k^2p^2\right)                          \nn \\
a_{74}(k,p) &=& a_{73}(p,k)                                  \nn \\
a_{75}(k,p) &=& (\xi-1)q^4                                   \nn \\
a_{76}(k,p) &=& m(\xi-1)\frac{\Delta^2}{\chi}\Bigg[ q^2
\Bigg\{(k^2+p^2)k\cdot p-2k^2p^2\Bigg\}m^2                   \nn \\
&&+2(k^4\!+\!p^4)(k\cdot p)^2\!-\!4k^2 p^2(k^2\!+\!p^2)k\cdot p
\!-\!k^2p^2(k^4\!+\!p^4\!-\!6k^2p^2)\Bigg]                               \nn \\
a_{81}(k,p) &=& -\eta_1\frac{(\xi+2)}{2}q^2(m^2+k\cdot p)    \nn \\
a_{82}(k,p) &=& a_{81}(p,k)                                 \nn \\
a_{83}(k,p) &=& 2(\xi+2)k\cdot q                             \nn \\
a_{84}(k,p) &=& a_{83}(p,k)                                  \nn \\
a_{85}(k,p) &=& -(\xi+2)q^2                                  \nn \\
a_{86}(k,p) &=& 0    \;.      \label{coeftau}
\eea

An important point to note is that these coefficients do not contain any trigonometric function, as it has been extracted out for raising the $\tau_i$ to a non-perturbative status. The $\tau_i$ have the required symmetry under the exchange of vectors $k$ and $p$. All the $\tau_i$ are symmetric except $\tau_4$ and $\tau_6$ which are antisymmetric. Note that the form in which we write the transverse vertex makes it clear that each term in all the $\tau_i$ is either proportional to $\alpha I(l)$ or $\alpha/(k^2 p^2)$. We shall see that this form provides us with a natural scheme to arrive at its simple non-perturbative extension.

A few comments in comparison with the work by Davydychev {\em et.
al.} \cite{davydychev}, are as follows:  (i) None of the $\tau_i$
we have calculated has kinematic singularity when $k^2 \to p^2$.
This clearly suggests that the choice of the basis $\{ T^\mu_i\}$
suggested by K{\i}z{\i}lers\"{u} {\it et. al.} is preferred over
the one of Ball and Chiu (in QED3 as well) used by Davydychev
{\em et. al.} \cite{davydychev}.  In particular our $\tau_4$ and
$\tau_7$ are independent of kinematic singularities.  (ii) In
three dimensions, their factorization of the common constant
factor in Eq.~(E.1) is singular.  However, as the divergences
completely cancel out, we find our expressions more suitable for
writing the transverse vertex in three dimensions.  (iii) With
the way we express $J_0$, all the $\tau_i$ are written in terms
of basic functions of $k$ and $p$ and a single trigonometric
function of the form $I(l)$.  This form plays a key role in
enabling us to make an easy transition to the possible
non-perturbative str ucture of the vertex, as explained in the
next section.  Moreover, with the given form of $J_0$, a direct
comparison can be made with the massless case.

\section{Non-perturbative Form of the
Vertex (Effective Transverse WTI)}

\subsection{On the Gauge Parameter Dependence of the Vertex}
\noindent
Let us first look at the $\tau_i$ in the simplified massless case, with the
notation  $k= \sqrt{-k^2}$, $p=\sqrt{-p^2}$ and $q=\sqrt{-q^2}$
\cite{BKP2,BKP3},
\bea
  \tau_2 &=& \frac{\alpha \pi}{4} \; \frac{1}{kp(k+p)(k+p+q)^2} \;
\left[  1 + (\xi-1) \, \frac{2k+2p+q}{q}  \right] \;, \\ \nn \\
   \tau_3 &=& \frac{\alpha \pi}{8} \; \frac{1}{kpq(k+p+q)^2} \;
 \left[ 4kp+3kq+3pq+2q^2 \right.\nn\\
 &&+\left. (\xi-1) \, (2k^2+2p^2+kq+pq)  \right] \;,
\nn\\
   \\
\tau_6 &=& \frac{\alpha \pi (2- \xi)}{8} \; \frac{k-p}{kp(k+p+q)^2} \;,
\label{masslesstau6} \\
   \nn \\
 \tau_8 &=& \frac{\alpha \pi (2+ \xi)}{2} \; \frac{1}{kp(k+p+q)} \;.
\eea
It is interesting to note that the existence of the factor
\begin{eqnarray*}
    \frac{k-p}{kp} = -\left( \frac{1}{k} \; - \; \frac{1}{p} \right)
\end{eqnarray*}
in eq.~(\ref{masslesstau6}) puts $\tau_6$ on a different footing as compared to the rest of the $\tau_i$. The reason is that in the massless limit, the Fermion Propagator is simply
\begin{eqnarray*}
   \frac{1}{F(p)} &=& 1 + \frac{\pi \alpha \xi}{4} \; \frac{1}{p} \;,
\end{eqnarray*}
implying
\begin{eqnarray*}
 \frac{1}{F(k)} -  \frac{1}{F(p)} \propto
\left[ \frac{1}{k} \; - \; \frac{1}{p} \right]
 \;.
\end{eqnarray*}
Therefore, the relation of $\tau_6$ with the Fermion Propagator of the type $[1/F(k^2)-1/F(p^2)]$ seems to arise rather naturally~:
\bea
  \tau_6 &=& - \frac{1}{2 \xi} \, \frac{2-\xi}{(k+p+q)^2} \;
\left[ \frac{1}{F(k)} - \frac{1}{F(p)} \right]\;,  \label{m0nptau6}
\eea
as noticed first by Curtis and Pennington \cite{CP1} (note however that their coefficient is not the same). In the rest of the $\tau_i$, the factor $1/k-1/p$ does not arise. However, one could
introduce it by hand to arrive at the following expressions~:
\bea
  \tau_2 &=& -\frac{1}{\xi} \; \frac{1}{(k^2-p^2)(k+p+q)^2}
\left(  1 + (\xi-1) \, \frac{2k+2p+q}{q}  \right) \nn\\
&&\times\left[ \frac{1}{F(k)} - \frac{1}{F(p)} \right]
 \\ \nn                                 \label{m0nptau2}
   \tau_3 &=& -\frac{1}{2 \xi} \;
   \frac{\left[ 4kp+3kq+3pq+2q^2 + (\xi-1) \, (2k^2+2p^2+kq+pq)  \right]}{q(k-p)(k+p+q)^2} \nn\\
%&&\times \left[ 4kp+3kq+3pq+2q^2 + (\xi-1) \, (2k^2+2p^2+kq+pq)  \right] \nn\\
%\nn\\ && \hspace{45mm} \times
&&\times\left[ \frac{1}{F(k)} - \frac{1}{F(p)} \right]
 \label{m0nptau3}  \\
 \tau_8 &=& -\frac{2 (2+ \xi)}{\xi} \; \frac{1}{(k-p)(k+p+q)}
\times\left[ \frac{1}{F(k)} - \frac{1}{F(p)} \right] \;. \label{m0nptau8}
\eea
Equations.~(\ref{m0nptau6}-\ref{m0nptau8}) represent a non-perturbative vertex which is in agreement with  its complete one-loop expansion. This vertex has been constructed in accordance with the form advocated, e.g., in \cite{BR1,CP1,BP1}. There are a couple of important points which need to be discussed here~:
\begin{itemize}

\item There is an explicit dependence on the gauge parameter,
$\xi$.  A widespread belief has been that the gauge dependence of
the vertex should solely arise through functions $F(k^2)$ and
$F(p^2)$, and there should be {\em no explicit appearance of the
gauge parameter $\xi$}.  Such a belief has been expressed (or is
reflected) in various works to date, e.g.,
\cite{BR1,Dong,CP1,BP1,BT1}.  Here we show that at least in
massless QED3, such a construction is not possible.

\item One of the main reasons that the transverse vertex was believed to be proportional to the factor
\begin{eqnarray*}
\left[ \frac{1}{F(k)} - \frac{1}{F(p)} \right]
\end{eqnarray*}
was the assumption that the transverse vertex vanishes in the Landau gauge, based upon perturbative results. A complete one-loop calculation reveals that the transverse vertex does not vanish in the Landau gauge. Moreover, an explicit presence of the gauge parameter in the non-perturbative form of the vertex tells us that the presence of the factor $ [1/F(k) - 1/F(p) ] $ is no longer a guarantee that the transverse vertex vanishes in the Landau gauge.
\end{itemize}

We now show that the explicit dependence of the vertex on the gauge parameter $\xi$ is unavoidable in massless QED3. We notice that at the one loop level, each of the $\tau_i$ can be written in
the following form~:
\begin{eqnarray*}
\tau_{i}(k, p,q) &=& \alpha \; a_{i}(k,p,q) + \alpha \xi\; b_{i}(k,p,q) \;.
\end{eqnarray*}
On the other hand, Eq.~(\ref{FMradial}) yields the following form for $F$~:
\begin{eqnarray*}
     \frac{1}{F(p)} &=& 1 + \alpha \xi \; c_{i}(p) \;.
\end{eqnarray*}
If we want to write the non-perturbative form of the $\tau_{i}$ in terms of $1/F(p)$ and $1/F(k)$ alone and we do not expect explicit presence of $\alpha$, the only way to get rid of $\xi$  dependence is to have
\begin{eqnarray*}
b_{2} \; T_{2}^\mu + b_{3} \; T_{3}^\mu + b_{6} \; T_{6}^\mu + b_{8} \;
T_{8}^\mu = 0
\;.
\end{eqnarray*}
It is not possible as $T_i^{\mu}$ form a linearly independent set of basis vectors. Therefore, any construction of the 3-point vertex will surely have an explicit dependence on the gauge parameter.
Owing to these reasons, we realize that to demand the transverse vertex to be proportional to $ [1/F(k) - 1/F(p) ] $ is artificial (apart from $\tau_6$) and is not required. Therefore, we do not pursue this line of action anymore. In the next section, we move on to construct the vertex for the massive case inspired from our perturbative results.

\subsection{Non-perturbative Vertex (Effective Transverse WTI)}

\noindent
As pointed out in the previous section, each  term in all the $\tau_i$ is either proportional to the trigonometric function $\alpha I(l)$ or $\alpha/(k^2 p^2)$. On the other hand, the perturbative expressions for $\M(p)$ and $F(p)$, Eqs.~(\ref{FMradial}), permit us to write~:
\bea
\frac{1}{F(k)} - \frac{1}{F(p)} &=&\frac{\alpha}{k^2 p^2} \, \frac{\xi}{2}
\, \left[ k^2 A(p) -
p^2 A(k) \right],   \label{nonpF}
\eea
where
\be
A(p)=\left\{ m- (m^2+p^2) I(p) \right\}
\ee
and
\bea
\hspace{-3mm}\frac{\xi}{2 (2\!+\!\xi) l^2 I(l)} \, \left[
\frac{{\cal M}(l)}{F(l)}
\!-\! m \right] \!-\! \left[  1 \!-\! \frac{1}{F(l)}  \right] &=&
\frac{\xi (m^2\!+\!l^2)}{2 l^2} \; \alpha I(l) \;.     \label{nonpM}
\eea
In the massless limit, Eq.~(\ref{nonpF}) simply reduces to
\begin{eqnarray*}
\frac{1}{F(k)} - \frac{1}{F(p)} &=& \frac{\alpha \pi \xi}{4} \;
\left[  \frac{1}{k} - \frac{1}{p}  \right]
\end{eqnarray*}
in the Euclidean space, as expected. It was in fact an analogous massless expression in the limit when $k >>p$ that inspired Curtis and Pennington, \cite{CP1}, to propose their famous vertex in QED4. Here, we are extending the reasoning to all the momentum regimes in the massive QED3. Fortunate simultaneouss occurrence of the factor $\alpha/(k^2 p^2)$ in all the 8 Eqs.~(\ref{taui}) and Eq.~(\ref{nonpF}), and the presence of the same trigonometric factor $I(l)$ in the expressions for the vertex as well as the propagator, one naturally arrives at the following non-perturbative form of $\tau_i$~:
\bea
\nn
\tau_i &=&
 g_i\Bigg\{ \sum_{j=1}^5 \left(
\frac{2a_{ij}(k,p)l_j^2}{\xi(m^2 \!+\! l_j^2)} \left[
\frac{\xi}{2  (\xi \!+\! 2) l_j^2  I(l_j)} \left(\frac{{\cal M}(l_j)}
{F(l_j)}
\!-\!m \right)
\!-\! \left( 1\!-\!\frac{1}{F(l_j)} \right) \right] \right)\\
&&+ \frac{2 a_{i6}(k,p)}{\xi\left[k^2A(p)-
p^2A(k)\right]}
\left[ \frac{1}{F(k)}-\frac{1}{F(p)} \right] \Bigg\} \;.
\label{npvertexangle}
\eea

By construction, in the weak coupling regime, this non-perturbative form of  the transverse vertex reduces to its corresponding Feynman expansion at the one loop level in an arbitrary covariant gauge and in all momentum regimes. Note that we have managed to write the transverse vertex solely as a function of the fermion propagator. Therefore, effectively, we have a WTI for this part of the vertex. We would like to emphasize that this is not a unique non-perturbative construction. However, it is probably the most natural and the simplest. A two loop calculation similar to the one presented in this chapter, and the LKF transformation for the vertex should serve as tests of Eq.~(\ref{npvertexangle}) or guides for improvement towards the hunt for the exact non-perturbative vertex. On practical side, the use of our perturbation theory motivated vertex in studies addressing important issues such as Dynamical Mass Generation for fundamental fermions should lead to more reliable results, attempting to preserve key fe
atures of gauge field theories, e.g., gauge independence of physical observables. A computational difficulty to use the above vertex  in such calculations could arise as the unknown functions $F$ and $\M$ depend on the angle between $k$ and $p$. This would make it impossible to carry out angular integration analytically in the SDE for the Fermion Propagator. This problem can be circumvented by defining an effective vertex which shifts the angular dependence from the unknown functions $F$ and ${\cal M}$ to the known basic functions of $k$ and $p$. This can be done by re-writing the perturbative results, Eq.~(\ref{taui}), as follows~:
\bea
\tau_i(k,p) &=& \alpha g_i \left[ b_{i1}(k,p) I(k) + b_{i2}(k,p) I(p)
   + \frac{a_{i6}(k,p)}{k^2 p^2}  \right]  \;,
\eea
where
\bea
b_{i1}(k,p)  &=& a_{i1}(k,p) \, \frac{I(l_1)}{I(l_3)} + a_{i3}(k,p)
+ \frac{1}{2} \, a_{i5}(k,p) \frac{I(l_5)}{I(l_3)} \;,    \\
 b_{i2}(k,p)  &=& a_{i2}(k,p) \, \frac{I(l_2)}{I(l_4)} + a_{i4}(k,p)
+ \frac{1}{2} \, a_{i5}(k,p) \frac{I(l_5^2)}{I(l_4)} \;.
\eea
This form can now be raised to a non-perturbative level exactly as before, with the only difference that the functions $F$ and ${\cal M}$ are independent of the angle between the momenta $k$ and $p$~:
\bea
\nn
\tau_i &=&
g_i\Bigg\{ \sum_{j=1}^2 \left(
\frac{2b_{ij}(k,p)\kappa_j^2}{\xi(m^2 + \kappa_{j}^2)} \left[
\frac{\xi}{2 (\xi + 2) \kappa_j^2 I(\kappa_j)  }
\left(\frac{{\cal M}(\kappa_j)}{F(\kappa_j)}-m \right)
\right.\right.\nn\\
&&\hspace{-9mm}
\left.- \left.\left( 1-\frac{1}{F(\kappa_j)} \right) \right] \right)
+ \frac{2  a_{i6}(k,p)}{\xi\left[k^2A(p)-
p^2A(k)\right]}
\left[ \frac{1}{F(k)}-\frac{1}{F(p)} \right] \Bigg\}  \;,
\eea
where $\kappa_1=k$ and $\kappa_2=p$.

\section{Comments on the $d-$dimensional Case}

\noindent
Following Davydychev \emph{et. al.}, \cite{davydychev}, two of the Master Integrals
needed to calculate are~:
\bea
I(\nu_1,\nu_2,\nu_3)&\equiv&\int\frac{d^dw}
{[(p-w)^2-m^2]^{\nu_1}[(k-w)^2-m^2]^{\nu_2}[w^2]^{\nu_3}},\\
{\cal I}(\nu_1,\nu_2,\nu_3)&\equiv&\int\frac{d^dw}
{[(p-w)^2]^{\nu_1}[(k-w)^2]^{\nu_2}[w^2-m^2]^{\nu_3}}\;.
\eea
In terms of these integrals, one can write out the Fermion Propagator and fermion-boson vertex in the context of QED in arbitrary dimensions.

\subsection{Fermion Propagator}
\noindent
Fermion Propagator in the $d$-dimensional case at the one-loop order can be written as~:
\bea
\frac{1}{F(p)}&=&\frac{e^2}{i(2\pi)^{d}}\frac{(d-2)\xi}{2p^2}
\left[(p^2+m^2){\cal I}(0,1,1)-{\cal I}(0,0,1)\right],\nn\\
\frac{{\cal M}(p)}{F(p)}&=&\frac{e^2}{i(2\pi)^{d}}m(d-1-\xi)
{\cal I}(0,1,1)\;.
\eea
The integral ${\cal I}(0,1,1)$ is the $d$-dimensional counterpart of our $I(y)$ function in the corresponding equation in three dimensions.

\subsection{The Transverse Vertex}
\noindent
The corresponding $d$-dimensional $\tau$'s of the transverse vertex can be expressed as,
\cite{davydychev}~:
\bea
\tau_i&=&\frac{e^2}{i(2\pi)^{d}}\Bigg\{t_{i,0}I(1,1,1)+t_{i,1}[-(k\cdot
q){\cal I}(0,1,1)+(p\cdot q){\cal I}(1,0,1)\nn\\
&+&q^2I(1,1,0)]
+t_{i,2}({\cal I}(0,1,1)+{\cal I}(1,0,1)-2I(1,1,0))\nn\\
&+&t_{i,3}\left({\cal I}(0,1,1)+
{\cal I}(1,0,1)-2\frac{{\cal I}(0,0,1)}{m^2}\right)\nn\\
&+&t_{i,4}({\cal I}(0,1,1)+{\cal I}(1,0,1))+t_{i,5}\frac{{\cal I}(0,1,1)
-{\cal I}(1,0,1)}
{k^2-p^2}\Bigg\}\;,
\eea
where the $t_{i,j}$, $j=1\ldots 5$ are basic functions of $k,p$ and $q$, listed in \cite{davydychev}. A straightforward procedure to obtain a non-perturbative vertex in this case is a natural thing to ask for. However, a dificulty arises since the function $I(1,1,1)$ in general cannot be rewritten as a linear combination of the integrals which appear in the Fermion Propagator in an obvious way. Note that in threee dimensions $I(1,1,1)$ has the same structure as the integrals which define the Fermion Propagator, though with more complicated arguments. Therefore, we have not yet been able to write a simple non-perturbative extension of the transverse vertex in arbitrary dimensions.

\section{Comments on the Two Loop Case}
\noindent
As we have mentioned earlier in this chapter, the three point vertex must reduce to its all order Feynman expansion in the weak coupling regime. What we have achieved in our work is the realization of this statement at the one loop level. To be able to do so beyond this order, we must know the perturbative vertex at the higher loops to have a guide for the construction of its non-perturbative counterpart. A good amount of progress has already been made in the calculation of necessary integrals involved. Most of these attempts surge from the precision analysis in particle physics phenomenology at the next-to-next-to-leading order calculations in perturbation theory. However, the scattering (see for example \cite{ME}) or decay processes (see for example \cite{FV}) do not in general require the calculation of vertex-type integrals with off-shell fermions and bosons. A complete two-loop calculation of the transverse vertex in arbitrary covariant gauge in arbitrary dimensions, to the best of out knowledge, still d
oes not exist for off-shell particles in the legs.

Using Dimensional regularization with $d=4-\epsilon$, the integrals appearing in the calculation of two-loop corrections take the generic form~\cite{2looprev}~:
\be
I(p_1,\ldots,p_n)=\int \frac{d^dw_1}{(2\pi)^d}\int\frac{d^dw_2}{(2\pi)^d} \frac{1}
{P_1^{m_1}\ldots P_t^{m_t}}S_1^{n_1}\ldots S_q^{n_q}\;,\label{2loopgen}
\ee
where $P_i$ are the propagators, depending on $w_1$, $w_2$ and the external momenta $p_1\ldots p_n$ while $S_i$ are scalar products of a loop momentum with an external momentum or of the two loop momenta.  The integral is specified by the powers $m_i$, $(m_i\ge 1)$ of all propagators $(P_1,\ldots,P_t)$ and by the section $(S_1,\ldots,S_q)$ of scalar products and their powers $(n_1\ldots
n_q)$, $(n_i\ge 0)$.

Progress in the calculation of two-loop corrections to four-point amplitudes was based on an efficient procedure to reduce the large number of different scalar integrals to a very limited number of so-called Master Integrals. The scalar integrals are related among each other by various identities. One class of identities are the so-called integration by parts (IBP) identities~:
\be
\int \frac{d^dw_1}{(2\pi)^ d}\int \frac{d^dw_2}{(2\pi)^d}
\frac{\partial}{\partial w_2^\mu}V(w,\ldots)=0\;,
\label{IBP}
\ee
where $V$ is any combination of propagators, scalar products and loop momentum vectors. $V$ can be a vector or tensor of any rank. IBP identities are a generalization of (\ref{IBP3d}) and they follow from the fact that the integral over the total derivative with respect to any loop
momentum vanishes.

Another class of identities is obtained from the fact that all integrals under consideration are Lorentz scalars, which are invariant under Lorentz transformations of the external momenta \cite{4ofrev}. These Lorentz invariance  (LI) identities are obtained from~:
\be
\left(p_1^\nu\frac{\partial}{\partial p_1^\mu}- p_1^\mu\frac{\partial}
{\partial p_1^\nu}+\ldots + p_n^\nu\frac{\partial}{\partial p_n^\mu}-
p_n^\mu\frac{\partial}{\partial p_n^\nu}\right)I(p_1,\ldots,p_n)=0\;.
\label{LI}
\ee

Master Integrals can be analytically expressed in series of the Dimensional Regulator parameter $\epsilon$ using Feynman or Schwinger parametrizations. However, sometimes using these parametric representations does not leave an integral that can be solved easily. Different techniques arose due to this situation, among others ~:

\begin{itemize}
\item The Mellin-Barnes method is based on the representation for a sum to some power, as a contour integral over a complex variable and the integration is then performed on straight contour lines parallel to the imaginary axis. After closing the contour, the result is the sum of all enclosed residues that might be expressed  as a hypergeometic series. For the two-loop boxes we have \cite{MBhistoricals}.

\item The Negative Dimensions technique consists in rewriting the integral over the parameters by introducing new ones through a multinomial expansion. Many conditions have to be satisfied among the parameters, which leads to the restriction that $d$ must be a negative integer. Some results are\cite{NDIM}.

\item Double-Integral representation, that consist on splitting the loop space into the parallel and perpendicular subspaces to the external momentum. Then, using cylindrical instead of spherical coordinates, one can calculate the real and the imaginary parts of the integrals separately, performing contour integrals. For some results, see \cite{DIR}.

\item Strategy of Expansion by regions \cite{hSER}, in which, instead of the integration over the whole space of loop momenta, the integration is performed only over some specific regions. The strategy consists in considering various regions of the loop momenta and expand, in every region, the integrand in Taylor series with respect to the parameters that are considered small in the
given region; integrate the integrand expanded in every region and; put to zero any scaleless integral.

\item Dispersion relations are used to calculate self-energies for the small external momentum behavior \cite{BBBB}.

\item Numerical strategies are used, where any loop integral is
stripped a\-na\-ly\-ti\-ca\-lly of its IR singularities so that the finite
integrals can be performed numerically.  \cite{me4}.

\item The analytic evaluation of MI can also be carried out without explicit integration over loop momenta by deriving differential equations for MI in internal propagator masses or external momenta. The equations can be solved with appropriate boundary conditions \cite{ME9}.

\end{itemize}

\subsection{Two-Loops Fermion Propagator}
\noindent
The two-loops contributions to the fermion propagator are depicted in the following Diagram~:
\vspace{-1.5cm}
\begin{center}
\begin{picture}(300,100)(0,0)
\SetScale{0.7}
%d-1
\ArrowLine(25,50)(150,50)
\PhotonArc(75,50)(25,180,0){-3}{5.5}
\PhotonArc(100,50)(25,0,180){-3}{5.5}
%d-2
\ArrowLine(175,50)(275,50)
\Vertex(200,50){1}
\Vertex(250,50){1}
\PhotonArc(225,50)(25,115,180){-3}{3.5}
\CArc(225,75)(10,0,360)
\PhotonArc(225,50)(25,0,70){-3}{3.5}
%d-3
\ArrowLine(300,50)(400,50)
\Vertex(325,50){1}
\Vertex(375,50){1}
\PhotonArc(350,50)(25,0,180){-3}{7.5}
\PhotonArc(350,50)(15,0,180){-3}{3.5}
\end{picture}\\
{\sl Diagram~9~: Two-Loop Corrections to the Fermion Propagator.}
\end{center}
Using the Double-Integral representation method, the Fermion Propagator has been calculated in \cite{DIR} for arbitrary masses.  Scalar self-energies have been calculated using dispersion relations \cite{BBBB}. Using the Pinch Technique \cite{BPPT}, off-shell self-energies have been calculated, in terms of Harmonic Polylogarithms. Also, an asymptotic expansion has been used to calculate the on-shell master diagram for the two-loop propagator\cite{CzS}. This expansion is taken in the limit of the external momenta, and therefore the fermion masses, tend to infinity. On-shell self-energies master integrals with one mass have  been calculated in \cite{FKK}. In the light-cone gauge, the fermion  propagator has been calculated in \cite{LW}.  Following the algorithm developed in \cite{Tar}, these corrections have been calculated in an arbitrary covariant gauge for off-shell massive fermions in arbitrary dimensions\cite{FJTV}.

\newpage
\subsection{Two-Loops Vertex}
\noindent
The two-loops corrections for the vertex are shown in Diagram~(10)~:
\vspace{-3cm}
\begin{center}
\begin{picture}(400,400)(0,0)
\SetScale{0.7}
%First Diagram
\Photon(100,350)(125,350){-3}{3}
\ArrowLine(75,375)(100,350)
\ArrowLine(75,325)(100,350)
\Vertex(75,375){1}
\Vertex(75,325){1}
\Photon(75,375)(75,325){-3}{3.5}
\ArrowLine(50,400)(75,375)
\ArrowLine(50,300)(75,325)
\Vertex(50,400){1}
\Vertex(50,300){1}
\Photon(50,400)(50,300){-3}{7.5}
\ArrowLine(25,425)(50,400)
\ArrowLine(25,275)(50,300)
\Text(75,275)[t]{(a)}
%Second Diagram
\Photon(250,350)(275,350){-3}{3}
\ArrowLine(225,375)(250,350)
\ArrowLine(225,325)(250,350)
\Vertex(225,375){1}
\Vertex(225,325){1}
\ArrowLine(200,400)(225,375)
\ArrowLine(200,300)(225,325)
\Vertex(200,400){1}
\Vertex(200,300){1}
\Photon(225,375)(200,300){-3}{7.5}
\Photon(225,325)(200,400){-3}{7.5}
\ArrowLine(175,425)(200,400)
\ArrowLine(175,275)(200,300)
\Text(200,275)[t]{(b)}
%Third Diagram
\Photon(400,350)(425,350){-3}{3}
\ArrowLine(375,325)(400,350)
\Photon(350,400)(350,300){-3}{5.5}
\Vertex(375,375){1}
\Vertex(375,325){1}
\ArrowLine(350,400)(400,350)
\ArrowLine(350,300)(375,325)
\Vertex(350,400){1}
\Vertex(350,300){1}
\ArrowLine(325,425)(350,400)
\ArrowLine(325,275)(350,300)
\PhotonArc(350,300)(25,225,42){-3}{5.5}
\Text(300,275)[t]{(c)}
%fourth diagram
\Photon(200,150)(225,150){-3}{3}
%\ArrowLine(150,200)(200,150)
\ArrowLine(150,200)(168,182)
\ArrowLine(168,182)(180,170)
\ArrowLine(180,170)(200,150)
\ArrowLine(150,100)(200,150)
\Vertex(150,200){1}
\Vertex(150,100){1}
\Photon(150,200)(150,100){-3}{7.5}
\ArrowLine(125,225)(150,200)
\ArrowLine(125,75)(150,100)
\PhotonArc(175,175)(10,140,320){-2}{3.5}
\Text(175,75)[t]{(d)}
%fifth diagram
\Photon(375,150)(400,150){-3}{3}
\ArrowLine(325,200)(375,150)
\ArrowLine(325,100)(375,150)
\Vertex(325,200){1}
\Vertex(325,100){1}
\ArrowLine(300,225)(325,200)
\ArrowLine(300,75)(325,100)
\ArrowArc(325,150)(20,270,90)
\ArrowArc(325,150)(20,90,270)
\Photon(325,200)(325,170){-3}{3.5}
\Photon(325,100)(325,130){-3}{3.5}
\Text(300,75)[t]{(e)}
\end{picture}\\
\vspace{-30pt}
{\sl Diagram~10~: Two-Loop Corrections to the Vertex.}
\end{center}

Some partial calculations to the full vertex using the double
integral re\-pre\-sen\-ta\-tion, planar vertex in four dimensions has
been presented in \cite{DIR}.  In Ref. \cite{BMR} a recent
on-shell calculation for the vertex has been done at arbitrary
momentum transfer in arbitrary dimensions.  The planar vertex
((a) in Figure 5) with essential on shell singularities has been
calculated in arbitrary dimensions \cite{DS} applying the
strategy of expansion by regions.  The calculation of a the
vertex function at zero momentum transfere and with space-like
external momentum \cite{Cz}.  Vertex with two legs on-shell in
arbitrary dimensions has been calculated in \cite{SS} using the
Negative Dimensions technique.  Numerically speaking, the
massless vertex has been calculated in the Landau gauge
\cite{CS}.  For non-planar vertex diagrams ((b) in the
corresponding figure), a numerical method is propossed in
\cite{FSKK}.  In the Sudakov limit, the Vertex has been
calculated using the Mellin-Barnes method in arbitrary dimensions
\cite{Sudakov}.  A full off shell calculation of the two-loops
vertex has not been carried out yet.  A programme towards the
construction of the full two-loop vertex has to take into account
the transverse part of it, so that one can check of modify the
co\-rres\-pon\-ding basis $\{T_i^\mu\}$.  With this basis one can look
for a nonperturbative vertex in arbitrary dimensions, or, at
least, in our three dimensional case.

 Once we are able to construct a three-point vertex which agrees with its perturbative expansion in the weak coupling regime to all orders, we shall rest assured that all the gauge identies are
satisfied at every order of perturbation theory. We can then hope that the physical observables associated with non-perturbative physics shall automatically be gauge independent.

\chapter{Discussion and Conclusions}
\pagestyle{myheadings}
\markboth{Discussion and Conclusions}{Discussion and Conclusions}
%{Gauge Invariance and Construction of the Fermion-Boson Vertex in QED3}
%{Discussion and %Conclusions}

In this thesis we have focused our atention in incorporating
gauge identities (Ward-Green-Takahashi identities and Landau-Khalatnikov-Fradkin transformations) to the non perturbative study of the Schwinger-Dyson Equations in QED3. We know that Perturbation Theory is the only scheme where in each order of approximation, such identities are satisfied, and also it verified that, indeed, physical observable quantities are independent of the gauge parameter. However, in a non perturbative analysis of the Schwinger-Dyson Equations it has been impossible the conjunction of these two facts.

A trace back for the origin of the gauge dependence points out to the following problems~:
\begin{enumerate}
\item Violaton of the Ward-Green-Takahashi Identity.
\item Incorrect gauge behavior of the Green's functions.
\item Use of wrong technical tools, like gauge dependent regularization schemes.
\end{enumerate}

In order to avoid that the technical details overlap the pure theoretical issues of the study, we choose QED3, because in a theory that lacks of ultraviolet divergences. Therefore, the source for the gauge dependence should be only due to the first two points above mentioned.

In Chapter~3 we considered the simplest scenario, and we calculated the mass and the chiral condensate by solving the Schwinger-Dyson Equation for the Fermion Propagator. We explicitly observed that these quantities depend upon the gauge parameter. Since an incorrect regularization cannot induce such dependence, the only room for improvement is in the three-point vertex ansatz.

An ansatz should be such that~:
\begin{itemize}
\item It satisfies the Ward-Green-Takahashi Identity,
\item The Fermion Propagator obtained from this ansatz and the vertex itself should satisfy their corresponding Landau-Khalatnikov-Fradkin transformation.
\end{itemize}

The incorporation of the Ward-Green-Takahashi Identity is straightforward. However, as we did in Chapter~4, even in the simplest case, the Landau-Khalatnikov-Fradkin transformation for the Fermion Propagator leads to a gauge behavior extremely complex for this Green's function, and in spite of this, we were able to extact valuable information about its structure beyond the tree level. The search for a non perturbative vertex which leads to this gauge bahavior for the Fermion Propagator is not at all a trivial task. It is here that Perturbation Theory plays its role. If we can put forward to a non perturbative ansatz which reduces to its Feynman expansion to all orders in the weak coupling regime, we can hope that it automatically incorporates the Landau-Khalatnikov-Fradkin transformations not only for the Fermion Propagator, but for the Vertex itself.

Therefore, in Chapter~5 we calculated the Fermion-Boson Vertex at
one-loop level outside of the quiral phase of QED3, previously
studied.  We decompose this vertex into its longitudinal and
transverse parts.  The former guarantees the validity of the
Ward-Green-Takahashi Identity, since it is related to the Fermion
Propagator calculated at the same order of approximation, while
the later ensures that the Landau-Khalatnikov-Fradkin
transformations are indeed satisfied.  Taking advantage of the
lack of additional constraints, we exploited the perturbative
form of the Fermion Propagator in such a way that we could write
an effective Ward-Green-Takahashi Identity for the transverse
vertex.  This identity allows us to have the first insight on the
non perturbative structure of the Fermion-Boson interaction,
since the transverse vertex, related in this way to the Fermion
Propagator does not depend explicitly on the electromagnetic
coupling.  We write, afterwards, the transverse vertex in a
convenient form that will allow, to corroborate its non
perturbative structure in a future work at the two-loop level.

 The systematic construction of the three point vertex guarantees
that~:

\begin{itemize}

\item The Ward-Green-Takahash identity, which relates the Fermion
Propagator to the Fermion-Boson vertex, is satisfied
non-perturbatively.

\item Both the vertex and the resulting Fermion Propagator satisfy
Landau-Khaltnikov-Fradkin transformations to ${\cal O}(\alpha)$
and ${\cal O}(\alpha^2)$ respectively.

\item The vertex does not contain any kinematic singularities when
$k^2 \rightarrow p^2$.

\item
     Most importantly, the vertex reduces to its correct perturbative
expansion at ${\cal O}(\alpha)$ in the weak coupling regime in an
arbitrary covariant gauge and for all momentum regimes.

\item The vertex has correct symmetry under the parity, charge
congugation
and time reversal operations.

\end{itemize}

This will allow that the technique as well as the reasoning used
in our cons\-truc\-tion could be applied in more complicated theories
like QCD in a suitable and realistic way.  Finally, it is also
possible to incorporate our ideas in some alternative scheme to
the Standard Model which pretends to give a solution to the
problem of the origin of the masses.  We extend the invitation to
whom whishes to do so.

\newpage

\pagestyle{myheadings}


\markboth{Bibliography}{Bibliography}
\begin{thebibliography}{555}
%primer cap\'{\i}tulo
\bibitem{ht4d1}\textsf{D. J. Gross, R. D., Pisarski, and L. G.Yaffe},  Rev.
Mod. Phys. {\bf 53} 43 (1981).
%
\bibitem{ht4d2}\textsf{R. Jackwin}.  Procc. \emph{ Artic School of
Physics} (1982).
%
\bibitem{review} \textsf{C.D. Roberts and A.G. Williams}, Prog. Part. Nucl. Phys.
{\bf 33} 477 (1994).
%
\bibitem{finap} \textsf{G. Scharf},  \emph{Finite Quantum Electrodynamics.
The Causal Approach.} Springer. Germany. (1995).
%
\bibitem{maris}\textsf{P. Maris}.  \emph{Nonperturbative Analysis of the
Fermion Propagator~:  Complex Singularities and Dynamical Mass
Generation}. Ph. D. Thesis. Neatherlands. (1992).
%
\bibitem{mavro} \textsf{N. Dorey and N. E. Mavromatos}, Nucl.  Phys.  {\bf
B386} (1992).
%
\bibitem{BR1} \textsf{C.J. Burden and C.D. Roberts}, Phys. Rev. {\bf D44} 540 (1991).
%
\bibitem{ryder} \textsf{L. H. Ryder},  \emph{Quantum Field Theory}.
Cambridge University Press. England. (1985).
%
\bibitem{hanmar} \textsf{F. Hanzel and A. D. Martin},  \emph{Quarks and
Leptons~:  An Introductory Course in Modern Particle Phisics.}
John Wiley \& Sons. Canada. (1984).
%
\bibitem{muta} \textsf{T. Muta}, \emph{Foundations of Quantum
Chromodynamics.}  World Scien\-ti\-fic. Singapour. (2000).
%
\bibitem{applequist} \textsf{T. W. Applequist, M. Bowick, D. Karabali and
L. C. R. Wijewardhana}, Phys. Rev. {\bf D33} 3704 (1986).
%
\bibitem{unquenched1} \textsf{R.D. Pisarski}, Phys. Rev. {\bf D29} 2423 (1984).
%
\bibitem{japon} K. \textsf{Shimizu}, Prog.  Theor.  Phys.  {\bf 74} 610
(1985).
%%%%%%%%%%%%%%%
%segundo cap\'itulo
%%%%%%%%%%%%%%%
\bibitem{esd} \textsf{J. S. Schwinger},  Proc.  Nat. Acad.  Sc. {\bf 37}
452 (1951). \textsf{F. J. Dyson},  Phys.  Rev. {\bf 75} 1736 (1949).
%
\bibitem{cdrnotes} \textsf{C. D. Roberts},  \emph{Notes on Quantum Field
Theory.} Rostok University, Germany,
given to the author of the thesis in USA (2002).
%
\bibitem{axodraw} \textsf{J. A. M. Vermaseren}, Comp.  Phys.  Comm.  {\bf
83} 45 (1994).
%
\bibitem{quenched1} \textsf{Y. Hoshino and T. Matsuyama}, Phys. Lett. {\bf B222}
493 (1989).
%
\bibitem{Dong} \textsf{Z. Dong, H.J. Munczek and C.D. Roberts}, Phys. Lett.
{\bf B333} 536 (1994).
%
\bibitem{BC} \textsf{J.S. Ball and T.-W. Chiu}, Phys. Rev. {\bf D22} 2542
(1980).
%
\bibitem{pen1enn} \textsf{M. R. Pennington and S. P. Webb}, Brookhaven Nat.
Lab. preprint, BNL-40886, (1988).
%
\bibitem{atk1enn} \textsf{D. Atkinson, P, Johnson and M. R.
Pennington},
Brookhaven Nat. Lab. preprint BNL-41615, (1988).
%
\bibitem{adnan1} \textsf{A. Bashir}, {\em ``Perturbation
Theory Constraints on
the 3-Point Vertex in massless QED3''}
Procc. Workshop on Light-Cone QCD
and Non Perturbative Hadron
Physics, World Scientific, University of Adelaide, Adelaide, Australia,
(227-232) 2000.
%
\bibitem{adnan2} \textsf{A. Bashir}, Phys. Lett. {\bf B491} 280 (2000).
%
\bibitem{BHR} \textsf{A. Bashir, A. Huet and A. Raya}, Phys.  Rev.
{\bf D66} 025029 (2002).
%
\bibitem{Salam1} \textsf{A. Salam}, Phys. Rev. {\bf 130} 1287 (1963).
%
\bibitem{SD1} \textsf{A. Salam and R. Delbourgo}, Phys. Rev. {\bf 135} 1398 (1964).
%
\bibitem{S1} \textsf{J. Strathdee}, Phys. Rev. {\bf 135} 1428 (1964).
%
\bibitem{DW1} \textsf{R. Delbourgo and P. West}, J. Phys. {\bf A10} 1049 (1977).
%
\bibitem{DW2} \textsf{R. Delbourgo and P. West}, Phys. Lett. {\bf B72} 96 (1977).
%
\bibitem{D1} \textsf{R. Delbourgo}, Nuovo Cimento {\bf A49} 484 (1979).
%
\bibitem{YH} \textsf{Y. Hoshino}, \emph{``The Gauge Technique in
$QED_{(2+1)}$''}
hep-th/0107219;
\emph{``A Gauge Covariant Approximation to QED"} hep-th/0202020.
%
\bibitem{Keck1} \textsf{R. Delbourgo and B.W. Keck}, J. Phys. {\bf G6} 275 (1980).
%
\bibitem{D2} \textsf{R. Delbourgo}, Austral. J. Phys. {\bf 52} 681 (1999).
%
\bibitem{W1} \textsf{J.C. Ward}, Phys. Rev. {\bf 78} (1950).
%
\bibitem{G1} \textsf{H.S. Green}, Proc. Phys. Soc. (London) {\bf A66} 873 (1953).
%
\bibitem{T1} \textsf{Y. Takahashi}, Nuovo Cimento {\bf 6} 371 (1957).
%
\bibitem{CP1} \textsf{D.C. Curtis and M.R. Pennington}, Phys. Rev. {\bf D42} 4165
(1990).
%
\bibitem{CP2} \textsf{D.C. Curtis and M.R. Pennington}, Phys. Rev. {\bf D44} 536
(1991).
%
\bibitem{CP3} \textsf{D.C. Curtis and M.R. Pennington}, Phys. Rev. {\bf D48} 4933
(1993).
%
\bibitem{ABGPR1} \textsf{D. Atkinson, J.C.R. Bloch, V.P. Gusynin, M.R. Pennington
and M. Reenders}, Phys. Lett. {\bf B329} 117 (1994).
%
\bibitem{AGM1} \textsf{D. Atkinson, V.P. Gusynin and P. Maris}, Phys. Lett. {\bf B303}
157 (1993).
%
\bibitem{BP1} \textsf{A. Bashir and M.R. Pennington}, Phys. Rev. {\bf D50} 7679 (1994).
%
\bibitem{BP2} \textsf{A. Bashir and M.R. Pennington}, Phys. Rev. {\bf D53} 4694 (1996).
%
\bibitem{LK1} \textsf{L.D. Landau and I.M. Khalatnikov}, Zh. Eksp. Teor. Fiz. {\bf 29}
89 (1956).
%
\bibitem{LK2} \textsf{L.D. Landau and I.M. Khalatnikov}, Sov. Phys. JETP {\bf 2} 69
(1956).
%
\bibitem{F1} \textsf{E.S. Fradkin}, Sov. Phys. JETP {\bf 2} 361 (1956).
%
\bibitem{JZ1} \textsf{K. Johnson and B. Zumino}, Phys. Rev. Lett. {\bf 3} 351 (1959).
%
\bibitem{Z1} \textsf{B. Zumino}, J. Math. Phys. {\bf 1} 1 (1960).
%
\bibitem{Fukuda} \textsf{T. Fukuda, R. Kubo and K. Yokoyama},
Prog. Theor. Phys. {\bf 63} 1384 (1980).
%%%%%%%%%%%%%%%%%%%%%%%%%%%%%%%%%%%%%%%%%
% cap\'itulo tercero
%%%%%%%%%%%%%%%%%%%%%%%%%%%%
\bibitem{GSSW} \textsf{V.P. Gusynin, A.W. Schreiber, T. Sizer and
A.G. Williams},
Phys. Rev. {\bf D60} 065007 (1999).
%
\bibitem{SSW} \textsf{A.W. Schreiber, T. Sizer and A.G. Williams}, Phys. Rev.
{\bf D58} 125014 (1998).
%
\bibitem{KSW} \textsf{A. K{\i}z{\i}lers\"{u}, A.W. Schreiber and
A.G. Williams},
Phys. Lett. {\bf B499} 261 (2001).
%
%\bibitem{BKP1} A. Bashir, A. K{\i}z{\i}lers\"{u} and M.R. Pennington,
%Phys. Rev. {\bf D57} 1242 (1998).
%\bibitem{AB1} A. Bashir, Phys. Lett. {\bf B491} 280 (2000).
\bibitem{Politzer} \textsf{H.D. Politzer}, Nucl. Phys. {\bf 117} 397 (1976).
%
\bibitem{Atkinson} \textsf{D. Atkinson, J.C.R. Bloch, V.P. Gusynin, M.R. Pennington
and M. Reenders}, Phys. Lett. {\bf B329} 117 (1994).
%
\bibitem{GR} \textsf{I.S. Gradshteyn and I.M. Ryzhik}, {\em Table of Integrals,
Series and Pro\-ducts.} Academic Press. USA. (2000).
%
\bibitem{BT1} \textsf{C.J. Burden and P.C. Tjiang}, Phys. Rev. {\bf D58} 085019 (1998).
%
\bibitem{BKP2} \textsf{A. Bashir, A. K{\i}z{\i}lers\"{u} and M.R.
Pennington},
Phys. Rev. {\bf D62} 085002 (2000).
%
\bibitem{BKP1} \textsf{A. Bashir, A. K{\i}z{\i}lers\"{u} and M.R.
Pennington},
Phys. Rev. {\bf D57} 1242 (1998).
%
%%%%%%%%%%%%%%%%%%%%%%%%%%%%%%%%%%%%
%LKF
%%%%%%%%%%%%%%%%%%%%%%%%%%%%%%%%%%%
\bibitem{Nielsen} \textsf{N.K. Nielsen}, Nucl. Phys. {\bf B101} 173 (1975).
%
\bibitem{Sibold} \textsf{O. Piguet and K. Sibold}, Nucl. Phys. {\bf B253} 517 (1985).
%
\bibitem{Steele} \textsf{J.C. Breckenridge, M.J. Lavelle and T.G.
Steele},
Z. Phys. {\bf C65} 155 (1995).
%
\bibitem{Grassi} \textsf{P. Gambino and P.A. Grassi}, Phys. Rev. {\bf D62}
076002 (2000).
%
\bibitem{H1} \textsf{B. Haeri}, Phys. Rev. {\bf D43} 2701 (1991).
%
\bibitem{Keck2} \textsf{R. Delbourgo and B.W. Keck}, J. Phys. {\bf A13} 701 (1980).
%
\bibitem{Keck3} \textsf{R. Delbourgo, B.W. Keck and C.N. Parker}, J. Phys. {\bf A14}
921 (1981).
%
\bibitem{Waites1} \textsf{A.B. Waites and R. Delbourgo}, Int. J. Mod. Phys. {\bf A7}
6857 (1992).
%
\bibitem{davydychev} \textsf{A.I. Davydychev, P. Osland and L. Saks},
Phys. Rev. {\bf D63} 014022 (2001).
%\bibitem{Ross} E.G. Floratos, D.A. Ross and C.T. Sachrajda,
%Nucl. Phys. {\bf B129} 66 (1977).

\bibitem{BKP3} \textsf{A. Bashir,
A. K{\i}z{\i}lers\"{u} and M.R. Pennington},
{\em ``Analytic Form of the One Loop Vertex y the Two
Loop Fermion Propagator in 3-Dimensional Massless QED''}
preprint no. ADP-99-8/T353 University of Adelaide, preprint
no. DTP-99/76  University of  Durham, hep-ph/9907418.
%
\bibitem{Moch} \textsf{S. Moch, P. Uwer and S. Weinzierl}, J. Math. Phys. {\bf
43} 3363 (2002).
%
\bibitem{Reenders} \textsf{A. K{\i}z{\i}lers\"{u}, M. Reenders and
M.R. Pennington}, Phys. Rev. {\bf D52} 1242 (1995).
%
\bibitem{BashirR1} \textsf{A. Bashir and A. Raya}, Phys. Rev. {\bf D64} 105001 (2001).
%
\bibitem{Zsums1} \textsf{L. Euler}, Novi Comm. Acad. Sci. Petropol. {\bf 20}
140 (1775).
%
\bibitem{Zsums2} \textsf{D. Zagier}, First European Congress of Mathematics,
Vol. II, Bir\-khau\-ser, Boston, 497 (1994).
%
\bibitem{BaRa4} \textsf{A. Raya}, Talk given at the XVII Annual Meeting of the Division of Particles and Fields of the Mexican Physics Societi (2003).
%
\bibitem{moretables} M. Abramowitz and I.A. Stegun,
{\em Handbook of Mathematical Functions.} Dover Publications.
USA. (1972).
%%%%%%%%%%%%%%%%%%%%%%%%%%%%%%
%Cap\'itulo~5
%%%%%%%%%%%%%%%%%%%%%%%%%%%%%%
\bibitem{unquenched3} \textsf{T. Applequist and D. Nash}, Phys. Rev. Lett. {\bf 60}
2575 (1988).
%
\bibitem{unquenched4} \textsf{M.R. Pennington and D. Walsh}, Phys. Lett. {\bf B253}
246 (1991).
%
\bibitem{unquenched5} \textsf{D.C. Curtis, M.R. Pennington and D.
Walsh}, Phys.  Lett. {\bf B295} 313 (1992).
%
\bibitem{unquenched6} \textsf{P. Maris}, Phys. Rev. {\bf D54} 4049 (1996).
%
\bibitem{GHR1} \textsf{V. P. Gusynin, A. H. Hams and M. Reenders}, Phys.
Rev.  {\bf D53}, 2227 (1996).
%%
\bibitem{GMS} \textsf{V. P. Gusynin, V. A. Miranski and A. V.
Shpagin},
Phys. Rev.  {\bf D58} 085023 (1998).
%
\bibitem{GHR2}\textsf{V. P. Gusynin, A. H. Hams and M. Reenders}, Phys.
Rev.  {\bf D63} 045025 (2001).
%
%
\bibitem{raya1} \textsf{A. Raya}, Talk given at the IX Universitary Meeting of Scientific, Technological and Humanistic Research of the
Universidad Michoacana de San Nicol\'as de
Hidalgo. Morelia, M\'exico. (1999).
%
%\bibitem{trans} C.J. Burden and C.D. Roberts, Phys. Rev. {\bf D47} 5581 (1993).
%
\bibitem{Wu} \textsf{H. Cheng and T. T. Wu}, \emph{Expanding Protons: Scattering at High
Energies.} MIT Press. USA. (1987); \textsf{K. S. Bjorkevoll, G
F\"aldt y P. Oslan}, Nucl. Phys. {\bf B386} 303 (1992).
%
%2-loop
\bibitem{ME} \textsf{M. E. Tejeda-Yeomans}, Talk given at the X Mexican School od Particles and Fields, M'exico (2002).
%
\bibitem{FV} \textsf{J. Fleischer and O. L. Veretin}, RADCOR98,
Barcelona, Spain (1988).
%
\bibitem{2looprev} \textsf{T. Gehrmann and E. Remddi}, RADCOR2000,
USA (2000).
%
\bibitem{4ofrev} \textsf{T. Gehrmann and E. Remiddi}, Nucl. Phys. {\bf B580}
485 (2000).
%
\bibitem{MBhistoricals} \textsf{V. A. Smirnov}, Phys. Lett. {\bf B460} 397
(1999); \textsf{J. B. Tausk}, Nucl. Phys. {\bf B580} 577 (2000);
\textsf{C. Anastasiou, J. B. Tausk, and M. E. Tejeda-Yeomans},
Nucl.  Phys.  Proc.  Suppl.  {\bf89} 262 (2000);
\textsf{V. A. Smirnov}, Phys. Lett. {\bf B495} 130 (2000); \textsf{V.
A. Smirnov}, Phys. Lett. {\bf B500} 330 (2001).
%
\bibitem{NDIM} \textsf{C. Anastasiou, E. Glover, and C. Oleari}, Nucl. Phys. {\bf
B572} 307 (2000).
%
\bibitem{DIR} \textsf{A. Czarnecki, U. Kilian, and D. Kreimer}, Nucl. Phys. {\bf
B433} 259 (1995).
%
\bibitem{hSER}\textsf{M. Beneke and V. A. Smirnov}, Nucl. Phys. {\bf B522} 321
(1998); V. A. Smirnov, Phys. Lett. {\bf B465} 226 (1999).
%
\bibitem{BBBB} \textsf{S. Bauberger, F. A. Berends, M. B\"om, and
M. Buza}, Nucl. Phys. {\bf B434} 383 (1995).
%
\bibitem{me4} \textsf{T. Binoth and G. Heinrich}, Nucl. Phys. {\bf B585} 741
(2000).
%
\bibitem{ME9} \textsf{T. Gehrmann and E. Remiddi}, Nucl. Phys. {\bf B580} 485
(2000);
\textsf{T. Gehrmann and E. Remiddi}, Nucl. Phys. {\bf B601} 248 (2001);
\textsf{T. Gehrmann and E. Remiddi}, Nucl. Phys. {\bf B601} 287 (2001).
%
\bibitem{BPPT} \textsf{D. Binosi and J. Papavassiliou}, Phys. Rev. {\bf D65}
085003 (2002).
%
\bibitem{CzS}  \textsf{A. Czarnecki and V. A. Smirnov}, Phys. Lett. {\bf B394}
211 (1997).
%
\bibitem{FKK} \textsf{J. Fleischer, M. Yu. Kalmykov, and A. V.
Kotikov}, Phys. Lett. {\bf  B462} 169 (1999).
%
\bibitem{LW} \textsf{G. Leibbrandt and J. D. Williams}, Nucl. Phys. {\bf  B566}
373 (2000).
%
\bibitem{Tar} \textsf{O. V. Tarasov}, Nucl. Phys. {\bf B502} 455 (1997).
%
\bibitem{FJTV} \textsf{J. Fleischer, F. Jegerlehner, O. V.
Tarasov, and O. L. Veretin},
Nucl. Phys. {\bf B539} 671 (1999);
\textsf{J. Fleischer, F. Jegerlehner, O. V.
Tarasov, and O. L. Veretin}, Nucl. Phys. {\bf B571} 511 (2000).
%
\bibitem{BMR} \textsf{R. Bonciani, P. Mastrolia, and E. Remiddi}, hep-ph/031170.
%
\bibitem{DS} \textsf{A. I. Davydychev and V. A. Smirnov}, Procc. of
ACAT2002, Russia. (2002).
%
\bibitem{Cz}\textsf{ A. Czarnecki}, hep-ph/9410332.
%
\bibitem{SS} \textsf{A. T.  Suzuki and A. G. M. Schmidt}, Phys. Rev. {\bf D58},
047701 (1998).
%
\bibitem{CS} \textsf{K. G. Chetyrkin and T. Seidensticker}, Phys. Lett. {\bf
B495} 747 (2000).
%
\bibitem{FSKK} \textsf{J. Fujimoto, Y. Shimizu, K. Kato, and T.
Kaneko}, Int. J. Mod. Phys. {\bf  C6} 255 (1995).
%
\bibitem{Sudakov} \textsf{V. A Smirnov and Rakhmetov}, Theor.
Math.  Phys.  {\bf 120} 870 (1999); Teor.  Mat. Fiz. {\bf 120} 64
(1999).

\end{thebibliography}
\end{document}